\documentclass[11pt,a4paper,nofootinbib,showpacs,preprint,utf8]{article}
\usepackage{jheppub}
\usepackage[english]{babel}
\usepackage{amsmath,amssymb,amsfonts, bm,bbm,slashed, subdepth}
\usepackage{graphicx}
\usepackage{enumerate}
\usepackage{setspace}
\usepackage{booktabs, tabularx}
\usepackage{units}
\usepackage{color}
\usepackage{float}
\usepackage{multirow}
\usepackage[dvipsnames]{xcolor}
\usepackage[pscoord]{eso-pic}
\usepackage[normalem]{ulem}
\usepackage{import}
\usepackage{url}
\usepackage{tikz}
\usetikzlibrary{snakes}
\usepackage{cancel}
\allowdisplaybreaks[1]
\usepackage{scalerel}
\usepackage{soul}
\usepackage{hyperref}
\usepackage{braket}
\usepackage{subcaption}
\hypersetup{
	setpagesize=false,
	bookmarksnumbered=true,
	colorlinks=true,
	linkcolor=blue,
	citecolor=red,
	hypertexnames=true
}
\usepackage{spverbatim}
\usepackage{upgreek}
\usepackage{subcaption}

\usepackage{tcolorbox}

\makeindex
\def\ba{\begin{eqnarray}}
\def\ea{\end{eqnarray}}
\def\br{\begin{array}}
\def\er{\end{array}}
\def\be{\begin{equation}}
\def\ee{\end{equation}}
\title{Minimal Dirac seesaw dark matter}
\author[a]{{Zafri Ahmed Borboruah}}
\author[b,c]{{,~Debasish Borah}}
\author[a]{{,~Lekhika Malhotra}}
\author[d]{{,~Utkarsh Patel}}

\affiliation[a]{Department of Physics, Indian Institute of Technology Bombay, Mumbai 400076, India}
\affiliation[b]{Department of Physics, Indian Institute of Technology Guwahati, Assam 781039, India}
\affiliation[c]{Pittsburgh Particle Physics, Astrophysics, and Cosmology Center, Department of Physics and Astronomy, University of Pittsburgh, Pittsburgh, PA 15260, USA}
\affiliation[d]{Department of Physics, Indian Institute of Technology Bhilai, Durg 491002, India}
\emailAdd{zafri123@iitb.ac.in}
\emailAdd{dborah@iitg.ac.in}
\emailAdd{lekhika.malhotra@iitb.ac.in}
\emailAdd{utkarshp@iitbhilai.ac.in}

\abstract{We propose a minimal Type-I Dirac seesaw which accommodates a thermal scalar dark matter (DM) candidate protected by a charge conjugation symmetry in dark sector $C_{\rm dark}$, without introducing any additional field beyond the ones taking part in the seesaw. A $Z_4$ symmetry is introduced to realise the tree level Dirac seesaw while the Majorana mass terms are prevented by an unbroken global lepton number symmetry. While the spontaneous $Z_4$ breaking together with electroweak symmetry breaking lead to the generation of light Dirac neutrino mass, it also results in the formation of domain walls. These cosmologically catastrophic walls can be made to annihilate away by introducing bias terms while also generating stochastic gravitational waves (GW) within reach of near future experiments like \texttt{LISA}, \texttt{BBO}, $\mu$-\texttt{ARES} etc. The scalar DM parameter space can be probed at direct and indirect search experiments. Light Dirac neutrinos also enhance the relativistic degrees of freedom $N_{\rm eff}$ within reach of future cosmic microwave background (CMB) experiments. The model can also explain the observed baryon asymmetry via Dirac leptogenesis.}

\makeatletter
\gdef\@fpheader{}
\makeatother

\begin{document}
\maketitle
\flushbottom

\section{Introduction}
The observation of non-zero neutrino mass and mixing \cite{ParticleDataGroup:2024cfk} has been one of the most crucial evidences for beyond standard model (BSM) physics. While the neutrino oscillation experiments have measured two mass squared differences and three mixing angles to a great accuracy, we still do not know whether neutrinos are Dirac or Majorana fermions. Although a positive signal at neutrinoless double beta decay (see \cite{Dolinski:2019nrj} for a review) experiments could have validated the Majorana nature, continued null results at such experiments (see \cite{Majorana:2022udl}, for example) have prompted the study of Dirac nature of light neutrinos. The canonical seesaw models for Majorana neutrino mass \cite{Minkowski:1977sc, Gell-Mann:1979vob, Mohapatra:1979ia,Sawada:1979dis,Yanagida:1980xy, Schechter:1980gr, Mohapatra:1980yp, Schechter:1981cv, Wetterich:1981bx, Lazarides:1980nt, Brahmachari:1997cq, Foot:1988aq} have been suitably extended to Dirac seesaw equivalents in several existing works in the literature \cite{Roncadelli:1983ty, Roy:1983be, Babu:1988yq,
	Peltoniemi:1992ss, Chulia:2016ngi, Aranda:2013gga, 
	Chen:2015jta, Ma:2015mjd, Reig:2016ewy, Wang:2016lve, Wang:2017mcy, Wang:2006jy, 
	Gabriel:2006ns, Davidson:2009ha, Davidson:2010sf, Bonilla:2016zef, 
	Farzan:2012sa, Bonilla:2016diq, Ma:2016mwh, Ma:2017kgb, Borah:2016lrl, 
	Borah:2016zbd, Borah:2016hqn, Borah:2017leo, CentellesChulia:2017koy, 
	Bonilla:2017ekt, Memenga:2013vc, Borah:2017dmk, CentellesChulia:2018gwr, 
	CentellesChulia:2018bkz, Han:2018zcn, Borah:2018gjk, Borah:2018nvu, Borah:2019bdi, CentellesChulia:2019xky, Jana:2019mgj, Nanda:2019nqy, Guo:2020qin, Bernal:2021ezl, Borah:2022obi, Li:2022chc, Dey:2024ctx, Singh:2024imk, Borah:2024gql}. While such Dirac seesaw models can be falsified by a positive signal at neutrinoless double beta decay experiments, such models often come with their own predictions such as enhanced effective relativistic degrees of freedom $N_{\rm eff}$ \cite{Abazajian:2019oqj, FileviezPerez:2019cyn, Nanda:2019nqy, Han:2020oet, Luo:2020sho, Borah:2020boy, Adshead:2020ekg, Luo:2020fdt, Mahanta:2021plx, Du:2021idh, Biswas:2021kio, Borah:2022obi, Borah:2022qln, Li:2022yna, Biswas:2022fga, Adshead:2022ovo, Borah:2023dhk, Borah:2022enh, Das:2023oph, Esseili:2023ldf, Das:2023yhv, Borah:2024twm} within reach of future CMB experiments, stochastic gravitational waves (GW) \cite{Barman:2022yos, Barman:2023fad} etc. Although lepton number is conserved in typical Dirac seesaw scenarios, it is still possible to realise baryogenesis via leptogenesis \cite{Fukugita:1986hr} in such models, often referred to as Dirac leptogenesis \cite{Dick:1999je, Murayama:2002je, Boz:2004ga,Thomas:2005rs,Gu:2006dc,Bechinger:2009qk,Chen:2011sb,Choi:2012ba,Borah:2016zbd,Gu:2016hxh,Narendra:2017uxl}. This makes Dirac seesaw an attractive alternative to canonical seesaw models for Majorana neutrino masses as it can solve the origin of neutrino mass and baryon asymmetry problems simultaneously while offering experimentally verifiable predictions.

In addition to the origin of neutrino mass and baryon asymmetry which the standard model (SM) can not explain, the nature of particle dark matter (DM) has been another longstanding puzzle. In spite of several astrophysics and cosmology based evidences suggesting the presence of DM in the Universe \cite{Zwicky:1933gu, Rubin:1970zza, Clowe:2006eq, Planck:2018vyg}, its particle origin is still unknown with none of SM particles being fit to be a DM candidate. Among several BSM proposals for particle DM, the weakly interacting massive particle (WIMP) remains the most widely studied one. In the WIMP paradigm, a particle DM candidate having mass and interaction strength typically around the electroweak ballpark can give rise to the observed DM abundance after thermal freeze-out, a remarkable coincidence often referred to as the \textit{WIMP Miracle}. Recent reviews of WIMP type models can be found in \cite{Arcadi:2017kky, Arcadi:2024ukq}. Due to sizable interaction rate of WIMP with the SM particles, it also remains verifiable at direct detection experiments \cite{LZ:2022lsv}.

Motivated by this, in this work we consider a dark matter embedded Dirac seesaw scenario and study the observational signatures. Interestingly, we can have a stable scalar DM candidate in the minimal tree level seesaw realisation for Dirac neutrinos without any additional fields. While the stability of DM is guaranteed by a charge conjugation symmetry in dark sector, a $Z_4$ symmetry is imposed to get the desired terms in the Lagrangian. An unbroken global lepton number symmetry keeps the lepton number violating Majorana mass terms away guaranteeing pure Dirac nature of light neutrinos. While non-zero Dirac neutrino mass relies upon spontaneous breaking of $Z_4$ symmetry, the latter also leads to the formation of domain walls (DW) which can be catastrophic in cosmology, if allowed to dominate. We introduce tiny bias terms which explicitly break $Z_4$ symmetry such that the DW annihilate away leaving a stochastic gravitational wave (GW) background within reach of near future experiments. On the other hand, the right chiral parts of light Dirac neutrinos, when thermalised, lead to enhancement of relativistic degrees of freedom $N_{\rm eff}$, which can be observed at future CMB experiments. The heavy singlet fermions introduced for seesaw can lead to successful Dirac leptogenesis explaining the observed baryon asymmetry. We perform a numerical analysis to show the possibility of explaining neutrino mass, dark matter and baryon asymmetry while being in agreement with all experimental bounds. We also point out the interesting detection prospects at GW frontiers due to annihilating $Z_4$ domain walls and CMB prospects of detecting enhanced $N_{\rm eff}$. While these new degrees of freedom like heavy singlet Dirac fermions have interesting phenomenology related to the origin of light Dirac neutrino mass and leptogenesis, they can potentially generate large radiative corrections to the scalar potential. We constrain the parameter space such that such corrections remain small guaranteeing the stability of the potential upto the Planck scale. While we still have the hierarchy problem associated with light scalars as in the SM, radiative corrections to scalar masses can be kept under control in well-motivated frameworks like supersymmetry, extra dimensions, new composite dynamics among others, a recent review of which can be found in \cite{Craig:2022eqo}. We stick to the minimal framework to be discussed below as we do not intend to provide a solution to the hierarchy problem in this work.

The outline of this paper is as follows. In sec.~\ref{sec:model}, we introduce the framework and discuss the generation of neutrino mass, baryon asymmetry from Dirac Leptogenesis and dark matter relic. In sec.~\ref{sec:grav}, we discuss domain wall formation from the spontaneous breaking of $Z_4$ symmetry and the corresponding gravitational wave detection prospects.  In sec. \ref{sec:Neff}, we discuss the possibility of enhanced $N_{\rm eff}$ from light Dirac neutrinos. We summarise our key results in sec.~\ref{sec:results} and finally conclude in sec.~\ref{sec:conclusion}.

\section{Model}
\label{sec:model}
We consider a minimal Dirac seesaw realization for light neutrino masses along with a scalar dark matter candidate by augmenting the SM gauge symmetry with a global $Z_4$ symmetry. We assume a global lepton number symmetry $U(1)_L$ in order to prevent Majorana mass terms at any order. We extend the SM field content by introducing three copies of heavy Dirac fermions $N_{L,R}$, the right-handed counterparts of active neutrinos $\nu_R$, a real scalar $\eta$ and a complex scalar $\rho$, all singlets under the SM gauge symmetry. The discrete $Z_4$ symmetry where $\nu_R$ is odd, prevents direct coupling of the SM lepton doublet $L$ with $\nu_R$ via the SM Higgs $H$. However $\nu_R$ can have Yukawa interaction with $N_L$ via $\eta$ in order to realise Dirac seesaw. Table~\ref{tab:particlecontent} summarizes the particle content of our model along with their charges under $SU(2)_L,U(1)_Y$, global $Z_4$ and $U(1)_L$ symmetries.

\begin{table}[h!]
\centering
\begin{tabular}{ccccc}
\hline\hline
 & $\text{SU(2)}_L$ & $\text{U(1)}_Y$  & $Z_4$ & $U(1)_L$\\
\hline Fermions \\
$Q =  \begin{pmatrix}
             u_L \\ d_L
            \end{pmatrix}$ & $\textbf{2}$ & $1/6$ & $+1$ & $0$	\\
$u_R$	& $\textbf{1}$ & $2/3$  & $+1$	& $0$\\
$d_R$	& $\textbf{1}$ & $-1/3$  & $+1$	& $0$\\
$L =  \begin{pmatrix}
             \nu_L \\ e_L
            \end{pmatrix}$ & $\textbf{2}$ & $-1/2$ & $+1$	& $1$\\
$e_R$	& $\textbf{1}$ & $-1$  & $+1$	& $1$\\
$\nu_R$	& $\textbf{1}$ & $0$  & $-1$	& $1$\\
$N_L$ & $\mathbf{1}$ & $0$ &  $+1$ & $1$\\
$N_R$ & $\mathbf{1}$ & $0$ &  $+1$ & $1$\\
\\[1mm] \hline
Scalar fields\\
$H$ & $\mathbf{2}$ & $+1/2$  & $+1$ & $0$\\
$\eta$ & $\mathbf{1}$ & $0$ & $-1$ & $0$\\
$\rho$ & $\mathbf{1}$ & $0$ & $ i$ & $0$\\
\hline\hline
\end{tabular}
\caption{\small Particle content with quantum numbers under the symmetries of the model. All the exotic fermions and scalars are color singlets.}
\label{tab:particlecontent}
\end{table}

\noindent The most general scalar potential for our model can be written as,
\begin{align}
    \label{eq:scalarpot}
    \nonumber
    V(H,\eta,\rho)=&-\mu_\rho^2\,\rho^\dag\rho+\mu_\eta^2\,\eta^2-\mu_H^2\,H^\dag H+\frac{\lambda_\rho}{4}(\rho^\dag\rho)^2+\frac{\lambda_\eta}{4}\eta^4+\frac{\lambda_H}{4}(H^\dag H)^2\\
   &+\lambda_{H\eta}\,\eta^2H^\dag H+\lambda_{H\rho}\,\rho^\dag\rho \,H^\dag H+\lambda_{\rho\eta}\,\eta^2\rho^\dag\rho-(\mu_1 \rho^2\eta+\lambda_1\rho^4+\text{h.c.})
\end{align}
This potential is invariant under $Z_4$ transformations $\rho\rightarrow i\rho,\,\eta\rightarrow-\eta$. Due to the charge assignments under $Z_4$, the real scalar $\eta$ transforms only in the subgroup $Z_2$ of the $Z_4$. 
We consider CP conserving limit such that the potential~\eqref{eq:scalarpot} is invariant under the transformation $\rho\leftrightarrow\rho^*$, i.e. $\mu_1$ and $\lambda_1$ are real parameters. That makes our model effectively symmetric under the dihedral group $D_4$ which is isomorphic to $Z_4\rtimes Z_2^{\rm CP}$~\cite{Wu:2022tpe}. Due to the CP symmetry, the imaginary part of $\rho$ behaves as a stable dark matter candidate \cite{Coito:2021fgo, Pham:2024vso}. While these earlier works used the nomenclature of CP symmetry, it is effectively a charge conjugation symmetry $C_{\rm dark}$ \cite{Sakurai:2021ipp}. The requirement of a stable DM also justifies the extension of simplest $Z_2$-symmetric version of Dirac seesaw studied earlier \cite{Barman:2022yos, Barman:2023fad}.

We require all the coefficients in the potential~\eqref{eq:scalarpot} to be positive and $\lambda_\rho/4>\lambda_1$ to ensure correct ground state structure of the potential. Additionally, we consider that the vacuum expectation value (VEV) of $\rho$, denoted by $v_\rho$, is at a much higher scale than the scale of electroweak symmetry breaking. We also consider $\mu_\eta,\mu_1\ll v_\rho$ and $\lambda_{\rho\eta}\sim\lambda_{H\eta}\sim0$ such that $\eta$ acquires a VEV, $v_\eta$ induced by $v_\rho$. After electroweak symmetry breaking, the scalar fields are parameterized as,
\begin{align}
H(x) &=
	\left(
		\begin{array}{c}
		G^+(x) \\
		\frac{1}{\sqrt{2}}\big(v+h(x)+iG^0(x)\big)
		\end{array}
	\right),\\
\rho(x) &= \frac{1}{\sqrt{2}}\big(v_\rho+ \tilde{\rho}(x)+i\chi(x) \big),\\
\eta(x) &= \frac{1}{\sqrt{2}}(v_\eta + \tilde{\eta}),
\end{align}
where $v(\simeq 246.22~\text{GeV})$, $v_\eta$ and $v_\rho$ are the VEVs of $H,\eta$ and $\rho$, respectively. The tadpole equations are given by $\left\langle\frac{\partial V}{\partial v}\right\rangle=\left\langle\frac{\partial V}{\partial v_\eta}\right\rangle=\left\langle\frac{\partial V}{\partial v_\rho}\right\rangle=0$ where $\big\langle...\big\rangle$ denotes that the field values are taken to be zero after the derivative. Solving the tadpole equations in the limit $\mu_\eta^2/v_\rho^2\rightarrow0$ and $\lambda_{\rho\eta}\sim\lambda_{H\eta}\rightarrow0$, one obtains the following relations,
\begin{align}\label{eq:veta}
    \nonumber v_\eta&\simeq\sqrt{2}\left(\frac{\mu_1 v_\rho^2}{\lambda_\eta}\right)^{1/3}\\
    \mu_H^2&\simeq\frac{1}{4}\lambda_H v^2+\frac{1}{2}\lambda_{H\rho}v_\rho^2\\
    \nonumber \mu_\rho^2&\simeq\left(\frac{\lambda_\rho}{4} - 2\lambda_1\right)v_\rho^2+\frac{1}{2}\lambda_{H\rho} v^2-\sqrt{2}\mu_1 v_\eta
\end{align}
Here $v_\eta$ is the induced VEV. The symmetry breaking pattern is as follows,
\begin{equation*}
   SU(2)_L  \otimes \, U(1)_{Y}\otimes Z_4\rtimes Z_2^{\rm CP}
   \overset{\braket{\rho},\braket{\eta}}{\xrightarrow{\hspace*{0.7cm}}}
    SU(2)_L  \otimes \,U(1)_{Y} \rtimes Z_2^{\rm CP}
    \overset{\braket{H}}{\xrightarrow{\hspace*{0.5cm}}} 
     U(1)_{\rm em}\rtimes Z_2^{\rm CP}
\end{equation*}
The details of the scalar masses namely $m_{h_1}, m_{h_2}, m_{h_3}, m_\chi$ are given in appendix \ref{app:scalarmasses}. We consider $ m_\chi, \lambda_\rho, \lambda_\eta, m_{h_3}, \mu_1, \lambda_{H \rho} $ to be the free parameters while fixing $\mu_\eta, \lambda_{H \eta}, \lambda_{\rho \eta}$ and the SM-like Higgs mass $m_{h_1}=125.32$ GeV in our numerical analysis. The other dependent parameters $\lambda_H, v_\rho, \lambda_1, v_\eta, \mu_H, \mu_\rho$ are derived from the free parameters. The boundedness from below and vacuum stability of the potential are discussed in appendix \ref{app:bounded} and appendix \ref{app:beta} respectively.

\subsection{Electroweak precision bounds}\label{sec:STU}
To explore new physics beyond the SM, electroweak radiative corrections have proven essential~\cite{Peskin:1990zt,Peskin:1991sw,Kennedy:1990ib}. A common method for parameterizing these effects at higher energy scales is the $STU$ formalism~\cite{Peskin:1991sw}. This framework introduces three oblique parameters: $S$, $T$, and $U$ which encapsulate deviations in self-energy corrections. These parameters are evaluated at two key energy scales, $m_Z$ and 0, and they are defined as~\cite{Han:2000gp},

\begin{equation}
\alpha\, S = 4 s^2 c^2\, \frac{\Pi_{Z}(m_Z^2)-\Pi_{Z}(0)}{m_Z^2},
\end{equation}

\begin{equation}
\alpha\, T = \frac{\Pi_{W}(0)}{m_W^2}-\frac{\Pi_{Z}(0)}{m_Z^2},
\end{equation}

\begin{equation}
\alpha\, (S+U) = 4 s^2 \, \frac{\Pi_{W}(m_W^2)-\Pi_{W}(0)}{m_W^2},
\end{equation}
where $s$ and $c$ are the sine and cosine of the weak mixing angle respectively, while $\alpha$ represents the fine structure constant, all evaluated at the energy scale $m_Z$.

The $S$ parameter quantifies the impact of new physics on neutral current processes across different energy scales, while the sum $(S + U)$ captures the corresponding effects on charged current processes. The $T$ parameter reflects the difference between new physics contributions to neutral and charged current processes at low energy, scaled by $\Delta \rho$. In most scenarios, the $U$ parameter remains small. According to the latest global fit from the Particle Data Group (2024 update)~\cite{ParticleDataGroup:2024cfk}, the current constraints on these parameters are:  

\begin{equation}
S = -0.04 \pm 0.10, \quad T = 0.01 \pm 0.12, \quad U = -0.01 \pm 0.09.  
\end{equation}

For our study, we compute the oblique parameters for various benchmark points using \texttt{SPheno}~\cite{Porod:2003um}. Our analysis focuses on the region of the model parameter space that remains consistent with the above bounds on $S$, $T$, and $U$ within a $1\sigma$ confidence level.

\subsection{Neutrino masses and leptogenesis}
\begin{figure}[ht]
\center
\includegraphics[width=0.4\linewidth]{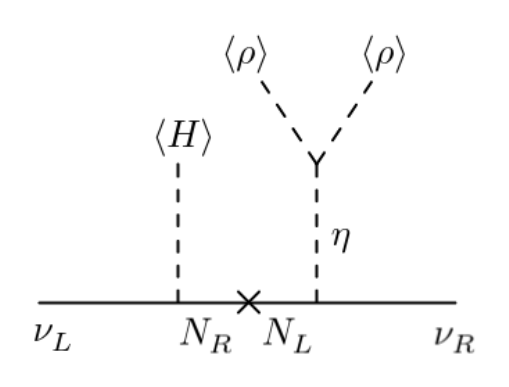} 
\caption{\small Feynman diagram contributing to light neutrino mass. Here $\eta$ acquires an induced VEV given by Eq.~\eqref{eq:veta}.
\label{fig:feynman}}
\end{figure}

The lepton number conserving Yukawa lagrangian for neutrinos is given by~\cite{Barman:2022yos},
\begin{align}\label{eq:LY}
    -\mathcal{L}_Y \supset Y_L\, \overline{L}\,\tilde{H}\,N_R + M_N\, \overline{N}\,N + Y_R\,\overline{N_L}\,\eta \,\nu_R + {\rm h.c.}
\end{align}
We consider only induced VEV of $\eta$ such that light Dirac neutrino mass arises due to the Feynman diagram shown in Fig.~\ref{fig:feynman}. Upon the acquisition of vacuum expectation values $v$, $v_\eta$ and $v_\rho$ by the neutral components of $H$, $\eta$ and $\rho$ respectively, the light Dirac neutrino mass arises from the Type-I seesaw as,
\begin{equation}\label{eq:Mnu}
    m_\nu = Y_L\,M^{-1}_N\,Y_R\, \frac{v \, v_\eta}{2}
\end{equation}
where $v_\eta$ is given in Eq.~\eqref{eq:veta} and the hierarchy $Y_L v, Y_R v_\eta \ll M_N$ is assumed in the spirit of seesaw.

While lepton number is conserved due to a global lepton number $U(1)_L$ symmetry, successful leptogenesis can occur via Dirac leptogenesis \cite{Dick:1999je, Murayama:2002je}. Equal and opposite amount of lepton asymmetries can be created in the left and right sectors via the CP violating out-of-equilibrium decays of $N\equiv N_L+N_R$ to $LH$ or $\nu_R \eta$ respectively. Total lepton asymmetry is zero due to lepton number conservation. The lepton asymmetries in the left and right handed sectors are prevented from equilibration due to the tiny effective Dirac Yukawa couplings thereby allowing the left sector asymmetry to get converted into baryon asymmetry by electroweak sphalerons. Various possible implementation of this idea can be found in Refs.~\cite{Boz:2004ga, Thomas:2005rs, Thomas:2006gr, Cerdeno:2006ha, Gu:2006dc, Gu:2007mc, Chun:2008pg, Bechinger:2009qk, Chen:2011sb, Choi:2012ba, Borah:2016zbd, Gu:2016hxh, Narendra:2017uxl, Babu:2024glr}. In a few related works~\cite{Heeck:2013vha, Gu:2019yvw, Mahanta:2021plx}, violation of $B-L$ symmetry was accommodated in a way to preserve the Dirac nature of light neutrinos, while generating lepton asymmetry simultaneously.

We consider a hierarchical mass structure $M_1\ll M_2<M_3$ for $N$ such that only $N_1$ decay is relevant for leptogenesis. The CP asymmetry parameter can be parameterised as~\cite{Cerdeno:2006ha},
\begin{equation}\label{eq:CPparam}
    \varepsilon\simeq\frac{1}{8\pi}\sum_{j \neq 1}\frac{M_1}{M_j}\frac{\text{Im}\left[(Y_R Y_R^\dagger)_{1j}(Y_L^\dagger Y_L)_{j1}\right]}{(Y_R Y_R^\dagger)_{11}+(Y_L^\dagger Y_L)_{11}}.
\end{equation}
The Yukawa matrices can be written according to the Casas-Ibarra parametrisation~\cite{Cerdeno:2006ha, Casas:2001sr},
\begin{equation}\label{eq:casas}
    Y_L=\frac{\sqrt{2}}{v}U^*_L\hat{m}_\nu^{1/2}\mathcal{R} M_N^{1/2},\quad Y_R=\frac{\sqrt{2}}{v_\eta}M_N^{1/2}X^\dagger \hat{m}_\nu^{1/2}U_R^\dagger,
\end{equation}
where $\mathcal{R},X$ are random $3\times3$ complex matrices satisfying $\mathcal{R}X^\dagger=1 \implies X^\dagger=\mathcal{R}^{-1}$. The matrix $\hat{m}_\nu$ is the diagonal active neutrino mass matrix. $U_L,U_R$ diagonalizes the neutrino mass matrix,
\begin{equation}
    \hat{m}_\nu=U_L^T m_\nu U_R.
\end{equation}
Without loss of generality, we consider $U_L=U_R\simeq U_{\rm PMNS}$. For numerical analysis we consider normal ordering of active neutrino masses $m_1<m_2<m_3$ and $m_1=10^{-6}$ eV, i.e.
\begin{equation}
    m_1=10^{-6}\text{ eV},\quad m_2=\sqrt{\Delta m_{21}^2+m_1^2}\sim0.0087\text{ eV},\quad m_3=\sqrt{\Delta m_{31}^2+m_2^2}\sim0.05\text{ eV},
\end{equation}
where $\Delta m_{21}^2,\Delta m_{31}^2$ are the solar and atmospheric squared mass differences respectively. The central values of the squared mass differences and other neutrino oscillation parameters are given in Table~\ref{tab:nuOsc} and they are taken as inputs to calculate $Y_L, Y_R$. We sample the elements of $\mathcal{R}$ randomly, keeping their magnitudes between $10^{-4}$ to $10$ for simplicity. It can be shown that the maximum value $\varepsilon$ can obtain is~\cite{Cerdeno:2006ha},
\begin{equation}
    \label{eq:epmax}
    \varepsilon_{\rm max}=\frac{1}{16\pi}\frac{M_1}{v\,v_\eta}(m_3-m_1).
\end{equation}

\begin{table}[ht!]
\centering
\begin{tabular}{|c|c|}
\hline \hline
 {\textbf{Oscillation}}   &  {\textbf{Numerical Input}}\\
 {{\bf parameters}} & {\tt (NH assumed)}   \\ \hline \hline
 $\Delta m_{21}^2 (10^{-5}~{\rm eV}^2$)  &    7.49    \\ 
 $\Delta m_{31}^2 (10^{-3}~{\rm eV}^2) $ &     2.534  \\ 
  $\sin^2{\theta_{12}}$   &      0.307   \\ 
 $\sin^2{\theta_{23}}$   &     0.561   \\ 
  $\sin^2{\theta_{13}}  $  &      0.02195    \\ 
  $\delta_{CP}/^\circ$  &  177  
  \\\hline \hline
\end{tabular}
\caption{The values of the neutrino oscillation parameters used in our numerical analysis correspond to the central values reported in a recent global fit~\cite{Esteban:2024eli}. We consider only the normal mass hierarchy (NH) scenario.}
\label{tab:nuOsc}
\end{table}

In addition to CP violation, an essential requirement for leptogenesis is out-of-equilibrium decays. The out-of-equilibrium condition for $N_1$ decay can be checked by defining a parameter $K$ such that, 
\begin{equation}\label{eq:K}
    K\equiv \frac{\Gamma (N_1\rightarrow L H) + \Gamma (N_1 \rightarrow \nu_R \eta)}{\mathcal{H}(T = M_1)} = \frac{\tilde{m}}{m_*}
\end{equation}
where the \textit{effective neutrino mass} is defined as~\cite{Barman:2022yos,Cerdeno:2006ha},
\begin{equation}
    \label{eq:mtilde}
    \tilde{m}= \frac{[(Y_L^\dag Y_L)_{11}+(Y_RY_R^\dag)_{11}]v\,v_\eta}{2M_1}.
\end{equation}
and $\mathcal{H}(T)$ is the Hubble rate in radiation dominated epoch,
\begin{equation}
    \mathcal{H}(T)=\sqrt{\frac{8\pi^3g_*}{90}}\frac{T^2}{M_{\rm Pl}}
\end{equation}
where, $g_* \sim 114$ is the total relativistic degrees of freedom and $M_{\rm Pl} = 1.22 \times 10^{19}$ GeV is the Plank mass. The \textit{equilibrium neutrino mass} $m_*$ is defined as~\cite{Cerdeno:2006ha},
\begin{equation}
    m_* = v \, v_\eta \frac{8 \pi}{M_{\rm Pl}} \sqrt{\frac{8 \pi^3 g_*}{90}}.
\end{equation}
$K\ll 1$ is the \textit{weak washout regime} which signifies that the $N_1$ decay is completely out-of-equilibrium and leptogenesis is efficient while $K\gg 1$ is called \textit{strong washout regime} where the final asymmetry depends on the interplay between decay, inverse decay and scatterings of $N_1$.

The sphaleron processes convert only the left sector lepton asymmetry to baryon asymmetry, provided the lepton asymmetries in the left and right sectors do not equilibrate upto the sphaleron decoupling epoch $T_{\rm sph} \sim 130$ GeV. The rate at which the lepton asymmetries in the left and right sectors equilibrate at high temperatures is given by~\cite{Barman:2022yos,Cerdeno:2006ha},
\begin{equation}
    \Gamma_{L-R}(T)\sim\frac{\left(|(Y_L)_{i1}||(Y_R)_{1j}|\right)^2 T^3}{M_1^2},
\end{equation}
where $i,j=e,\mu,\tau$ are the lepton favour index. This equilibration rate should be less than the Hubble rate at around $T \sim M_1$, which implies,
\begin{equation}\label{eq:baryo_condition}
    \frac{\left(|(Y_L)_{i1}||(Y_R)_{1j}|\right)^2}{M_1}\lesssim \frac{1}{M_{\rm Pl}}\sqrt{\frac{8\pi^3g_*}{90}}.
\end{equation}
The evolution and the final amount of asymmetry can be calculated by solving the relevant Boltzmann equations. The final baryon asymmetry is given by~\cite{Buchmuller:2004nz},
\begin{equation}\label{eq:final asymmetry}
    \eta_B=0.96\times10^{-2}\varepsilon\, \kappa_f
\end{equation}
where $\kappa_f$ is an efficiency parameter. This can be compared with the observed baryon-to-photon ratio $\eta_B = \frac{n_{B}-n_{\bar{B}}}{n_{\gamma}} = 6.1 \times 10^{-10}$ \cite{Planck:2018vyg} to constrain the model parameters. For thermal initial abundance of $N_1$ and weak washout scenarios, i.e. $K\ll1$, efficiency parameter $\kappa_f$ is normalized to unity. 
For $K\gg1$, the efficiency factor can be estimated as~\cite{Cerdeno:2006ha},
\begin{equation}\label{eq:kappaf}
    \kappa_f\simeq\frac{0.12}{K^{1.1}}
\end{equation}
For numerical analysis, we simply consider $\kappa_f=1$ for $K\leq1$ and $\kappa_f=0.12/K^{1.1}$ for $K>1$ and calculate the final asymmetry using Eq.~\eqref{eq:final asymmetry}. Finally we compare the asymmetry to observed value of $6.1\times10^{-10}$. For $K\sim1$, the minimum $\varepsilon$ value required to obtain observed baryon asymmetry is around $6.3\times10^{-8}$ which gives a Davidson-Ibarra~\cite{Davidson:2002qv} type lower bound on $M_1$ using Eq.~\eqref{eq:epmax},
\begin{equation}\label{eq:seesawscalenaive}
    M_1\gtrsim 1.6\times 10^9\left(\frac{v_\eta}{100\,\text{GeV}}\right) \, {\rm GeV}.
\end{equation}

\subsection{Stability of radiative corrections}
The presence of heavy Dirac fermions $N_{L,R}$ in our setup can lead to large quantum corrections rendering the scalar potential unstable. In the SM, the absolute stability of the Higgs vacuum can be ensured by demanding a positive definite self-quartic coupling $\lambda_H (\mu) > 0$ for any $\mu$ assuming the electroweak vacuum to be a global minima. In presence of an additional deeper minimum other than the electroweak one, the Higgs vacuum can be mestastable if its decay lifetime is more than the age of the Universe. In the SM, the top quark Yukawa drives the Higgs quartic coupling negative around $\mu \sim 10^{10}$ GeV making the electroweak vacuum mestastable \cite{Degrassi:2012ry}. While a large coupling of heavy fermions to the SM Higgs can drive $\lambda_H (\mu)$ negative at a lower energy scale, the additional scalars present in our setup can help in stabilising the potential due to their quartic couplings with the SM Higgs \cite{Gonderinger:2009jp}. We solve the renormalisation group evolution (RGE) equations of all the relevant parameters of the model and constrain the parameter space by demanding the stability of the scalar potential all the way till the Planck scale. The details of the RGE equations are given in appendix \ref{app:beta}. Therefore, the present scenario ensures the stability of the electroweak vacuum all the way till Planck scale, in contrast to the mestasble minima in the SM. 

While the potential is made stable by suitable choice of parameters, the minimal setup discussed here can not keep the radiative corrections to scalar masses under control. Suitable extension of the model incorporating supersymmetry, extra dimension or new composite dynamics can help in keeping such radiative corrections finite solving the hierarchy problem, which is beyond the scope of the present work.




\subsection{Dark matter}

\begin{figure}[ht!]
    \centering
    \begin{subfigure}[t]{0.32\textwidth}
        \centering
        \includegraphics[width=\linewidth]{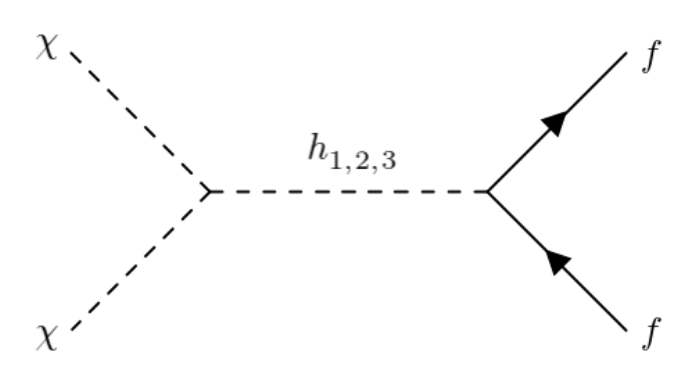}
        \caption{}
    \end{subfigure}%
    ~ 
    \begin{subfigure}[t]{0.32\textwidth}
        \centering
        \includegraphics[width=\linewidth]{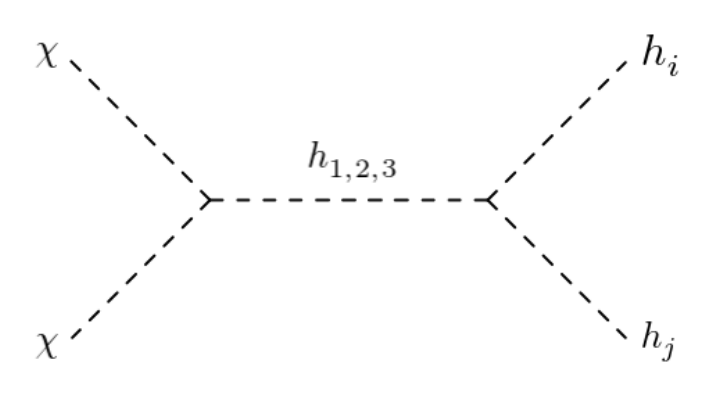}
        \caption{}
    \end{subfigure}%
    ~
    \begin{subfigure}[t]{0.32\textwidth}
        \centering
        \includegraphics[width=\linewidth]{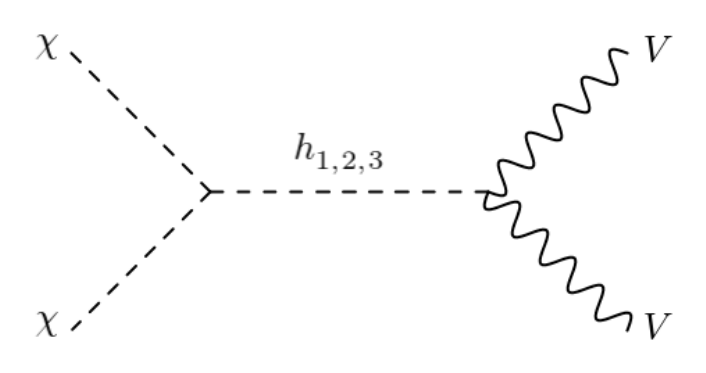}
        \caption{}
    \end{subfigure}\\
    ~ 
    \begin{subfigure}[t]{0.32\textwidth}
        \centering
        \includegraphics[width=\linewidth]{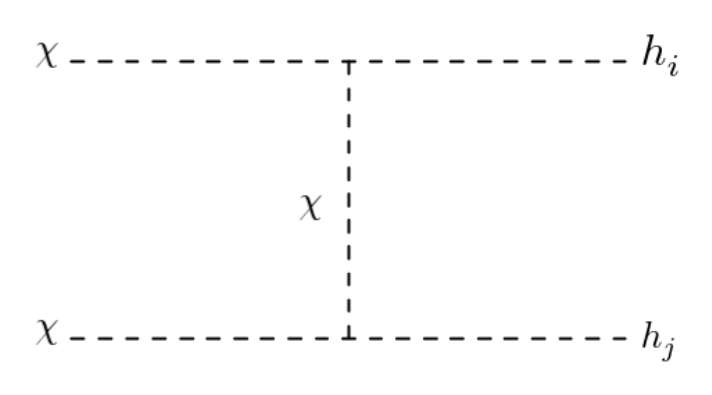}
        \caption{}
    \end{subfigure}%
    \begin{subfigure}[t]{0.20\textwidth}
        \centering
        \includegraphics[width=\linewidth]{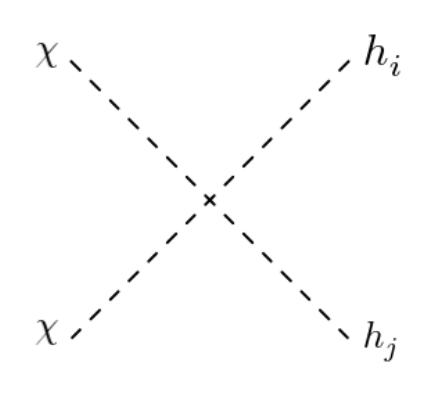}
        \caption{}
    \end{subfigure}
    \caption{\small The DM annihilation processes, where $f$ denote the SM fermions, while $V$ represent $W^\pm$ and $Z$ bosons. The three scalars in our model are denoted by $h_i$, described in Appendix~\ref{app:scalarmasses}.}
\label{fig:DM}
\end{figure}

One can see that the scalar potential in Eq.~\eqref{eq:scalarpot} is invariant under $\chi(x)\leftrightarrow -\chi(x)$ due to the imposed CP invariance, where $\chi$ is the imaginary component of the complex field $\rho$. Hence $\chi(x)$ can play a role of DM. The DM mass in the limit $\mu_\eta^2\sim\lambda_{\rho\eta}\sim\lambda_{H\eta}\rightarrow0$ is given by,  
\begin{equation}\label{mDM_tree}
    m_\chi^2\simeq 8\,\lambda_1 v_\rho^2 + 2\sqrt{2}\mu_1 v_\eta
\end{equation}
where $v_\eta$ is the induced VEV given in Eq.~\eqref{eq:veta}. In Fig.~\ref{fig:DM}, we show the annihilation channels of the DM~\cite{Pham:2024vso}, assuming it to be produced via the usual freeze-out mechanism. The relic abundance of the DM candidate $\chi$ is calculated by solving the Boltzmann equation~\cite{Borah:2016zbd},
\begin{equation}
    \label{eq:DMboltz}
    \frac{dn_\chi}{dt}+3\mathcal{H}n_\chi=-\braket{\sigma v_{\rm rel}}\left(n_\chi^2-\left(n_\chi^{\rm eq}\right)^2\right)
\end{equation}

\begin{figure}[th!]
    \centering
    \begin{subfigure}[t]{0.49\textwidth}
        \centering
        \includegraphics[width=\linewidth]{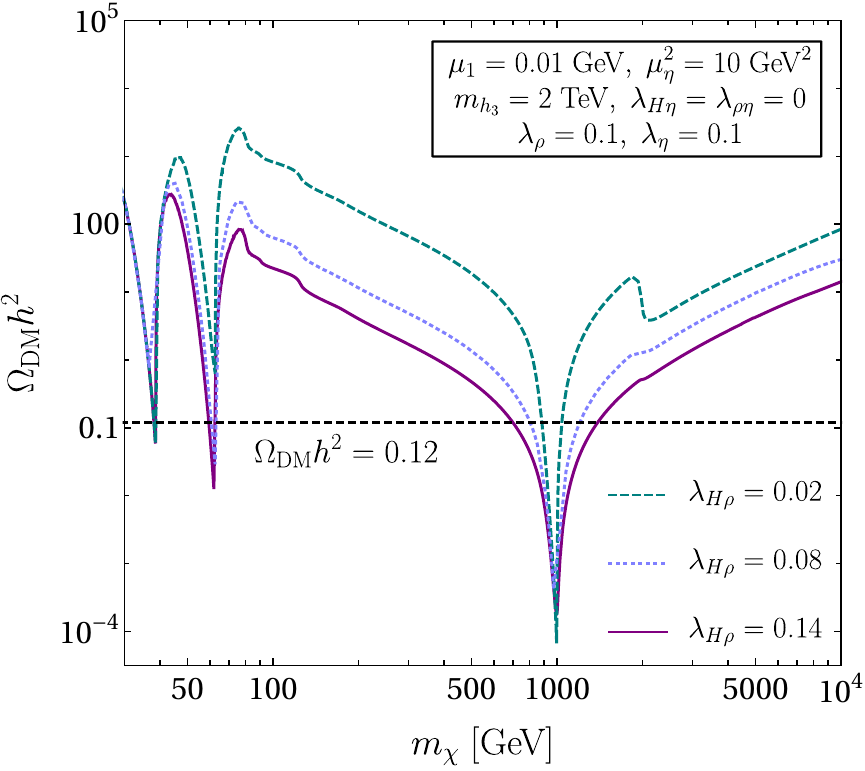}
        \caption{}
    \end{subfigure}%
    ~ 
    \begin{subfigure}[t]{0.49\textwidth}
        \centering
        \includegraphics[width=\linewidth]{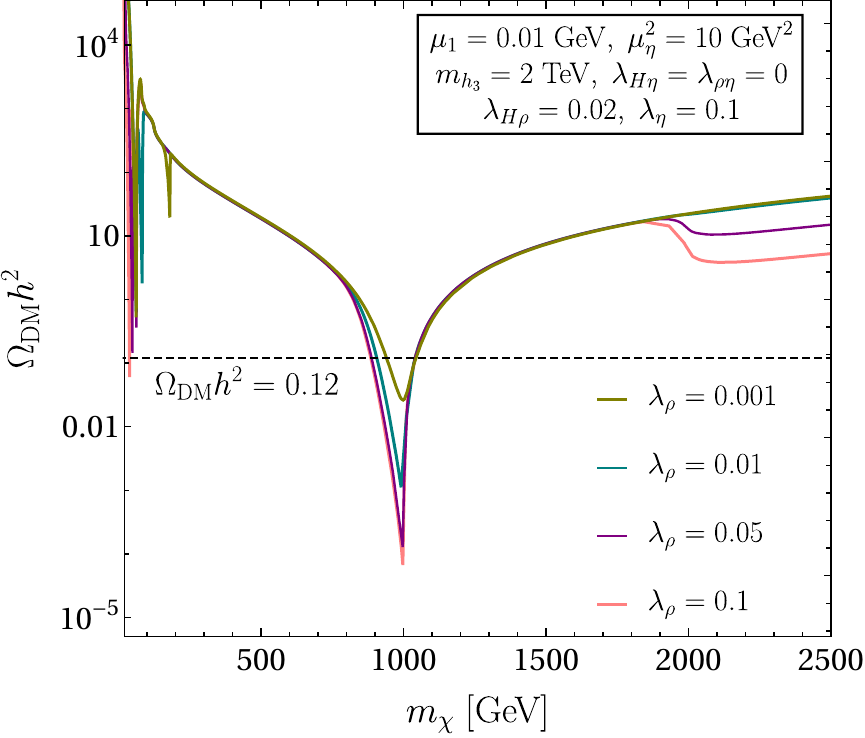}
        \caption{}
    \end{subfigure}\\
    ~
    \begin{subfigure}[t]{0.49\textwidth}
        \centering
        \includegraphics[width=\linewidth]{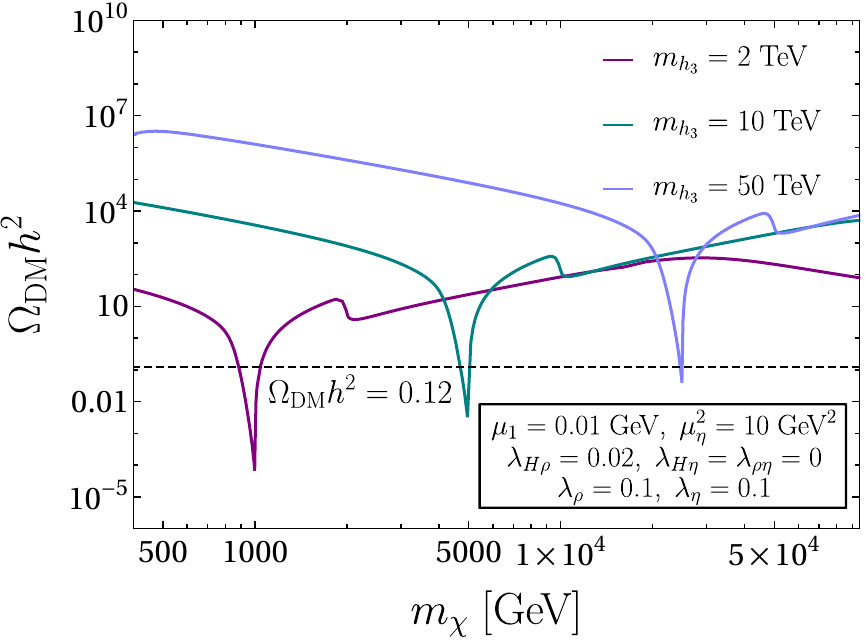}
        \caption{}
    \end{subfigure}
    \caption{\small DM relic density with DM mass $m_\chi$. The dashed horizontal black line corresponds to the observed value of DM abundance. The plots (a) and (b) show the dependence on the portal coupling $\lambda_{H\rho}$ and quartic coupling $\lambda_\rho$ of scalar $\rho$ respectively, defined in Eq.~\eqref{eq:scalarpot}. For the plot (c), we consider $m_{h_3}=2,10$ and $50$ TeV.}
\label{fig:DMrelic}
\end{figure}

\noindent Approximate analytical solution of Eq.~\eqref{eq:DMboltz} is given by~\cite{Gondolo:1990dk,Duerr:2015aka},
\begin{equation}
    \Omega_{\rm DM}h^2\sim\frac{1.07\times10^9\text{ GeV}^{-1}}{J(x_f)\sqrt{g_*}M_{\rm Pl}}
\end{equation}
where $M_{\rm Pl}=1.22\times10^{19}$ GeV is the Planck mass, $g_*\sim114$  is the total effective relativistic degrees of freedom at the time of freeze-out and the function $J(x_f)$ is given by,
\begin{equation}
    J(x_f)=\int_{x_f}^\infty\frac{\braket{\sigma v_{\rm rel}}(x)}{x^2}dx
\end{equation}
where $x_f=m_\chi/T_f$ with $T_f$ being the freeze-out temperature. It can be calculated from the iterative relation,
\begin{equation}
    x_f=\ln\frac{0.038\,g\,M_{\rm Pl}\,m_\chi\braket{\sigma v_{\rm rel}}(x_f)}{\sqrt{g_*\,x_f}}
\end{equation}
where $g=1$ is the degree of freedom of the scalar DM particle. The thermally averaged annihilation cross section multiplied by the relative velocity is given by \cite{Gondolo:1990dk},
\begin{equation}\label{eq:sigmavthermal}
    \braket{\sigma v_{\rm rel}}=\frac{1}{8m_\chi^4TK_2^2(m_\chi/T)}\int_{4m_\chi^2}^\infty\sigma(s-4m_\chi^2)\sqrt{s}K_1(\sqrt{s}/T)ds
\end{equation}
where $K_i$ represent the modified Bessel functions of order $i$. Here $s$ is the Mandelstam variable defined as the square of the total energy of the 2-to-2 annihilation process in the center-of-mass frame~\cite{Pham:2024vso, Binder:2021bmg},
\begin{equation}
    s=\frac{4m_\chi^2}{1-v_{\rm rel}^2/4}
\end{equation}
with $v_{\rm rel}=0.3$. We implement our model in \texttt{SARAH}~\cite{Staub:2015kfa}, \texttt{SPheno}~\cite{Porod:2003um}, \texttt{CalcHEP}~\cite{Belyaev:2012qa} and use \texttt{micrOMEGAs}~\cite{Belanger:2013ywg} to calculate the relic density. In Fig.~\ref{fig:DMrelic}, we show the variation of relic density with DM mass and the dependence on different parameters. The dashed horizontal black line corresponds to observed DM relic density $\Omega_{\rm DM}h^2\sim 0.12$~\cite{Planck:2018vyg}. As seen in the plots (a) and (b) of Fig.~\ref{fig:DMrelic}, there are at least three large dips in the relic density curve at the resonance regions, $m_\chi=m_{h_i}/2,\,i\in\{1,2,3\}$, corresponding to the three scalars in the model, described in appendix~\ref{app:scalarmasses}, where $h_1$ is identified as the SM-like Higgs. This resonant enhancement of the annihilation cross section can be understood from the fact that when $m_\chi=m_{h_i}/2$, the dark matter particles can annihilate through s-channel process shown in Fig.~\ref{fig:DM} (b), mediated by on-shell Higgs boson $h_i$. The term proportional to the propagator goes as,
\begin{equation}\label{eq:DMpropagator}
    \frac{1}{\left(s-m_{h_i}^2\right)^2+m_{h_i}^2\Gamma_{h_i}^2}
\end{equation}
where $\Gamma_{h_i}$ is the $h_i$ decay width. In the resonant regime, $s\sim m_{h_i}^2$, this expression peaks, leading to a sharp annihilation rate of DM particle and hence low relic density.

In Fig.~\ref{fig:DMrelic} (a), we show the effect of the portal coupling $\lambda_{H\rho}=0.02,0.08$ and $0.14$ on the relic density. We notice that decreasing the amount of mixing of the SM Higgs with $\rho$ narrows down the resonance region. It is expected because smaller mixing leads to smaller decay width $\Gamma_{h_i}$ which controls the width of the resonance dip for $h_i$ mediated annihilation channels. Fig.~\ref{fig:DMrelic} (b) demonstrates the dependence of the relic density on the quartic coupling $\lambda_\rho$ of the scalar $\rho$. We consider $\lambda_\rho=0.001,0.01,0.05$ and $0.1$. It is evident that a smaller quartic coupling increases the relic density in the resonance region, which is expected as the vertex $\chi\chi\,h_3$ in Fig.~\ref{fig:DM} is proportional to $\lambda_\rho$. Fig.~\ref{fig:DMrelic} (c) shows the dip corresponding to $m_\chi=m_{h_3}/2$ for the scalar masses $m_{h_3}=2, 10$ and $50$ TeV. We see that the relic density in the resonance region goes up for larger $m_{h_3}$. This behavior can also be understood from Eq.~\eqref{eq:DMpropagator}. As $m_{h_3}$ increases, the overall annihilation cross section decreases, leading to higher relic density. For the given parameter values in Fig.~\ref{fig:DMrelic} (c), DM becomes overabundant even in the resonance region if $m_{h_3}\gtrsim 50$ TeV, which puts an upper bound on the DM mass $m_\chi\lesssim25$ TeV for the given parameter values in the plot. From here onward we will only consider DM in the resonance region $m_\chi=m_{h_i}/2$, such that the DM relic density is never overabundant. This resonance condition transcribes to a relation between the couplings given in Eq.~\eqref{eq:resonancecond}.

\subsubsection{Direct Detection of DM}
\label{ssubsec:dd}
In this subsection, we focus on the constraints that DM direct detection (\texttt{DD}) experimental data place on the parameter space of our model. These experiments operate under the premise that DM can interact directly with the material of the detector. This interaction typically manifests as scattering events with nuclei, leading to observable nuclear recoil. Consequently, we can calculate the DM-nucleon scattering cross-section, particularly with protons and neutrons. 


\begin{figure}[h!]
\center
\includegraphics[width=0.8\linewidth]{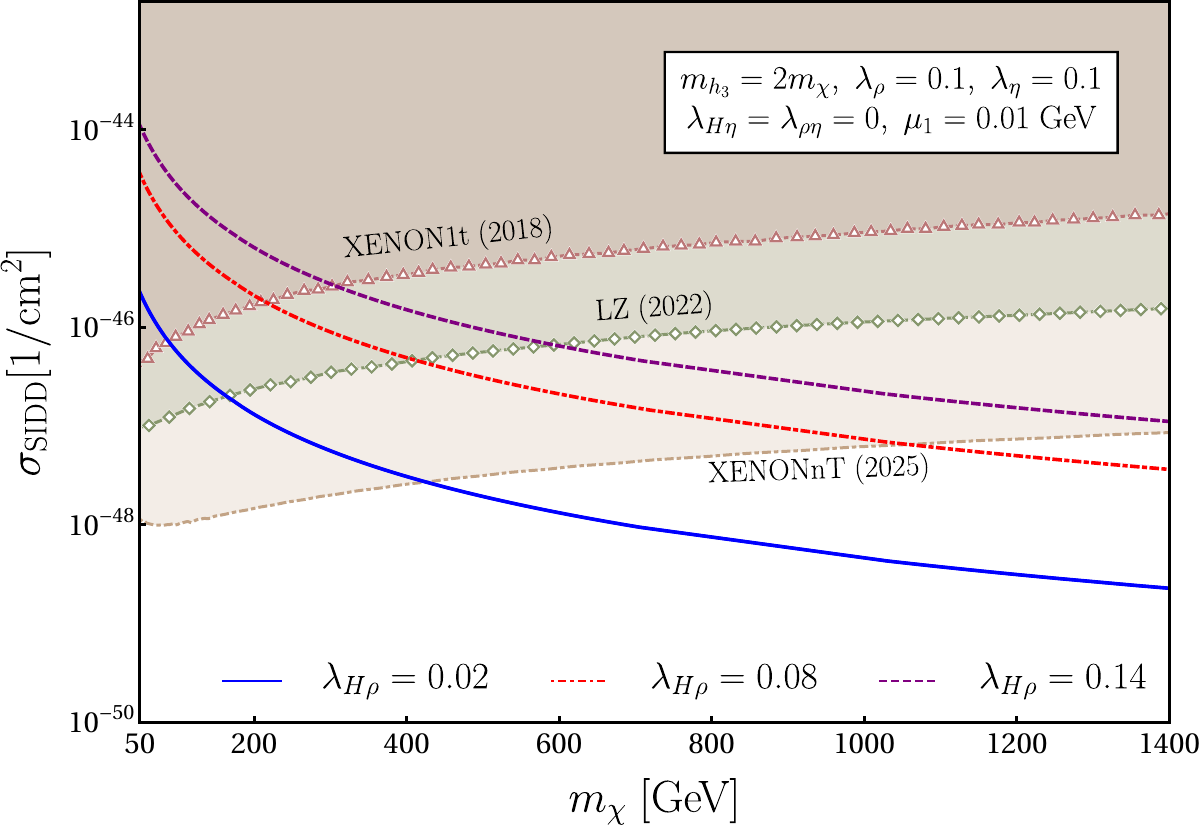}
\caption{\small \texttt{SIDD} cross-section as a function of DM mass in the Higgs resonance regime $m_\chi=m_{h_3}/2$. Blue solid/red dot-dashed/purple dashed lines correspond to portal coupling $\lambda_{H\rho}=0.02,0.08$ and $0.14$ respectively. The shaded colored regions correspond to the current bounds from the \texttt{XENON1T}~\cite{XENON:2018voc} and \texttt{LUX-ZEPLIN (LZ)}~\cite{LZ:2022lsv} experiments and expected 90\% C.L. sensitivity constraints from the future experiment \texttt{XENONnT}~(2025)~\cite{XENON:2020kmp}.}
\label{fig:SIDD_p2}
\end{figure}
In the context of our model, the scalar DM exhibits spin-independent interactions with nuclei, which occur at tree level and are mediated by 
the three scalar mass states, $h_i$ (with $i = \{1,2,3\}$). 
The detailed expressions for the tree level spin-independent direct detection (\texttt{SIDD}) cross-section for a scalar DM can be found in Refs.~\cite{Goodman:1984dc, Duerr:2015aka}. For the purpose of this work, we rely on \texttt{micrOMEGAs} to obtain the \texttt{SIDD} cross-section~$(\sigma_{\texttt{SIDD}})$ as a function of DM mass~$(m_{\chi})$ for three different values of the Higgs portal parameter $\lambda_{H\rho}$. The results have been shown in Fig.~\ref{fig:SIDD_p2}. The values of all the other relevant parameters are mentioned within the plot. Evidently, $\sigma_{\texttt{SIDD}}$ increases with increase in portal couplings. We find that the most stringent lower limit on the DM mass is set by the results of \texttt{LUX-ZEPLIN (LZ)} data~\cite{LZ:2022lsv}. Based on the plot results, we provide a table~\ref{tab:bench1} showing the lower limit on allowed DM mass from the current \texttt{SIDD} constraints for three considered benchmark values of $\lambda_{H\rho}$ parameter. From the table, one may see that for $\lambda_{H\rho}$ varying from 0.02 to 0.14, the lower bound on DM mass due to \texttt{SIDD} results constraints changes from 161 to 593 GeV for the considered parameter space.
\begin{table}[H]
\centering
\begin{tabular}{|>{\centering\arraybackslash}p{1.5cm}|>{\centering\arraybackslash}p{4.5cm}|>{\centering\arraybackslash}p{4cm}|} 
\hline
\multicolumn{3}{|c|}{
\begin{tabular}{c}
$m_{h_3}=2m_{\chi}$ ~~
$\lambda_{\eta}=0.1$ ~~
$\lambda_{\rho}=0.1$ ~~
$\lambda_{H\eta}=\lambda_{\rho\eta}=0 ~~ \mu_1=0.01\text{ GeV}$
\end{tabular}} \\
\hline
\hline
S. No. & $\lambda_{H\rho}$ & $(m_{\chi})_{\text{min}}~[\text{GeV}]$ \\
\hline
1 & 0.02 & 161.26 \\
\hline
2 & 0.08 & 407.16 \\
\hline
3 & 0.14 & 592.87 \\
\hline
\end{tabular}
\caption{\small Lower limit on DM mass from the current best \texttt{SIDD} results of \texttt{LUX-ZEPLIN (LZ)} data~\cite{LZ:2022lsv} for the three benchmark values of $\lambda_{H\rho}$ parameter. Based on the results from Fig.~\ref{fig:SIDD_p2}. The values of all the relevant input parameters are provided in the first row.}
\label{tab:bench1}
\end{table}

\subsubsection{Indirect Detection of DM}
\label{sec:idd}
\begin{figure}[ht!]
\centering
    \begin{tikzpicture}[line width=0.5 pt, scale=0.85]
          \draw[dashed] (-3.0,1.0)--(-1.5,0.0);
        \draw[dashed] (-3.0,-1.0)--(-1.5,0.0);
         \draw[dashed] (-1.5,0.0)--(0.8,0.0);
        \draw[solid] (0.8,0.0)--(2.5,1.0);
        \draw[snake] (0.8,0.0)--(2.5,1.0);
        \draw[snake] (1.59,0.5)--(2.5,0.0);
         \draw[solid] (0.8,0.0)--(2.5,-1.0);
         \draw[snake] (0.8,0.0)--(2.5,-1.0);
         \node at (-3.3,1.0) {${\chi}$};
         \node at (-3.3,-1.0) {${\chi}$};
         \node [above] at (0.0,0.02) {$h_i$};
        \node at (3.2,1.0) {$f/W^\pm$};
        \node at (2.7,0.0) {$\gamma$};
        \node at (3.2,-1.0) {$\overline{f}/W^\mp$};
     \end{tikzpicture}
\caption{DM annihilation into secondary photons, also known as final state radiation~(FSR).}
\label{fig:fig18}
\end{figure}
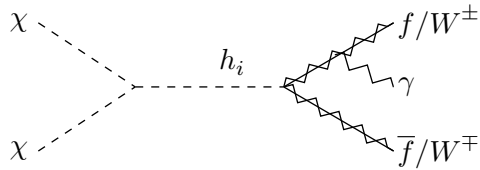

DM parameter space can also be constrained by data from indirect detection experiments looking for DM annihilation into SM particles \cite{Conrad:2014tla}, such as photons. Excess of gamma-rays, either monochromatic or diffuse, can be constrained from such astrophysical observations. While DM annihilation to monochromatic photons is loop-suppressed, tree-level DM annihilation into different charged particles (as shown in Fig. \ref{fig:fig18}) can contribute to diffuse gamma-rays which can be constrained by existing data. While \texttt{Fermi-LAT} \cite{Fermi-LAT:2015bhf, Fermi-LAT:2015kyq, Foster:2022nva} and \texttt{H.E.S.S.} \cite{HESS:2016mib, HESS:2018cbt} data provide stringent constraint at present, future experiments like \texttt{CTA}~\cite{CTAConsortium:2017dvg, CTAO:2024wvb} can probe such DM signatures even further.

\begin{figure}[h!]
    \centering
    \includegraphics[width=0.49\textwidth]{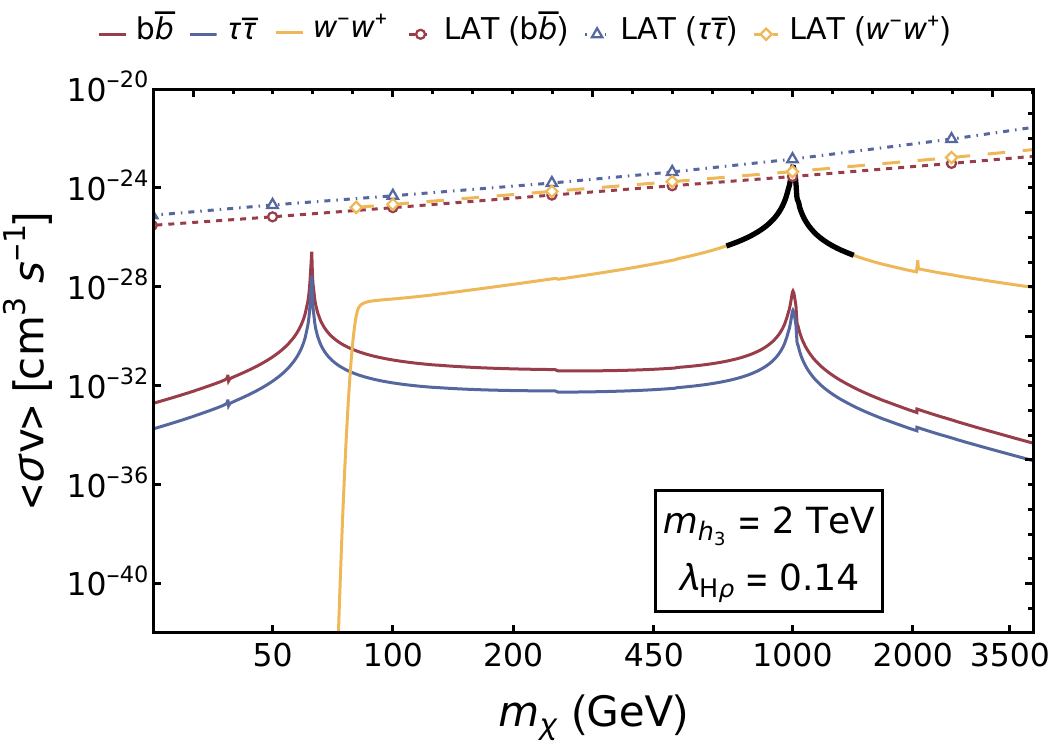} 
    \hfill
    \includegraphics[width=0.49\textwidth]{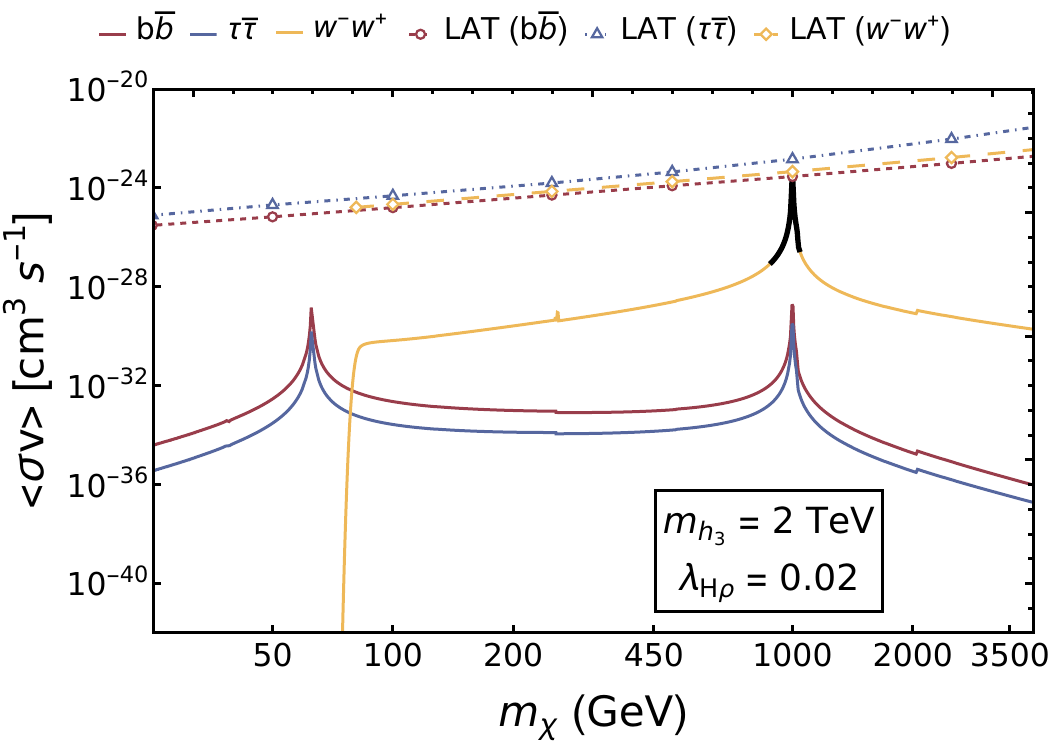} 
    
     \vspace{0.4cm} 
   \includegraphics[width=0.49\textwidth]{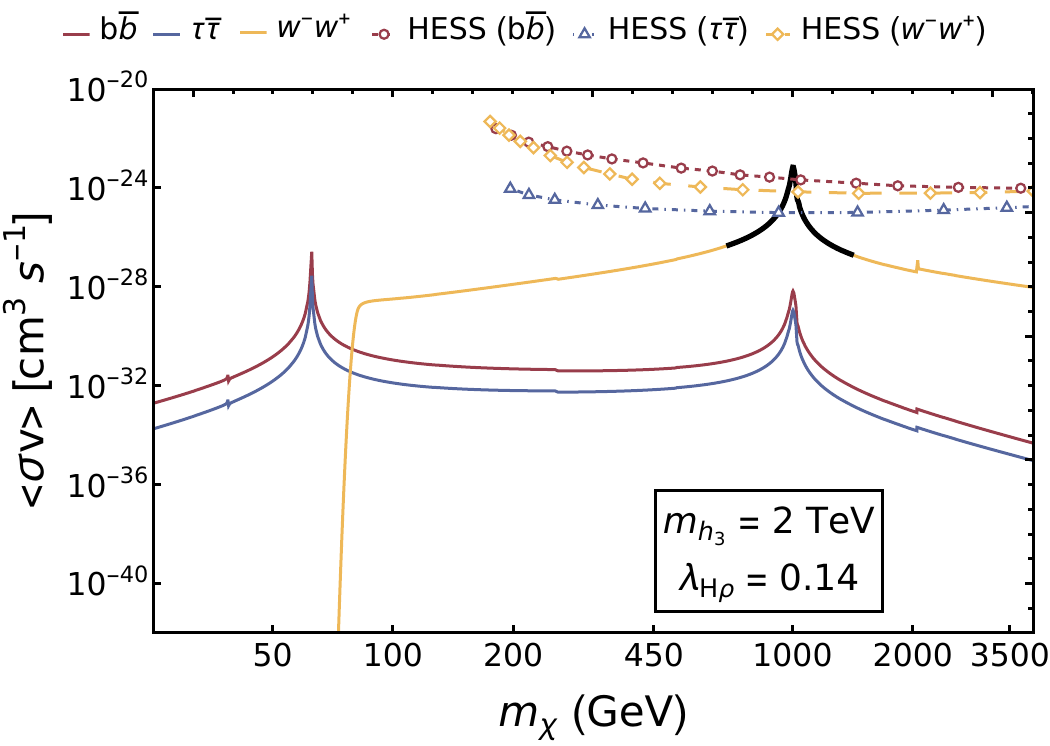} 
    \hfill
    \includegraphics[width=0.49\textwidth]{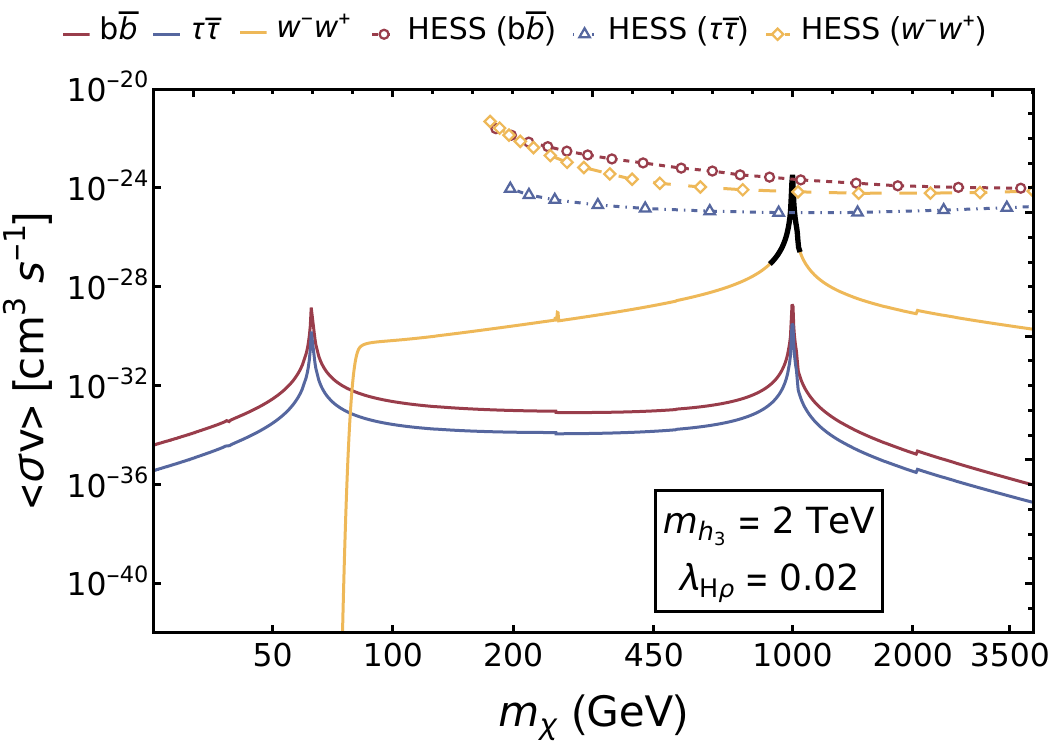} 
    
   \vspace{0.4cm}
   \includegraphics[width=0.49\textwidth]{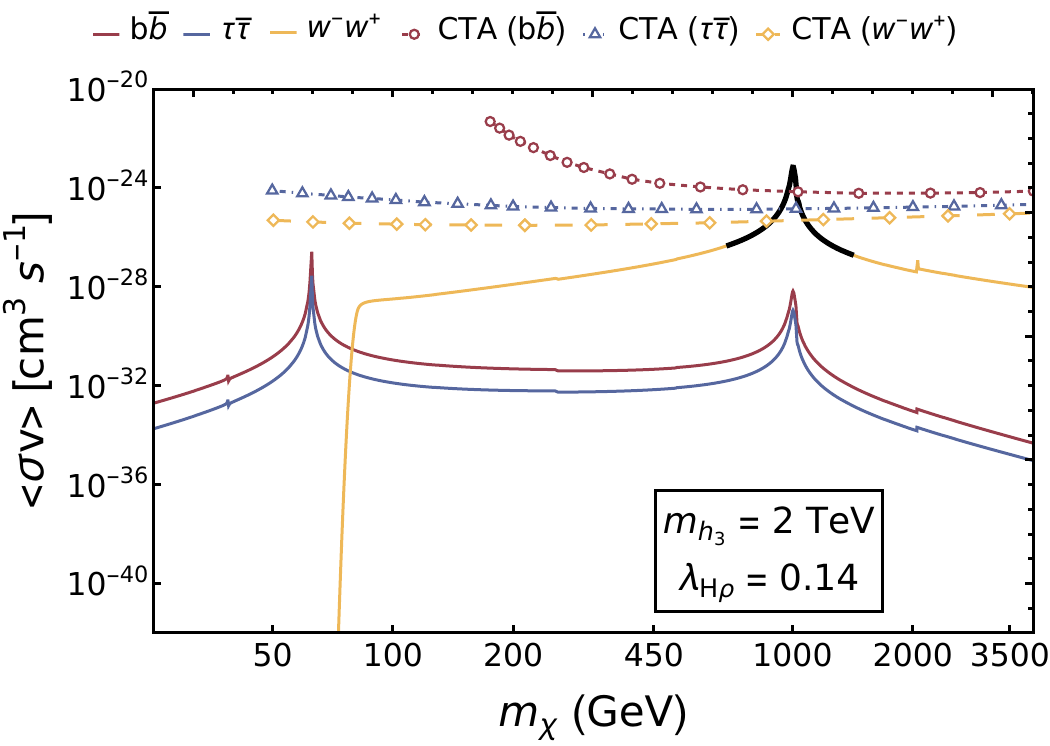} 
   \hfill
    \includegraphics[width=0.49\textwidth]{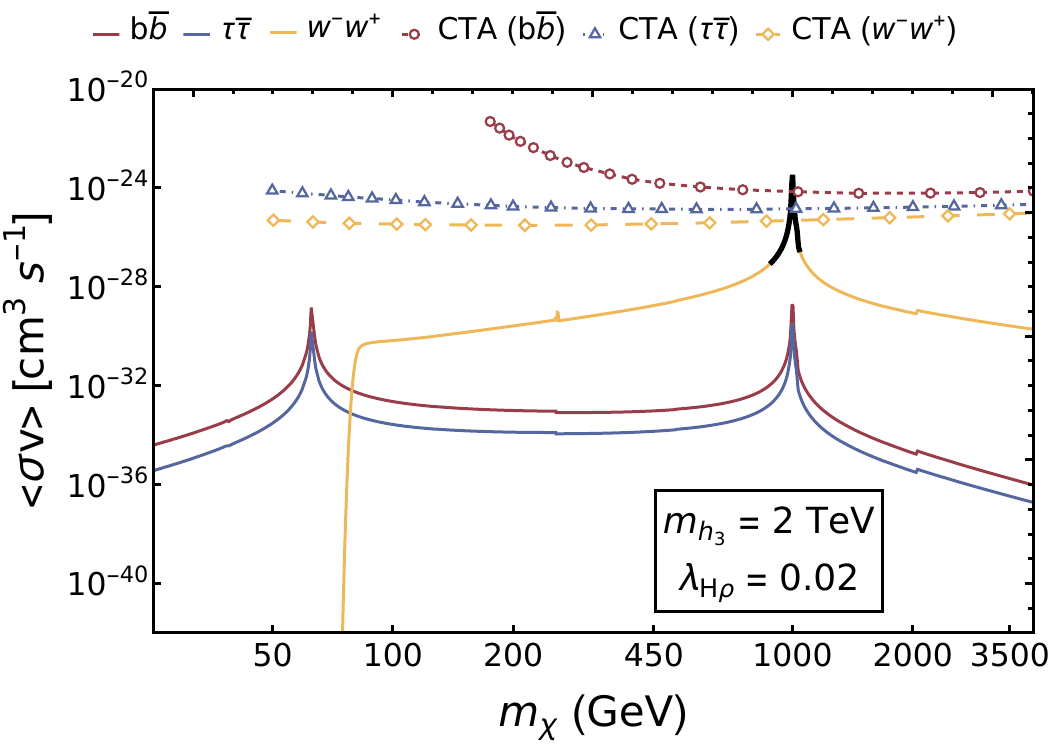}
    \caption{Indirect detection annihilation cross-section,~$\langle\sigma v\rangle$~[cm\(^3\)s\(^{-1}\)] as a function of DM mass,~$m_\chi$ from $10$ GeV to $3500$ GeV for two distinct sets of input parameters shown in left and right panel plots respectively. Existing constraints on diffuse gamma-rays from \texttt{Fermi-LAT}~\cite{Fermi-LAT:2011vow}, \texttt{H.E.S.S.}~\cite{HESS:2016mib} are shown while also projecting future sensitivity of \texttt{CTA}~\cite{CTAConsortium:2017dvg}. Black colored segments in \(W^+W^-\) channels show the region of parameter space where DM is not overproduced and allowed by \texttt{SIDD} constraints.}
    \label{fig:IDD}
    \end{figure}

In Fig.~\ref{fig:IDD}, we show the DM annihilation cross-sections for various SM final states namely~($b\overline{b}$, $\tau\overline{\tau}$, $W^+W^-$), alongside the latest constraints from the \texttt{Fermi-LAT}~\cite{Fermi-LAT:2015bhf}, \texttt{H.E.S.S.}~\cite{HESS:2016mib} and with future sensitivities of \texttt{CTA}~\cite{CTAConsortium:2017dvg} experiment. The plots in the left column correspond to $m_{h_3}=2$~TeV and $\lambda_{H\rho}=0.14$, clearly illustrating resonance enhancement of $\langle \sigma v \rangle$ at the SM Higgs pole $m_\chi = m_{h_1}/2$ and $m_\chi = m_{h_3}/2$. Furthermore, the $W^+W^-$ annihilation channel (depicted by the solid yellow curve) becomes accessible for $m_\chi > m_W$, imposing stringent constraints on the parameter space from the \texttt{Fermi-LAT} and \texttt{H.E.S.S.} data. Nevertheless, the black segments of the curve indicate regions of parameter space where the DM relic is not overproduced and remains consistent with \texttt{SIDD} constraints. These regions, therefore, offer a promising testbed for our framework in future indirect detection (\texttt{IDD}) experiments such as \texttt{CTA}.

A similar plot behavior is observed in the right-column plots, where $\lambda_{H\rho}$ is set to 0.02. In this case, the region of parameter space satisfying relic density requirements is smaller compared to the $\lambda_{H\rho}=0.14$ case, as can also be seen from Fig.~\ref{fig:DMrelic}. For a more conclusive analysis of indirect constraints, a detailed treatment of the annihilation cross-section around the $h_3$ resonance is necessary. However, this lies beyond the scope of the present work. In conclusion, a large part of the DM parameter space in our model is not constrained by the existing indirect detection bounds while some allowed region for the channel~$\chi\overline{\chi}\rightarrow W^+W^-$ face further scrutiny at upcoming \texttt{IDD} experiments.
 
\begin{figure}[ht!]
    \centering
    \includegraphics[width=0.49\linewidth]{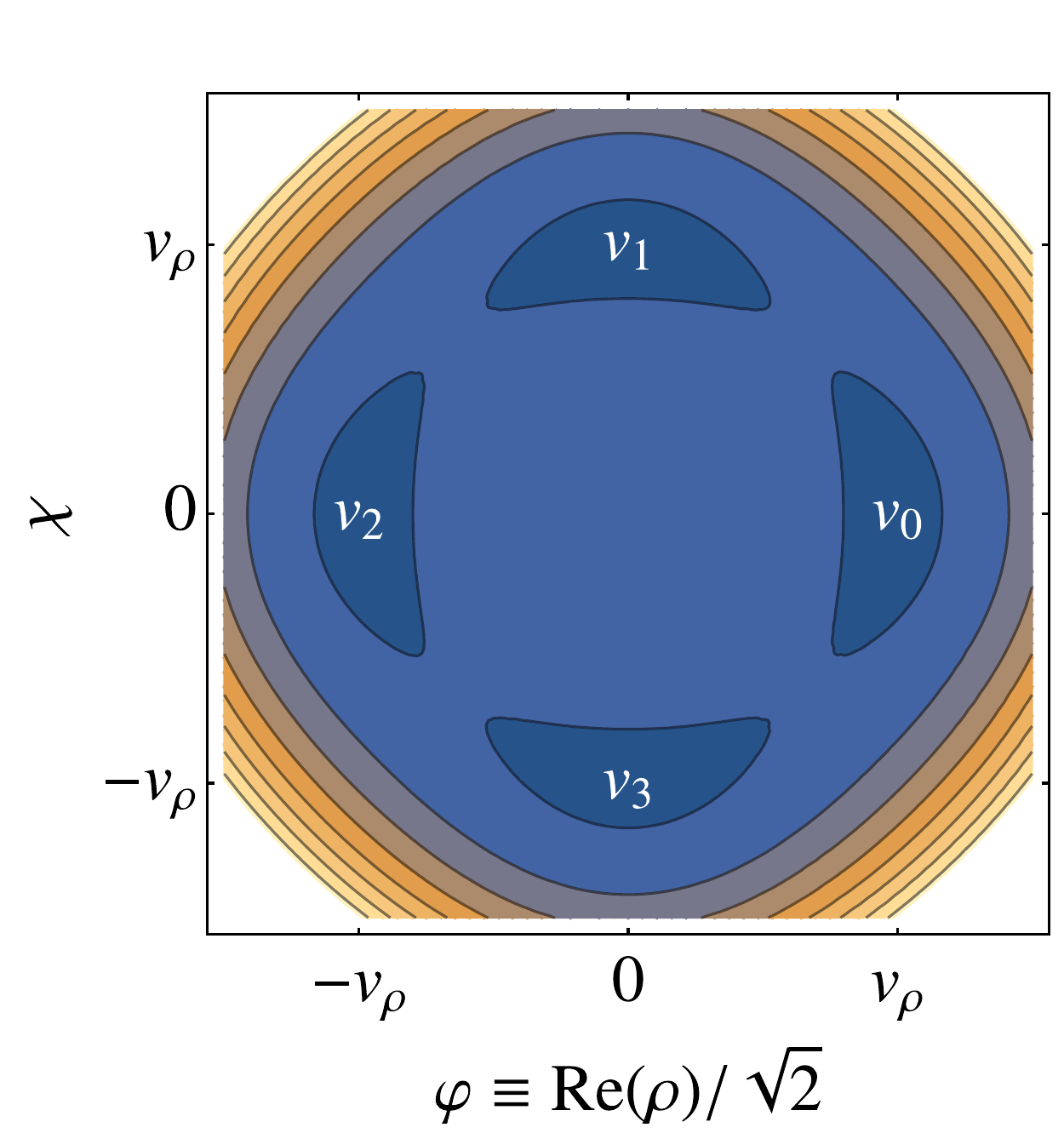}
    \includegraphics[width=0.49\linewidth]{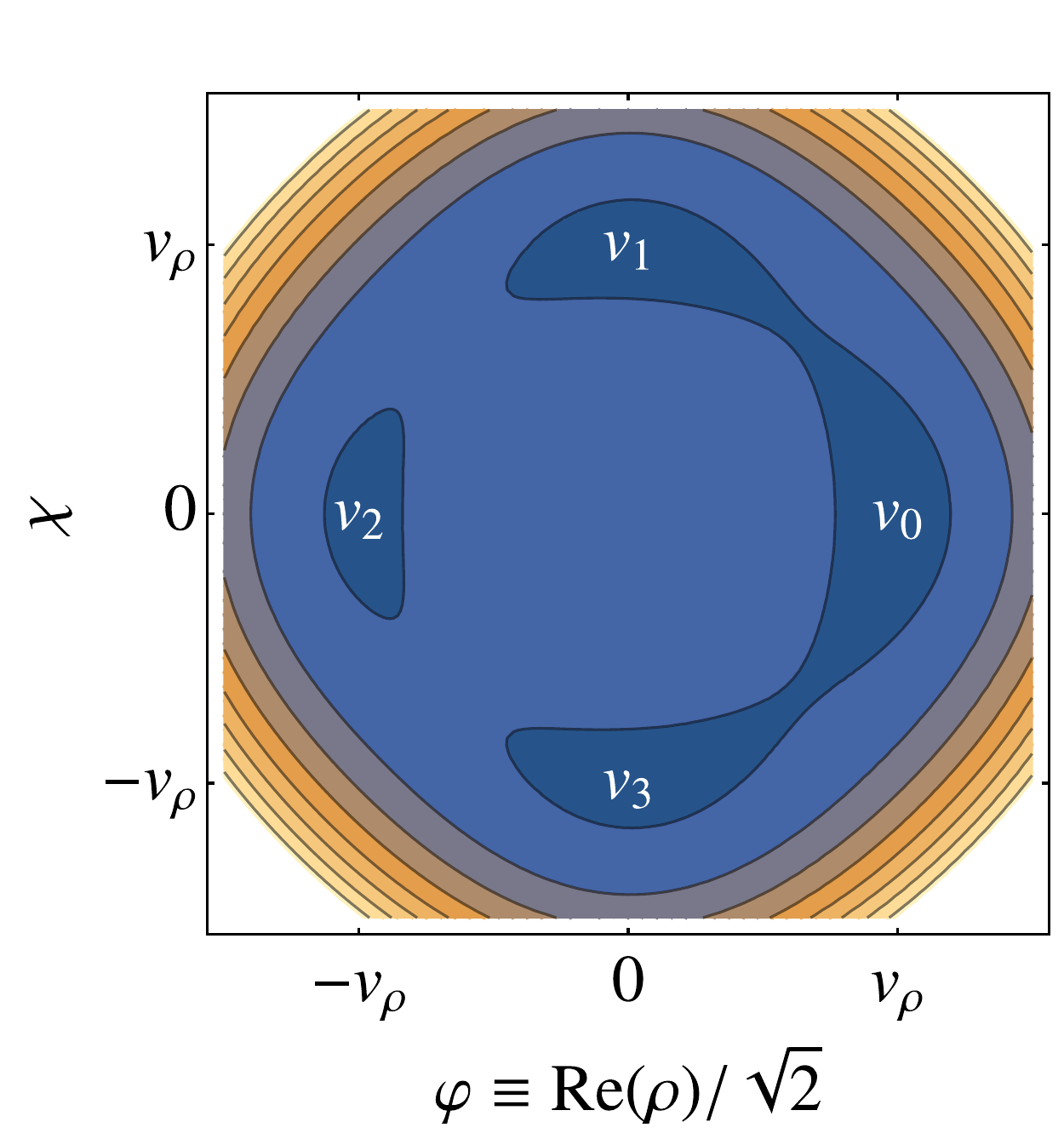}
    \caption{\small Contour plots of the $Z_4$ symmetric potential for no bias $\epsilon=0$ (\textbf{Left}) and with a bias $\epsilon=0.0002$ (\textbf{Right}), where $\epsilon$ is defined in Eq.~\eqref{eq:bias}. The $x$ and $y$-axes correspond to the real and imaginary parts of $\rho$ respectively. The \textbf{Left} plot shows the four degenerate minima denoted by $v_{0,1,2,3}$ defined in Eq~\eqref{eq:4minima}. In the \textbf{Right} plot, the degeneracy is broken by the bias.}
    \label{fig:potential_plot}
\end{figure}


\section{Gravitational waves from domain walls}
\label{sec:grav}
Domain walls (DW) are two-dimensional topological defects arising from spontaneous breaking of discrete symmetries \cite{Zeldovich:1974uw, Kibble:1976sj, Vilenkin:1981zs}. With the expansion of our Universe, the DW energy density falls slower compared to that of radiation or ordinary matter, and hence they could potentially start dominating the energy density to ruin the successful predictions of standard cosmology. Such a disastrous situation can be prevented if DW are made unstable or diluted or if they have asymmetric initial field fluctuations ~\cite{Coulson:1995nv, Krajewski:2021jje}.

In our setup, domain walls are formed as soon as $Z_4$ symmetry is spontaneously broken by the $\rho$ field acquiring a VEV. Since $\eta$ does not acquire a VEV until $\rho$ does, there are four degenerate minima of the potential given in Eq.~\eqref{eq:scalarpot} in the complex plane of $\rho$, denoted by $v_k$ where $k=0,1,2,3$ as shown in Fig.~\ref{fig:potential_plot} (\textbf{Left}). These minima are positioned at,
\begin{align}\label{eq:4minima}
    \nonumber v_0&\equiv\left(\braket{\varphi},\braket{\chi}\right)=(v_\rho,0)\\
    \nonumber v_1&\equiv\left(\braket{\varphi},\braket{\chi}\right)=(0,v_\rho)\\
    \nonumber v_2&\equiv\left(\braket{\varphi},\braket{\chi}\right)=(-v_\rho,0)\\
    v_3&\equiv\left(\braket{\varphi},\braket{\chi}\right)=(0,-v_\rho)
\end{align}
where $\varphi\equiv \text{Re}(\rho)/\sqrt{2}$ is the real component and $\chi$ is the imaginary component of $\rho$. In general, $\rho$ can assume any one of the four VEVs with equal probability. Notice that in minima $v_1$ and $v_3$, the $\chi$ field acquires a VEV, hence it can not be a stable dark matter candidate. 

A domain wall forms the boundary between two spatial regions where the field $\rho$ assumes different minima. Thus there are six possible configurations of domain walls in our model. The profile for a domain wall perpendicular to $z$-axis at position $z_0$, separating two adjacent minima is similar to axionic domain walls with $N=4$, given by~\cite{Borboruah:2022eex,Yajnik:1998sw,Hiramatsu:2012sc},
\begin{equation}
    \theta(z)=\tan^{-1}\exp\left(m_\theta(z-z_0)\right)
\end{equation}
where $\theta=\tan^{-1}(\chi/\varphi)$ and $m_\theta=\sqrt{8\lambda_1}v_\rho$. The wall width of such domain wall is approximately $\delta_{\rm adj.}\sim m_\theta^{-1}$ and the wall tension is given by~\cite{Borboruah:2022eex},
\begin{equation}
    \sigma_{\rm adj.}=\frac{m_\theta v_\rho^2}{2}.
\end{equation}
On the other hand, the profile of the non-adjacent domain walls separating $v_0$ and $v_2$ or $v_1$ and $v_3$ is given (similar to $Z_2$ domain walls) by~\cite{Borah:2022wdy},
\begin{equation}
    \phi(z)=v_\rho\tanh\left(\sqrt{\frac{\lambda_\rho}{2}}v_\rho(z-z_0)\right).
\end{equation}
The wall width is $\delta_{\rm non-adj.}\sim\left(\sqrt{2\lambda_\rho}v_\rho\right)^{-1}$ and the tension is given by,
\begin{equation}\label{eq:sigmanonadj}
    \sigma_{\rm non-adj.}=\frac{2\sqrt{2}}{3}\sqrt{\lambda_\rho}v_\rho^3.
\end{equation}
After formation, domain walls evolve under three forces: the tension force $ p_T = \sigma / L$ (where $\sigma$ is the wall tension and $L$ is the wall size), which tends to straighten and shrink the wall; the friction force from particle reflections in the plasma; and the pressure difference $ p_V = \delta V $ if there is an energy difference between the two minima that the wall separates. Assuming the walls to be formed after inflation, the simplest way to make them disappear is to introduce a small pressure difference \cite{Zeldovich:1974uw, Vilenkin:1981zs, Sikivie:1982qv, Gelmini:1988sf, Larsson:1996sp}, also known as the bias. Without any bias, the domain walls never cease to exist and eventually they dominate the energy density of the Universe, which can cause problems in late time evolution of the Universe and structure formation. However, neglecting friction, in presence of a bias, domain walls initially grow due to the Universe's expansion and then collapse and annihilate when $ p_T $ equals $ p_V $ \cite{Bai:2023cqj}, producing stochastic gravitational waves. One requires that the domain walls annihilate away before the big bang nucleosysthesis (BBN) commences at around a few MeV.

The domain walls in our model can be rendered unstable by adding a small bias term in the scalar potential given in Eq.~\eqref{eq:scalarpot},
\begin{equation}\label{eq:bias}
    V_{\rm bias}=-\epsilon\, v_\rho^3(\rho+\rho^*)=-\sqrt{2}\,\epsilon\, v_\rho^4\cos \theta
\end{equation}
where $\epsilon\ll1$ is the dimensionless bias parameter. Such bias terms can naturally arise due to quantum gravity (QG) scale suppressed operators like $(\rho^5+\eta^4 \rho+ (H^\dagger H)^2 \rho)/\Lambda_{\rm QG}$ given the fact that any theory of QG is expected to violate global symmetries explicitly \cite{Kallosh:1995hi, Witten:2017hdv, Rai:1992xw}. However, similar operators are also likely to allow DM decay if the latter is protected by a global symmetry, leading to interesting indirect detection aspects \cite{King:2023ayw, King:2023ztb, Borah:2024kfn}. While the scale of QG $\Lambda_{\rm QG}$ associated with higher dimensional operators breaking global symmetry explicitly is often taken to be the Planck scale $M_{\rm Pl}$, presence of non-perturbative effects can further lower the explicit symmetry breaking effects. As argued in \cite{King:2023ayw, King:2023ztb, Gouttenoire:2025ofv} and references therein, this can effectively lead to a much larger $\Lambda_{\rm QG}$ compared to $M_{\rm Pl}$. This can result in a much smaller bias parameter $\epsilon$ (compared to what dimension-5 Planck scale suppressed operators can give) as we use in remainder of our analysis. Another possibility is to consider additional gauge symmetries that prevent such explicit global symmetry breaking operators up to certain dimensions. Similar ideas have already been adopted in the context of the axion quality problem \cite{Barr:1992qq, Dias:2002gg, Carpenter:2009zs}. In such a case, higher powers of the QG or Planck scale leads to a smaller bias term. However, we remain agnostic about such UV completions behind the origin of explicit $Z_4$-breaking terms and consider the bias term as a free parameter in our analysis. For illustrative purposes, we show the details of dimension five QG scale suppressed operator origin of bias in appendix \ref{app:D} and compare it with the constraints from dark matter lifetime.

The term given in Eq. \eqref{eq:bias} introduces potential differences among the four minima,
\begin{equation}
\delta V_{\rm adj.}\equiv\delta V_{10}=\delta V_{21}=\delta V_{30}=\delta V_{23}=\sqrt{2}\,\epsilon\,v_\rho^4,\quad
\delta V_{\rm non-adj.}\equiv\delta V_{20}=2\sqrt{2}\,\epsilon\,v_\rho^4,\quad   \delta V_{13}=0 
\end{equation}
where $\delta V_{ij}=V_i-V_j$ is the difference of potential energy at minima $v_i$ and $v_j$. This is seen in Fig.~\ref{fig:potential_plot} (\textbf{Right}) for $\epsilon = 0.0002$. The domain walls in our model annihilate away when $p_V = p_T$, i.e. on a time scale $1/L\sim \epsilon\, v_\rho$ due to the vacuum pressure. However, the annihilation time scale of domain walls between adjacent and non-adjacent minima are slightly different due to the difference in the pressure and tension.

The requirement that the domain walls would scale at least to the Hubble size before collapsing provides an upper bound on the bias~\cite{Borboruah:2022eex,Hiramatsu:2010yz,Hiramatsu:2013qaa},
\begin{equation}\label{eq:scalingbound}
    \delta V<\sigma \,\pi\sqrt{\frac{g_*}{90}}\frac{v_\rho^2}{M_{\rm Pl}}
\end{equation}
where $\sigma$ is the wall tension. We will consider the tension of the adjacent walls $\sigma_{\rm adj.}$ for this bound. On the other hand, bias can not be arbitrarily small lest the domain walls dominate the energy density of the Universe. This condition puts a lower bound on the bias~\cite{Borah:2022wdy,Borboruah:2022eex,Hiramatsu:2010yz,Hiramatsu:2013qaa},
\begin{equation}
    \label{eq:walldominationbound}
    \delta V>\frac{\sigma^2}{M_{\rm Pl}^2}
\end{equation}
for which we will consider the tension of the non-adjacent walls $\sigma_{\rm non-adj.}$.

Assuming the DW to vanish within the radiation-dominated era, the peak amplitude and the peak frequency of the GWs produced, as seen at the present epoch $t_0$, are given by \cite{Kadota:2015dza,Hiramatsu:2013qaa,Borboruah:2022eex},
\begin{align}
    \label{eq:GWpeakAmp}
    \Omega_{\mathrm{gw}} h^{2}\left(t_{0}\right)_{\text {peak }} &\simeq 5.20 \times 10^{-20} \times \tilde{\epsilon}_{\mathrm{gw}} \mathcal{A}^{4}\left(\frac{10.75}{g_{*}}\right)^{1 / 3}\left(\frac{\sigma_{\text {wall }}}{1 \mathrm{TeV}^{3}}\right)^{4}\left(\frac{1 \mathrm{MeV}^{4}}{\delta V}\right)^{2},\\
    \label{eq:GWpeakFreq}
    f_{\text{peak}} &\simeq 3.99 \times 10^{-9} \mathcal{A}^{-1 / 2}\left(\frac{1 \mathrm{TeV}^{3}}{\sigma_{\text {wall }}}\right)^{1 / 2}\left(\frac{\delta V}{1 \mathrm{MeV}^{4}}\right)^{1 / 2}\mathrm{~Hz},
\end{align}
where $\mathcal{A}$ is the area parameter \cite{Hiramatsu:2012sc,Hiramatsu:2013qaa, Paul:2020wbz} and $\tilde{\epsilon}_{\text{gw}}$ is called the efficiency parameter $\sim0.7$~\cite{Hiramatsu:2013qaa}. $g_*$ is the relativistic degrees of freedom at the time of the DW decay and we take $g_*\sim114$, appropriate for our model. The GW spectrum is approximately given by,
\begin{equation}
    \label{eq:spectrum}
    \left.\Omega_{\mathrm{GW}} \simeq \Omega_{\mathrm{GW}}\right|_{\text {peak }} \times\left\{\begin{array}{ll}
\left(\frac{f_{\text {peak }}}{f}\right) & \text { for } f>f_{\text {peak }} \\
\left(\frac{f}{f_{\text {peak }}}\right)^{3} & \text { for } f<f_{\text {peak }}
\end{array}\right.
\end{equation}

\begin{figure}[ht!]
    \centering
    \includegraphics[width=0.48\linewidth]{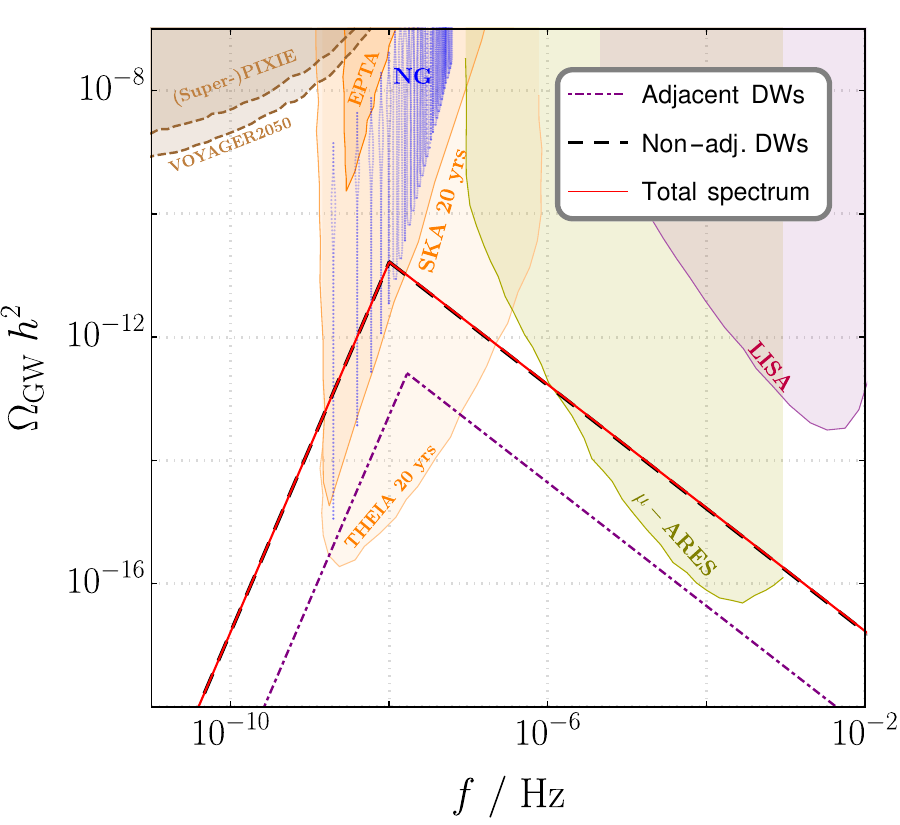}
    \includegraphics[width=0.48\linewidth]{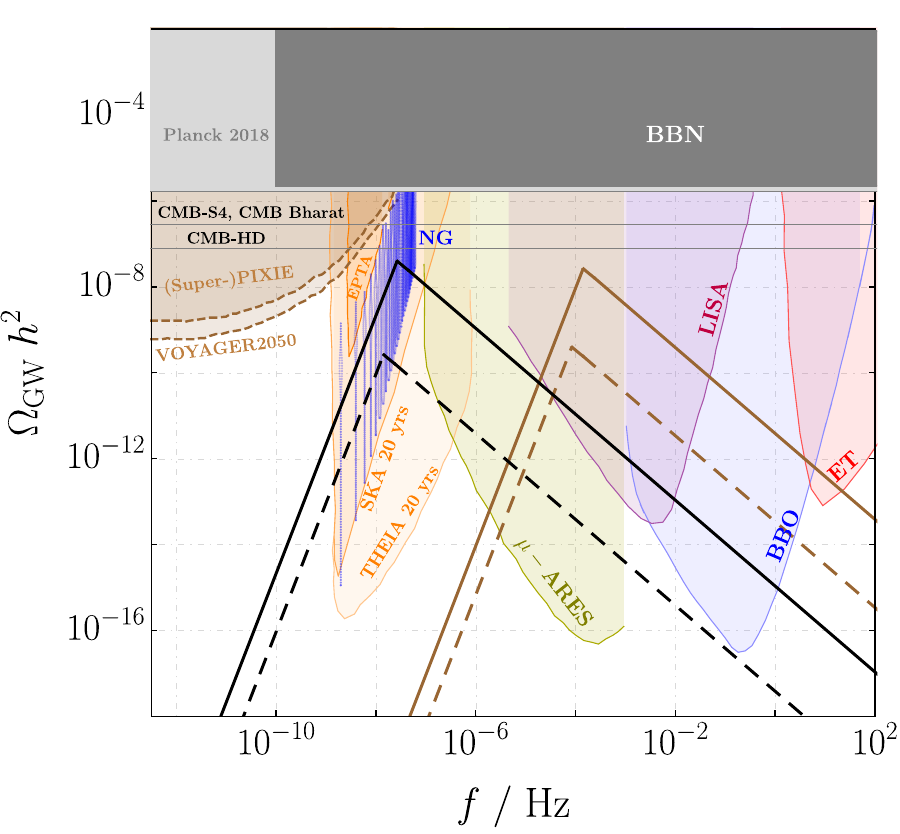}
    \caption{\small Example GW spectra assuming Higgs resonance $m_\chi=m_{h_3}/2$. (\textbf{Left}) Spectra for $\epsilon=10^{-26}$  and $v_\rho = 10^5 $ GeV, where dashed black curve is the contribution from non-adjacent DWs, dot-dashed purple curve is from adjacent DWs and solid red curve is the final spectrum (sum of black and purple). (\textbf{Right}) Total GW spectra for $\epsilon = 10^{-26}$ with $v_\rho = 7 \times 10^5$ GeV (black solid curve) and $v_\rho = 2 \times 10^5$ GeV (black dashed curve), and for $\epsilon = 10^{-21}$ with $v_\rho = 2 \times 10^8$ GeV (brown solid curve) and $v_\rho = 7 \times 10^7$ GeV (brown dashed curve), highlighting signals in the NANOGrav and LISA sensitivity ranges. For both plots, we take $ \lambda_\rho = 0.1, \lambda_1\sim\lambda_\rho/72 $ assuming $\mu_1/v_\rho \ll 1$ [see Eq.~\eqref{eq:resonancecond}].}
    \label{fig:Spectrum}
\end{figure} 

Starting from a homogeneous medium, one needs to calculate the statistical distribution of adjacent and non-adjacent domain walls in the Universe after spontaneous breaking of $Z_4$ symmetry. Initially, domains labeled by $v_i$ (with $i=0,1,2,3$) are created with equal probabilities. The domain wall between $v_1$ and $v_3$ is excluded from our analysis since the associated bias, $\delta V_{13}=0$, and its collapse dynamics is more complex due to the influence of surrounding domain walls. The remaining biases, $\delta V_{ij}$, consist of four biases corresponding to adjacent domain walls and one bias for non-adjacent domain walls. As a result, approximately 4/5th of the domain walls evolve under the influence of the adjacent bias, $\delta V_{\rm adj.}$, while 1/5th experience the non-adjacent bias, $\delta V_{\rm non-adj.}$~\cite{Bai:2023cqj}. For these two different biases we can calculate the GW spectra separately and take a weighted sum to obtain the final spectrum, where the weights are the respective fractions of the adjacent and non-adjacent domain walls. For $Z_4$ domain walls we take the area parameter $\mathcal{A}\sim1.5$ according to the results from~\cite{Hiramatsu:2012sc}.

\begin{figure}[ht!]
    \centering
    \includegraphics[width=0.6\linewidth]{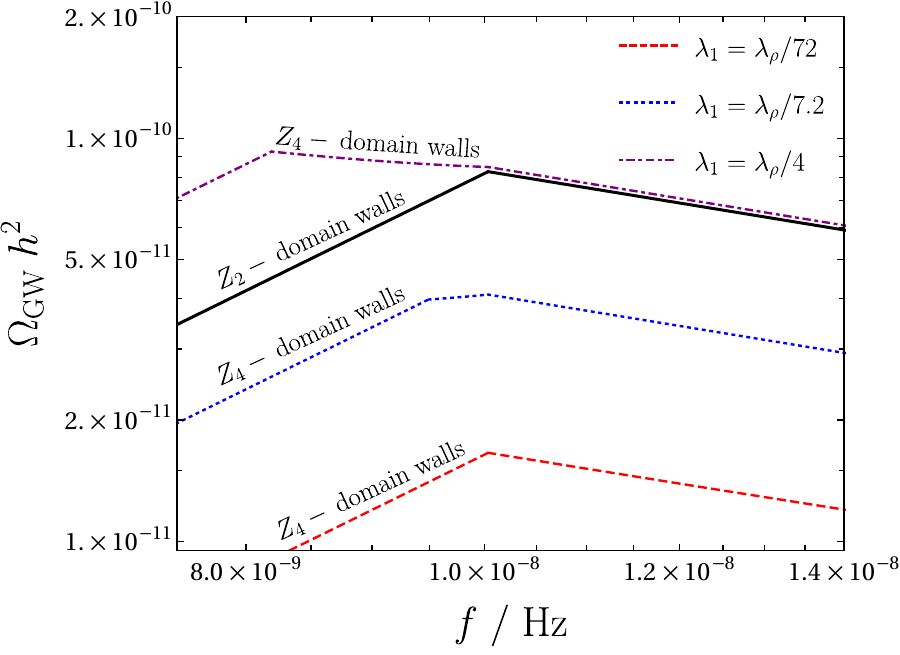}
    \caption{\small Comparison of GW spectra from $Z_4$ domain walls in our model to that from generic $Z_2$ domain walls. The solid black curve represents the standard spectra from $Z_2$ domain walls for $\lambda_\rho=0.1, v_\rho=10^5$ GeV and $\epsilon=10^{-26}$. The red dashed, blue dotted, and purple dot-dashed curves represent the total spectra from the $Z_4$ domain walls for the same parameter values as above, but for $\lambda_1=\lambda_\rho/72,\lambda_\rho/7.2$ and $\lambda_\rho/4$, respectively.}
    \label{fig:spectraCompare}
\end{figure} 

The gravitational wave energy density must satisfy the bound on dark radiation, parameterized by the effective number of neutrino species, $\Delta N_{\rm eff}$~\cite{Luo:2020fdt,Maggiore:1999vm}:
\begin{equation}
    \int_{f_{\rm min}}^{\infty} \frac{df}{f} \, \Omega_{\rm GW}(f) h^2 \leq 5.6 \times 10^{-6} \Delta N_{\rm eff}
\end{equation}
We approximate this by setting $\Omega_{\rm GW} \leq 5.6 \times 10^{-6} \Delta N_{\rm eff}$ in our spectra. Current bounds are $\Delta N_{\rm eff}^{\rm BBN} \simeq 0.4$ from BBN~\cite{Cyburt:2015mya} and $\Delta N_{\rm eff}^{\rm Planck+BAO} \simeq 0.28$ from Planck+BAO~\cite{Planck:2018vyg}. Future projections include $\Delta N_{\rm eff}^{\rm Proj.} = 0.014$ (CMB-HD)~\cite{CMB-HD:2022bsz}, $\Delta N_{\rm eff}^{\rm Proj.} = 0.05$ (CMB-Bharat)~\cite{CMB-bharat}, $\Delta N_{\rm eff}^{\rm Proj.} = 0.06$ (CMB-S4, PICO)~\cite{doi:10.1146/annurev-nucl-102014-021908,Alvarez:2019rhd}, and $\Delta N_{\rm eff}^{\rm Proj.} \lesssim 0.12$ (CORE, SPT, Simons)~\cite{CORE:2017oje,SPT-3G:2014dbx,SimonsObservatory:2018koc}.

In Fig.~\ref{fig:Spectrum} we show example GW spectra with parameter values $\epsilon=10^{-26}, \lambda_\rho = 0.1$ along with the Higgs resonance condition~$\lambda_1\sim\lambda_\rho/72 $, given in Eq.~\eqref{eq:resonancecond}, ignoring $\mu_1$. Shaded colored regions show the noise curves of various current and future GW experiments like BBO~\cite{Yagi:2011wg}, LISA~\cite{2017arXiv170200786A}, ET~\cite{Punturo_2010}, THEIA~\cite{Garcia-Bellido:2021zgu}, $\mu$ARES~\cite{Sesana:2019vho}, EPTA \cite{Antoniadis:2023ott} and SKA~\cite{Weltman:2018zrl} together with the NANOGrav (NG) results \cite{NANOGrav:2023gor}. Future experiments like Super-PIXIE~\cite{Kogut:2024vbi} and VOYAGER 2050~\cite{Basu:2019rzm} are aimed at detecting gravitational wave signatures through spectral distortions in the cosmic microwave background. The gray regions represent the $\Delta N_{\rm eff}$ bounds from BBN and Planck while and horizontal lines are the projections of $\Delta N_{\rm eff}$ sensitivity from CMB-S4, CMB Bharat and CMB-HD. The \textbf{Left} plot shows the contributions to the GW spectra from non-adjacent (red dashed curve) and adjacent (dot-dashed purple curve) domain walls. The solid black curve is the total spectra (sum of red and purple curves). We see that in the Higgs resonant region, the major contribution to the total GW spectrum comes from non-adjacent walls due to their high tension. For this plot we considered $Z_4$ breaking scale $v_\rho = 10^5$ GeV. 
The \textbf{Right} plot demonstrates the total GW spectra for $\epsilon = 10^{-26}$ but with $v_\rho = 7 \times 10^5$ GeV (black solid curve) and $v_\rho = 2 \times 10^5$ GeV (black dashed curve), and for $\epsilon = 10^{-21}$ with $v_\rho = 2 \times 10^8$ GeV (brown solid curve) and $v_\rho = 7 \times 10^7$ GeV (brown dashed curve), demonstrating sensitivity in NANOGrav and LISA. For a fixed bias, the peak amplitude and peak frequency of the GW spectrum increases with increasing $v_\rho$. We see that for $v_\rho$ of the order of $10^5$ GeV, the peak of the spectra generally lies within the sensitivity region of NANOGrav and other PTA experiments, as well as LISA, $\mu$ARES, SKA and THEIA.

For completeness, we also demonstrate the differences in the GW spectra originating from $Z_4$ domain walls in our model and from a generic $Z_2$ domain wall scenario of earlier works \cite{Barman:2022yos, Barman:2023fad} in Fig.~\ref{fig:spectraCompare}. The $Z_2$ domain walls are identical to the non-adjacent walls in our model. However, they contribute 100\% to the total spectrum of $Z_2$-symmetric models, in contrast to the 1/5th fraction contributed by the non-adjacent walls in our model. In Fig.~\ref{fig:spectraCompare}, the solid black curve represents the standard spectra from $Z_2$ domain walls for $\lambda_\rho=0.1, v_\rho=10^5$ GeV and $\epsilon=10^{-26}$. The red dashed, blue dotted, and purple dot-dashed curves represent the total spectra from the $Z_4$ domain walls for the same parameter values as above, but for $\lambda_1=\lambda_\rho/72,\lambda_\rho/7.2$ and $\lambda_\rho/4$, respectively. As expected, for small $\lambda_1\lesssim\lambda_\rho/72$ values, the contribution from adjacent domain walls is subdominant; therefore, the amplitude of the spectra from $Z_4$ walls is just 1/5th of that from $Z_2$ walls. The peak frequency remains the same in this case. However, for large $\lambda_1$ values, the contribution of adjacent domain walls increases and changes the total spectra for $Z_4$ walls. For large enough $\lambda_1$, adjacent walls can dominate the total spectra and change the peak frequency and amplitude compared to the $Z_2$ case. 

An important fact to consider here is that the analytical expressions in Eq.~\eqref{eq:GWpeakAmp} and Eq.~\eqref{eq:GWpeakFreq} used to calculate the GW spectra are derived from numerical simulations assuming instantaneous annihilation of domain walls in the scaling regime. However, in realistic scenarios, DW annihilation is delayed depending on the balance between the bias and the tension. It has been mentioned that the delayed annihilation of DWs shifts the peak frequency of the spectrum to a lower value as compared to the frequency corresponding to Hubble while late time production of gravitational waves during annihilation can change the shape of the spectrum around the peak and at higher frequencies~\cite{Kitajima:2023cek,Ferreira:2023jbu,Dankovsky:2024zvs}. Here we do not consider such nuances while adding the spectra from annihilation of adjacent and non-adjacent DWs in Fig.~\ref{fig:spectraCompare} and simply consider that both types of DWs annihilate instantaneously.

Interferometers measure gravitational wave displacements in terms of strain noise, denoted as $ h_{\text{GW}}(f) $, which relates to the GW amplitude and converts to an energy density via,

\begin{equation}
\Omega_{\text{exp}}(f) h^2 = \frac{2 \pi^2 f^2}{3 \mathcal{H}_0^2} h_{\text{GW}}(f)^2 h^2
\end{equation}
where $ \mathcal{H}_0 = h \times 100 \, \text{km/s/Mpc} $ is the Hubble constant. To evaluate detectability, we calculate the signal-to-noise ratio (SNR) \cite{Dunsky:2021tih, Schmitz:2020syl} based on experimental sensitivity $ \Omega_{\text{exp}}(f)h^2 $ as
\begin{equation}
\text{SNR} = \sqrt{\tau \int_{f_{\text{min}}}^{f_{\text{max}}} df \left(\frac{\Omega_{\text{GW}}(f) h^2}{\Omega_{\text{exp}}(f) h^2}\right)^2}
\end{equation}
where we take $ h = 0.7 $, observation time $ \tau = 4 $ years, and set the detection threshold at $ \text{SNR} \geq 10 $.




\section{CMB prospects of detecting enhanced $N_{\rm eff}$ from Dirac neutrinos}
\label{sec:Neff}
In addition to gravitational waves, light Dirac neutrino $\nu_R$ can also behave like dark radiation and contribute to $\Delta N_{\rm eff}$. Excessive amount of $\nu_R$ in thermal equilibrium in the early Universe could alter the BBN or the CMB, rendering our model incompatible with standard cosmology. However, if the right-handed neutrinos decouple from the thermal bath much earlier than their left-handed counterparts, they contribute a fixed values $\Delta N_{\rm eff}=0.14$ for three generations of $\nu_R$~\cite{Borah:2024gql, Heeck:2023soj}. We can calculate the freeze-out temperature $T_{\rm f.o.}^{\nu_R}$ of $\nu_R$ with the condition,
\begin{equation}
    \label{eq:Tfo_condition}
    \Gamma_{\rm sc}\left(T_{\rm f.o.}^{\nu_R}\right)\sim\mathcal{H}\left(T_{\rm f.o.}^{\nu_R}\right),
\end{equation}
where $\Gamma_{\rm sc}\left(T\right)$ is the rate of 2-to-2 elastic scattering process $\nu_R\eta\leftrightarrow\nu_R\eta$ via exchange of $N_L$, that is primarily responsible for bringing $\nu_R$ into kinetic equilibrium with the SM bath, assuming $\eta$ is in thermal equilibrium. At high temperatures, the scattering rate can be estimated as,
\begin{equation}
    \label{eq:nuRthermarate}
    \Gamma_{\rm sc}\left(T\right)\sim \left(|(Y_R)_{1i}||(Y_R)_{1j}|\right)^2\frac{T^3}{M_1^2}
\end{equation}

\begin{figure}[ht!]
\center
\includegraphics[width=0.455\linewidth]{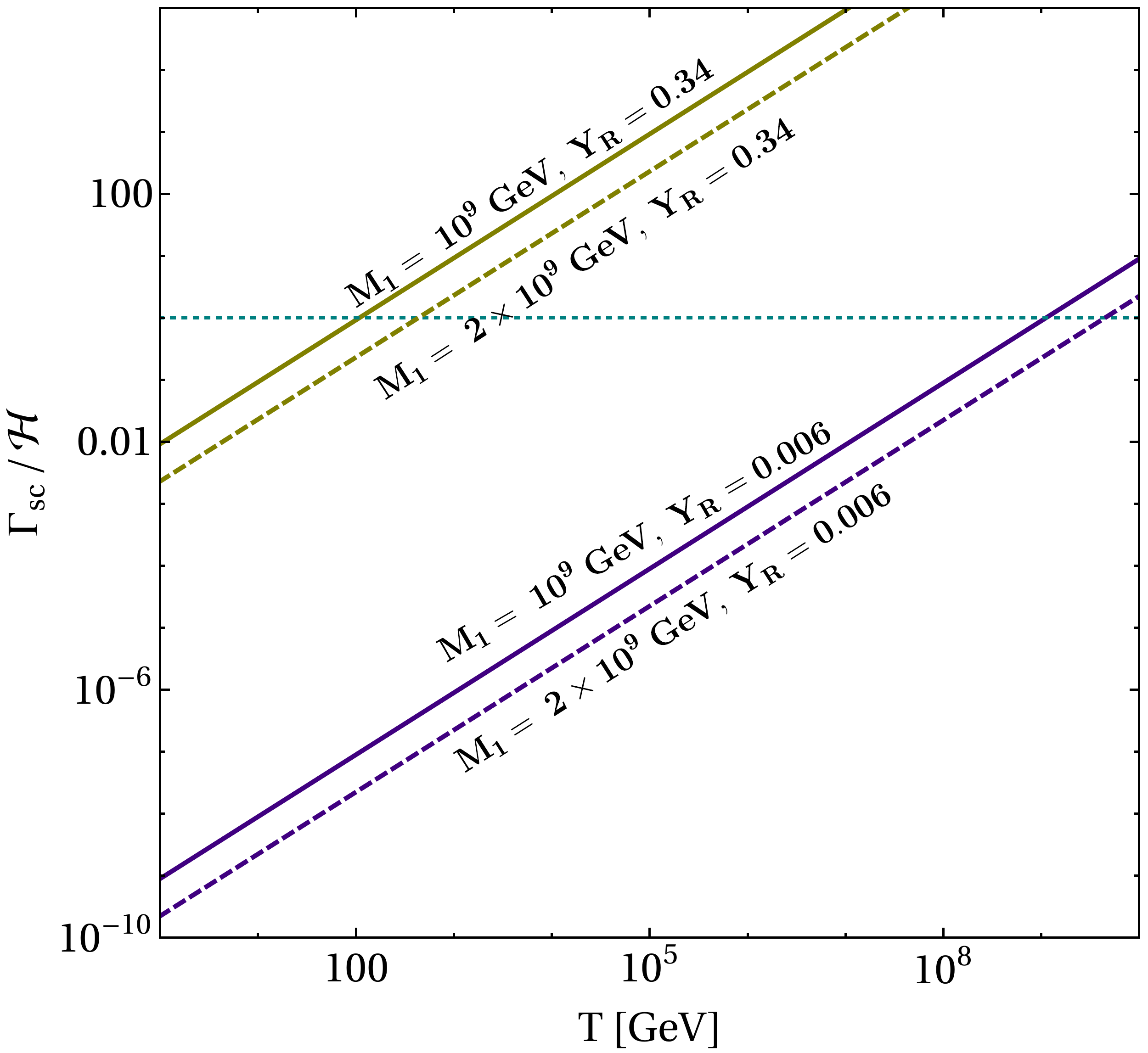}
\includegraphics[width=0.48\linewidth]{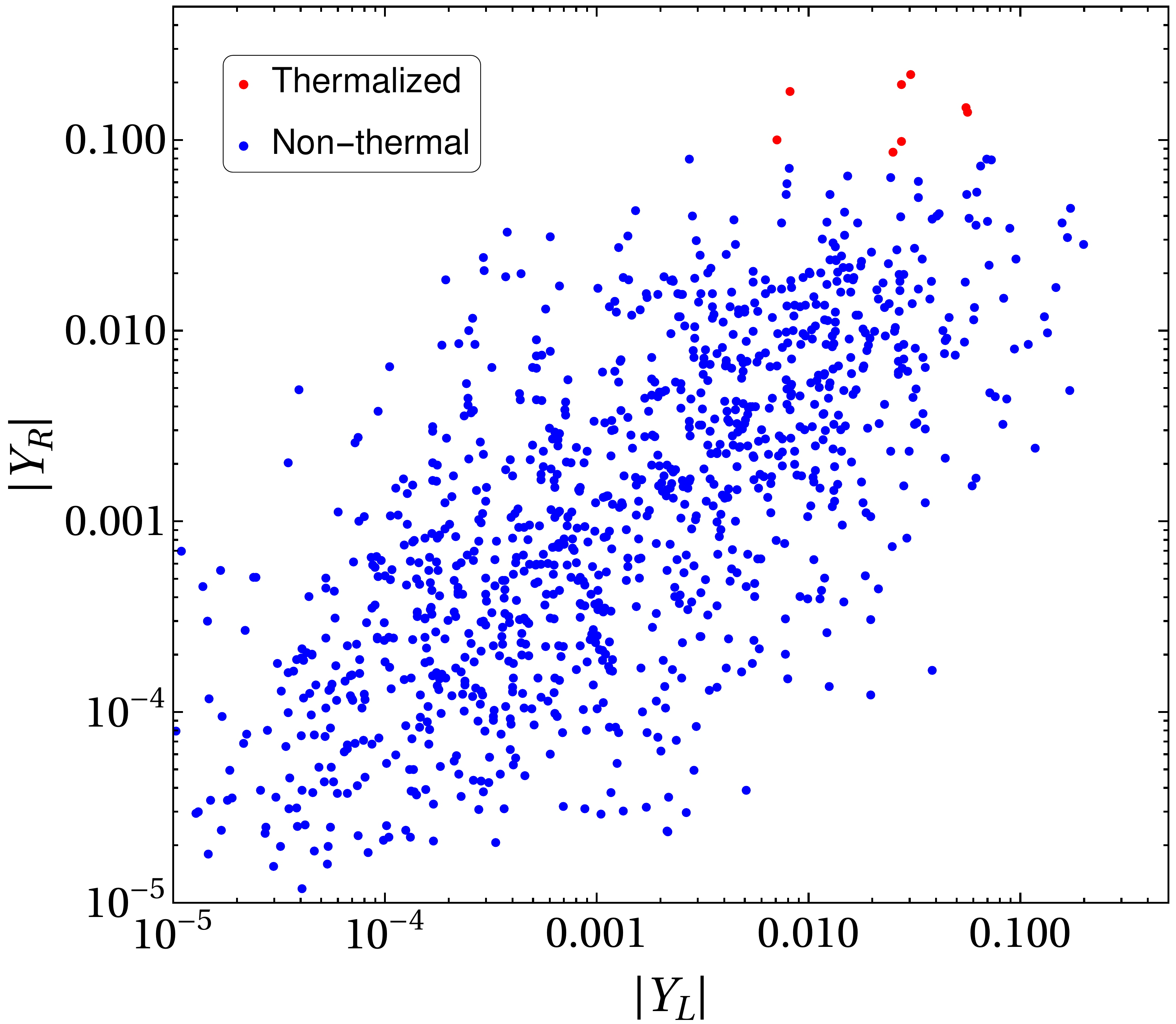}
\caption{\small \textbf{(Left)} Ratio of scattering rate to Hubble rate for $\nu_R$ as a function of temperature, assuming elastic scattering process $\nu_R\eta\leftrightarrow\nu_R\eta$ via exchange of Dirac fermion $N_L$. \textbf{(Right)} Scatter plot of largest elements of $Y_L,Y_R$ for the points in Fig.~\ref{fig:leptoDM} as a result of our random scan. The red points indicate that $\nu_R$ gets thermalized at high temperatures and then they decouple. The blue points indicate that $\nu_R$ are never thermalized. Detailed discussion of parameter scan is given in Sec.~\ref{sec:results}.}
\label{fig:nuRtherm}
\end{figure}

In Fig.~\ref{fig:nuRtherm} \textbf{(Left)} plot, we show the ratio $\Gamma_{\rm sc}/\mathcal{H}$ as a function of $T$ for different benchmark values of $|(Y_R)_{1i}|\sim|(Y_R)_{1j}|$ and $M_1$. We see that in the considered parameter space for leptogenesis, i.e. $M_1\gtrsim 10^9$ GeV, $\nu_R$ either never thermalizes or decouples from the thermal bath much earlier than electroweak scale.
If they thermalise and decouple before the electroweak scale, the contribution of $\nu_R$ to $\Delta N_{\rm eff}$ is 0.14 which is well within current bounds from CMB and BBN and can be probed by future experiments such as CMB-S4 or CMB-HD. 
From Eq.~\eqref{eq:Tfo_condition}, we can estimate an upper bound on $|\left(Y_R\right)_{1i}|$ that makes the $\nu_R$ to decouple before electroweak scale $\sim100$ GeV,
\begin{equation}\label{eq:Neff condition}
    |\left(Y_R\right)_{1i}|\lesssim0.34\left(\frac{M_1}{10^9\,\text{GeV}}\right)^{1/2}.
\end{equation}
The condition that makes sure that $\nu_R$ thermalizes at temperatures $T\geq M_1$ can be easily estimated as,
\begin{equation}
    |\left(Y_R\right)_{1i}|\gtrsim0.006\left(\frac{M_1}{10^9\text{ GeV}}\right)^{1/4}.
\end{equation}
In Fig.~\ref{fig:nuRtherm} \textbf{(Right)} plot, we show the largest values of elements of $Y_L,Y_R$ matrices obtained from our parameter scan discussed in Sec.~\ref{sec:results}. The red points indicate thermalized $\nu_R$ while blue points indicate that $\nu_R$ never gets thermalized.

\section{Results and discussion}
\label{sec:results}
In this section we discuss the results of our parameter scan and explore the connections of neutrino masses and leptogenesis with DM phenomenology and GW. We perform the parameter scan by fixing $\mu_\eta^2=10$ GeV$^2$, $\lambda_{H\eta}=\lambda_{\rho\eta}=0, \lambda_{H \rho} = 0.02$ and generating 5000 random parameter points within the ranges,
\begin{align}
    \nonumber m_\chi&\in[1,10^5]\,\text{GeV},\,~~\lambda_\rho\in[10^{-4},1],\,~~\lambda_\eta\in[10^{-4},1], \\ & v_\rho\in[10^3,10^8]\,\text{GeV},\,~~\mu_1\in[10^{-4},10^3]\,\text{GeV}
\end{align}
where we have considered the Higgs resonance condition $m_\chi=m_{h_3}/2$ which transcribes to Eq.~\eqref{eq:resonancecond}. We calculate the DM relic abundance, $\sigma_{\texttt{SIDD}}$, $\langle\sigma v\rangle_{\text{ann.}}$ as-well-as the $S,T,U$ parameters using \texttt{micrOmegas} and \texttt{SPheno} for each of these points.

\begin{figure}[ht!]
\center
\includegraphics[width=\linewidth]{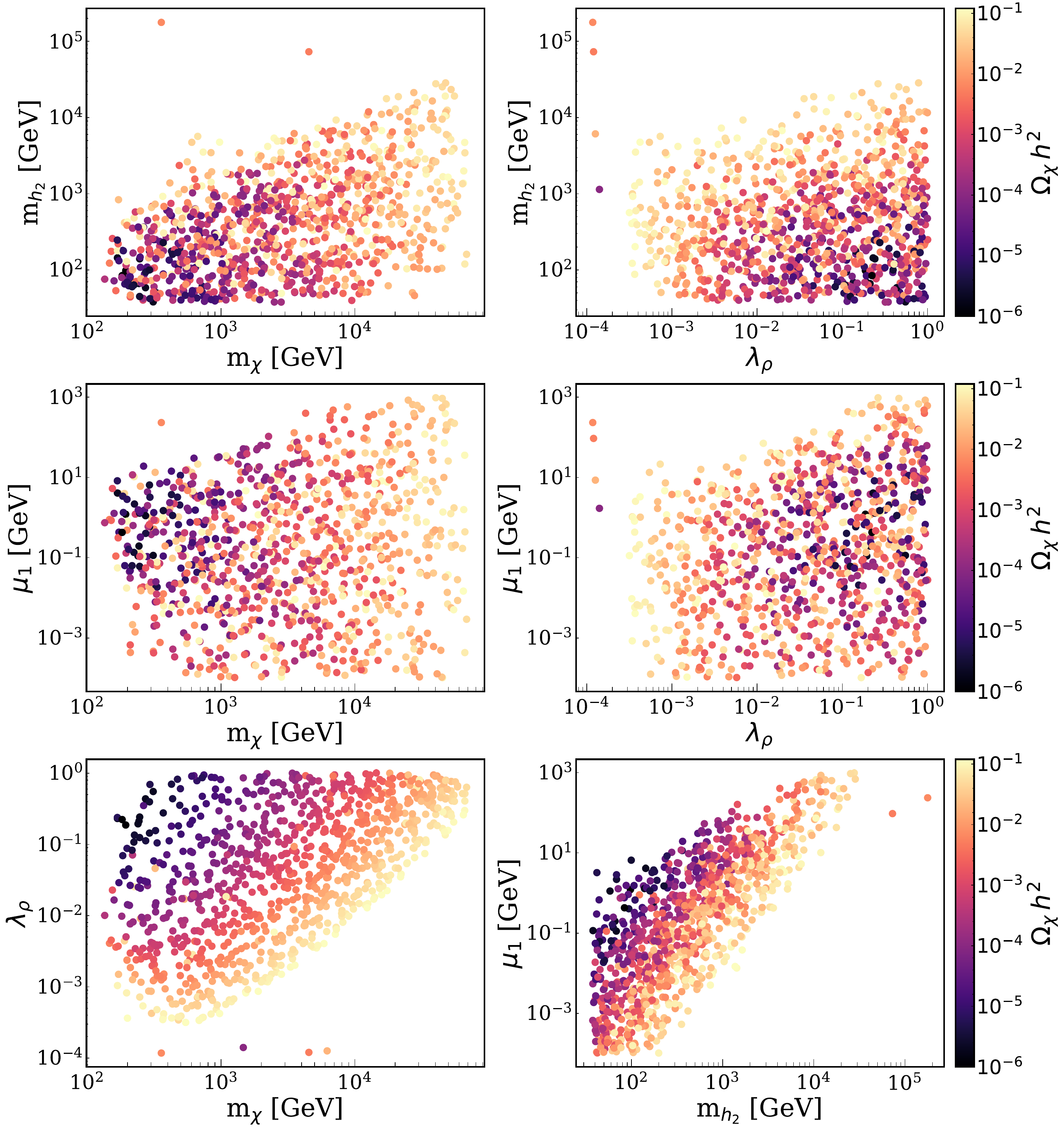}
\caption{\small Parameter space consistent with DM relic density bound $\Omega_\chi\,h^2\lesssim0.12$, DM direct detection bound (see sec.~\ref{ssubsec:dd}) and electroweak precision bounds at $1\sigma$ confidence level (see sec.~\ref{sec:STU}). Here we have taken $\lambda_{H\rho}=0.02$ and assumed the Higgs resonance condition~\eqref{eq:resonancecond}. Details of the random parameter scan is described in sec.~\ref{sec:results}. Correlations among relevant physical masses, scales and parameters are shown. The color-bar signifies the relic density.}
\label{fig:corr_lamHRho0.02}
\end{figure}

In Fig.~\ref{fig:corr_lamHRho0.02}, we show the result of this random parameter scan. Every colored point in the plot satisfies the correct DM relic density condition $\Omega_{\chi}h^2\lesssim0.12$, direct detection bound and the electroweak precision bounds on the oblique parameters $S,T,U$ within $1\sigma$ confidence level. The correlations among different physical masses ($m_{h_{2,3}},m_\chi$), scales ($\mu_1,v_\rho$) and dimensionless parameter $\lambda_\rho$ are shown. Here we considered the resonance condition given in Eq.~\eqref{eq:resonancecond}. One can see the viable parameter space of the model from this plot. 

The DM sector described above is connected to leptogenesis and neutrino masses via the induced VEV, $v_\eta$ given in Eq.~\eqref{eq:veta}. From Fig.~\ref{fig:corr_lamHRho0.02} we find that the viable range of $v_\eta$ is approximately $(50,10^6)$ GeV. Hence we perform a second scan involving $v_\eta, Y_L,Y_R$ and $M_N$ to show the parameter space that fits neutrino oscillation data and generates enough baryon asymmetry. We assume the structure of the mass matrix $M_N=\text{diag}[M_1,10M_1,20M_1]$ to be diagonal without loss of generality. The free parameters are $v_\eta, M_1$ and elements of the complex matrix $\mathcal{R}$ defined in Eq.~\eqref{eq:casas}. We randomly sample values of $v_\eta,M_1$ in the range,
\begin{equation}
    v_\eta\in[50,10^6]\;\text{GeV},\;M_1\in[10^9,10^{15}]\;\text{GeV},
\end{equation}
and elements of the complex matrix $\mathcal{R}$ with magnituded in the range $[10^{-4},10]$. As discussed earlier, we consider only normal hierarchy (NH) of neutrino masses with lightest neutrino mass $m_1=10^{-6}$ eV and the central values for oscillation parameters given in Table~\ref{tab:nuOsc}. Then we calculate $Y_L,Y_R$ using the Casas-Ibarra parametrization with $U_L=U_R\simeq U_{\rm PMNS}$ together with the CP asymmetry parameter using Eq.~\eqref{eq:CPparam} and the decay parameter $K$ using Eq.~\eqref{eq:K}. With the values of $\varepsilon$ and $K$, we calculate the final baryon asymmetry $\eta_B$ using Eq.~\eqref{eq:final asymmetry} and~\eqref{eq:kappaf} and choose the points that generate correct baryon asymmetry of the Universe in 2$\sigma$ range imposed by BBN, $5.8\times10^{-10}\leq\eta_B\leq6.3\times 10^{-10}$~\cite{ParticleDataGroup:2024cfk}. Additionnaly we make sure that each chosen point satisfies the conditions in Eq.~\eqref{eq:baryo_condition} and~\eqref{eq:Neff condition}. Moreover we also make sure the elements of the Yukawa matrices satisfy the condition $|Y_L|<0.3,|Y_R|<0.5$ from the requirement of stability of the scalar potential described in Appendix~\ref{app:beta}.

\begin{figure}[ht!]
    \centering
    \includegraphics[width=0.7\linewidth]{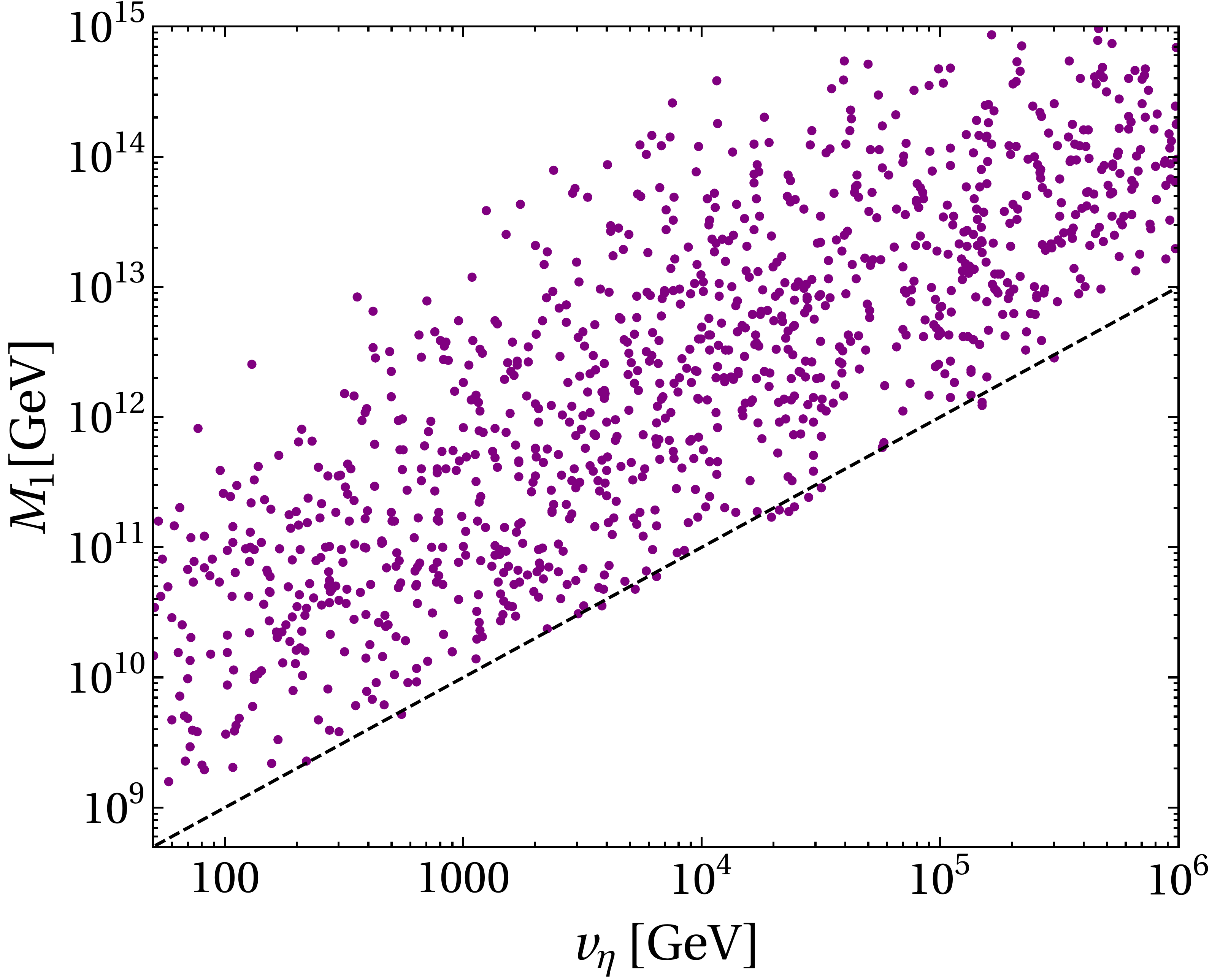}
    \caption{\small Correlation of seesaw/leptogenesis scale $M_1$ with $v_\eta$. The black dashed line corresponds to Davidson-Ibarra type bound given by Eq.~\eqref{eq:seesawscale}. See main text for details.}
    \label{fig:leptoDM}
\end{figure}

The results of this scan is shown in Fig.~\ref{fig:leptoDM}. Each point in this plot fits neutrino oscillation data and generates correct baryon asymmetry. From this plot, we find the approximate lower bound on $M_1$ as shown by the black dashed line in the plot,
\begin{equation}\label{eq:seesawscale}
    M_1\gtrsim 10^{9}\,\text{GeV}\left(\frac{v_\eta}{100\,\text{GeV}}\right),
\end{equation}
which matches with our estimate given in Eq.~\eqref{eq:seesawscalenaive}. 
For fixed values of $\mu_1,\lambda_\eta$ and $\lambda_1$, the VEV $v_\eta$ is directly correlated to the DM mass $m_\chi$  via $v_\rho$.

\begin{figure}[ht!]
    \centering
    \includegraphics[width=1.0\linewidth]{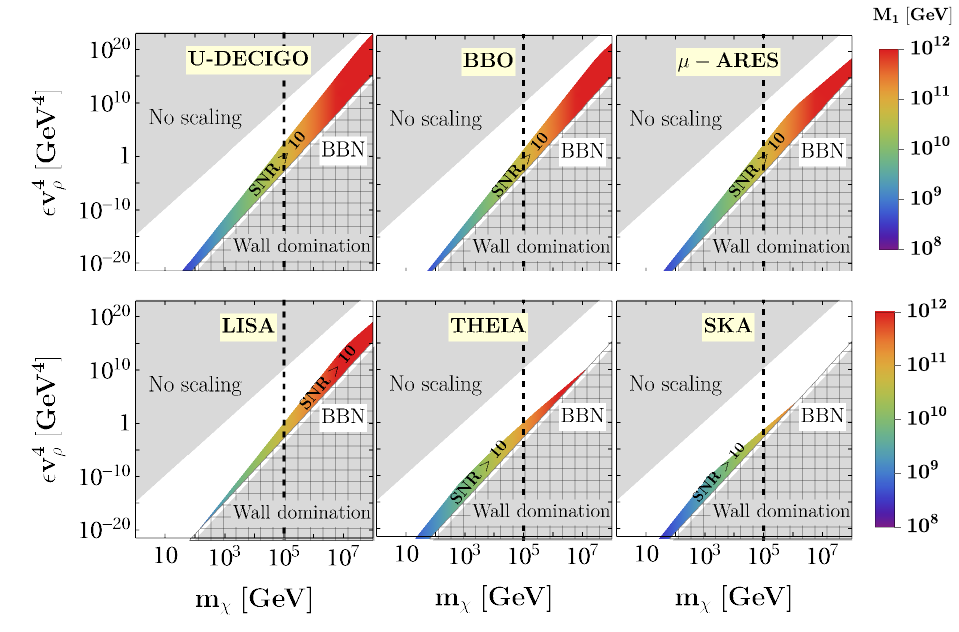}
    \caption{Projected sensitivity (with SNR $>10$) of future GW experiments in the parameter space of DM mass $m_\chi$ given in Eq.~\eqref{mDM_tree} and the bias parameter $\epsilon$ given in~\eqref{eq:bias}, multiplied by $v_\rho^4$. For this plot, we consider the benchmark point $\lambda_\rho=0.1, \lambda_1 = \lambda_\rho/72, \mu_1=0.01 \rm{GeV}, \lambda_\eta=0.1$ and assumed the Higgs resonance condition given in Eq.~\eqref{eq:resonancecond}. The upper and lower gray regions are excluded by scaling requirements and wall domination respectively. The meshed region signifies the $\Delta N_{\rm eff}$ bound from BBN. The color bar shows the leptogenesis scale $M_1$ given in Eq.~\eqref{eq:seesawscale}. The vertical dashed line shows the unitarity bound on thermal dark matter given in Eq.~\eqref{eq:unitarity bound}.}
    \label{fig:SNRmDM}
\end{figure}

Finally we summarize our GW discussion in Fig.~\ref{fig:SNRmDM} where we show the projected sensitivity of various future GW experiments to probe the parameter space of our model for a benchmark point: $\lambda_\rho=0.1, \lambda_1 = \lambda_\rho/72, \mu_1=0.01$ GeV, $\lambda_\eta=0.1$. The x-axis is the DM mass $m_\chi$ while the y-axis is depicting the required bias parameter $\epsilon$ defined in Eq.~\eqref{eq:bias}. The colored region in each plot generates the correct BAU and is sensitive to future GW experiments with SNR $>10$. The DM mass is calculated in the Higgs resonance region given in Eq.~\eqref{eq:resonancecond}. The color bar indicates the minimum scale of leptogenesis $M_1$, defined in Eq.~\eqref{eq:seesawscale}. The meshed region is discarded by $\Delta N_{\rm eff}$ bound from BBN on gravitational waves amplitude. The upper and lower gray regions are excluded by requirements of domain wall scaling and decay before domination respectively. The upper limit on DM mass, $m_\chi\lesssim25$ TeV coming from relic density considerations (see Fig.~\ref{fig:DMrelic} (c)) and the lower limit coming from direct detection experiments discussed in sec.~\ref{ssubsec:dd} are not shown in this plot as they depend on parameter choices. The vertical dashed line corresponds to the unitarity bound on thermal dark matter~\cite{Griest:1989wd},
\begin{equation}
    \label{eq:unitarity bound}
    m_\chi\lesssim 100 \text{ TeV.}
\end{equation}

We end this section by mentioning that there are two possible sources that can change the DM relic abundance within this model. Similar to the discussion in~\cite{Pham:2024vso}, the first source is dilution of energy densities of various particle species, including the DM candidate, due to entropy injection when the domain walls decay~\cite{Hattori:2015xla,Kawasaki:2004rx}. This will reduce the relic abundance of DM which freezes out prior to the decay of the domain walls. Since scalar DM is usually overabundant except in the resonance region, such dilution effects can expand the viable parameter space. However, in the parameter space of our work, the domain wall energy density is sub-dominant keeping the entropy injection is negligible.

The second source is the non-thermal production of DM from domain walls. It can happen a) indirectly via DW $\rightarrow h_3\rightarrow\chi\chi$ or b) directly from domain walls, i.e. radiation of the DM particles throughout the evolution of the walls~\cite{Hiramatsu:2012sc}. The indirect channel will not increase DM relic abundance since the decay $h_3\rightarrow\chi\chi$ is kinematically suppressed near the resonance $m_\chi \approx m_{h_3}/2$ considered in our analysis. The direct channel can however, in general, increase the DM relic density depending on the model parameters and the efficiency of domain wall energy transfer to DM. To give a naive estimate, consider a typical domain wall with tension $\sigma=10^6$ GeV$^3$ and annihilation temperature $T_{\rm ann}=1$ MeV. The Hubble rate is $\mathcal{H}=\sqrt{\frac{8\pi^3g_*}{90}} \frac{T_{\rm ann}^2}{M_{\rm Pl}}=4.4\times10^{-25}$ GeV, therefore, the energy density of the wall is $\rho_{\rm wall}= \sigma\, \mathcal{H}=4.4\times10^{-19}$ GeV$^4$. If we assume that all the energy of the domain wall gets transferred to dark matter at $T=T_{\rm ann}$, the resulting DM relic is $\rho_\chi=\rho_{\rm wall}$. The comoving energy density is $\rho_\chi/s=10^{-10}$ GeV where $s=\frac{2\pi^2}{45}g_{*s} T_{\rm ann}^3$. The present relic is $\Omega_\chi h^2 = 2.7\times10^8\frac{\rho_\chi}{s} \sim 2.7 \times 10^{-2}$ which is smaller than the observed DM relic. However, for larger DW tension $\sigma=1$ TeV$^3$ and same $T_{\rm ann}=1$ MeV, DM can be overproduced $\Omega_\chi h^2 \sim 27$.

On the other hand, if DW annihilates at a temperature above the freeze-out temperature of DM, then DW-generated DM does not affect the final abundance of the latter. In such a scenario, the DM particles produced from DW immediately enter thermal equilibrium with the bath and other DM particles followed by thermal freeze-out which eventually decides DM relic. Considering $\epsilon=10^{-21}, v_\rho=2 \times 10^8$ GeV (as considered in Fig. \ref{fig:Spectrum}), we get $T_{\rm ann} \sim \sqrt{V_{\rm bias} M_{\rm Pl}/\sigma} \sim \sqrt{\epsilon v_{\rho} M_{\rm Pl}} \sim 1$ TeV. Assuming freeze-out temperature of thermal DM to be $\sim m_\chi/25$, this leads to an upper bound of DM mass $m_\chi \leq 25$ TeV.

A robust estimate of DM production directly from DW annihilation will require complete understanding of DW dynamics and solution of coupled Boltzmann equations including different final states like radiation, gravitational waves, dark matter and other light degrees of freedom. Such detailed calculations are outside of the scope of present work and left for future studies.

\section{Conclusion} \label{sec:conclusion}
We have proposed a minimal Dirac seesaw model accommodating a thermal dark matter and study the phenomenology in details. We consider a $Z_4$-symmetric version of Type-I Dirac seesaw such that the imaginary part of a complex scalar singlet taking part in the seesaw become the DM candidate, stabilised by a charge conjugation symmetry in dark sector. We study the details of DM phenomenology namely, relic and detection aspects at direct and indirect detection frontiers. We also show the consistency of the model with observed neutrino data while explaining the observed baryon asymmetry via Dirac leptogenesis. While the existence of heavy fermions does not lead to large quantum corrections to the potential due to the chosen Yukawa couplings, presence of new scalar degrees of freedom ensures the stability of the electroweak vacuum upto the Planck scale, which is not possible in the SM alone.

While spontaneous $Z_4$ symmetry breaking plays a crucial role in generating light Dirac neutrino mass, it also leads to the formation of domain walls. We consider suitable bias terms which break $Z_4$ explicitly in order to make the walls disappear via annihilation while also generating stochastic gravitational waves. After pointing out the details of $Z_4$-walls and their difference compared to $Z_2$-walls studied earlier in Dirac seesaw, we find the GW spectrum for a few benchmark points consistent with rest of the phenomenology. We also perform a numerical scan of the parameter space to show the projected sensitivity of future GW experiments such as UDECIGO (UDECIGO-corr) \cite{Sato:2017dkf, Ishikawa:2020hlo}, BBO, LISA, $\mu-$ARES, THEIA and SKA with signal-to-noise ratio $>$ 10. We see that these GW experiments will be able to probe DM of mass $m_\chi$ in the range $\mathcal{O}\left(10^2-10^{5}\right)$ GeV and seesaw scale $M_1$ in the range $\mathcal{O}\left(10^9-10^{12}\right)$ GeV. Interestingly, thermal Dirac leptogenesis in this model also leads to thermalisation of right chiral parts of light Dirac neutrinos enhancing the effective relativistic degrees of freedom $N_{\rm eff}$ within reach of future CMB experiments. The complementary detection prospects of the model at dark matter, gravitational wave, CMB experiments keep the model verifiable in near future. Additionally, future observation of neutrinoless double beta decay provides a way of falsifying such a setup.

\section{Acknowledgement}
Z.A.B. is supported by a DST-INSPIRE fellowship. The work of D.B. is supported by the Science and Engineering Research Board (SERB), Government of India grants MTR/2022/000575 and CRG/2022/000603. D.B. also acknowledges the support from the Fulbright-Nehru Academic and Professional Excellence Award 2024-25. L.M. is supported by a UGC fellowship.  U.P. would like to acknowledge the financial support from the Indian Institute of Technology Bhilai, and the Ministry of Education, Government of India, for conducting the research work.

\appendix
\section{Scalar masses}\label{app:scalarmasses}
The scalar mass matrix after EWSB in the basis $(H,\rho,\eta)$ in the limit $\mu_\eta^2\sim\lambda_{H\eta}\sim\lambda_{\rho\eta}\rightarrow0$ is given by,
\begin{align}
\nonumber\mathcal{M}_S^2 &= 
\left(
	\begin{array}{ccc}
	\frac{3\lambda_H v^2}{4}+\frac{\lambda_{H\rho} v_\rho^2}{2}-\mu_H^2 &
	\lambda_{H\rho}vv_\rho & 0 \\
	  \lambda_{H\rho}vv_\rho & \frac{\lambda_{H\rho}v^2}{2}+\frac{3\lambda_\rho v_\rho^2}{4}-6\lambda_1 v_\rho^2-\sqrt{2}\mu_1 v_\eta - \mu_\rho^2 & -\sqrt{2}\mu_1v_\rho\\
    0 & -\sqrt{2}\mu_1v_\rho & \frac{3}{4}\lambda_\eta v_\eta^2
	\end{array}
\right) \\
&= 
\left(
\begin{array}{ccc}
	\frac{\lambda_H}{2}v^2 & \lambda_{H\rho}vv_\rho & 0 \\
	\lambda_{H\rho}vv_\rho & \frac{1}{2}(\lambda_\rho-8\lambda_1)v_\rho^2 & -\sqrt{2}\mu_1v_\rho\\
    0 & -\sqrt{2}\mu_1v_\rho & \frac{3}{2}\left(\lambda_\eta \mu_1^2 v_\rho^4\right)^{1/3}\\
\end{array}
\right),\label{Ms_tree}
\end{align}
where the second equality comes after applying the solutions of the tadpole equations given in Eq.~\eqref{eq:veta}. We can divide this matrix into 4 blocks as,
\begin{align}
\mathcal{M}_{S}^2=\left(
\begin{array}{c|cc}
	\frac{\lambda_H}{2}v^2 & \lambda_{H\rho}vv_\rho & 0 \\
    \hline
	\lambda_{H\rho}vv_\rho & \frac{1}{2}(\lambda_\rho-8\lambda_1)v_\rho^2 & -\sqrt{2}\mu_1v_\rho\\
    0 & -\sqrt{2}\mu_1v_\rho & \frac{3}{2}\left(\lambda_\eta \mu_1^2 v_\rho^4\right)^{1/3}
\end{array}
\right)=
\left(
\begin{array}{cc}
	M_H^2 & \left(M_{H\rho}^2\right)^T \\
	M_{H\rho}^2 & M_\rho^2
\end{array}
\right)
\end{align}
where $M_\rho^2\gg M_{H\rho}^2\gg M_H^2$. The SM Higgs squared mass $m_{h_1}^2$ can be calculated with a Type-I seesaw formula,
\begin{align}
    \nonumber m_{h_1}^2&\simeq M_H^2-\left(M_{H_\rho}^2\right)^T \cdot \left(M_\rho^2\right)^{-1} \cdot M_{H\rho}^2\\
    \nonumber &\simeq\frac{\lambda_H}{2}v^2+\frac{6\lambda_{H\rho}^2\lambda_\eta^\frac{1}{3}v^2}{3\lambda_\eta^\frac{1}{3}(8\lambda_1-\lambda_\rho)+8\left(\frac{\mu_1}{v_\rho}\right)^\frac{4}{3}}\\
    &\simeq v^2\left(\frac{\lambda_H}{2}+\frac{2\lambda_{H\rho}^2}{8\lambda_1-\lambda_\rho}\right)
\end{align}
where in the last equation, we assume $\mu_1/v_\rho\ll1$. We can calculate the value of the coupling $\lambda_H$ by equating $m_{h_1}$ with the observed SM Higgs mass of 125 GeV, which gives,
\begin{equation}\label{eq:lamH}
    \lambda_H\simeq\frac{2.5\times10^5\lambda_1-3.125\times10^4\lambda_\rho-4\lambda_{H\rho}^2v^2}{(8\lambda_1-\lambda_\rho)v^2}
\end{equation}
The squared masses of the other two scalars in the theory can be approximated by diagonalizing $M_\rho^2$ above,
\begin{equation}
    m_{h_2}^2\simeq \frac{3}{2}\lambda_\eta^\frac{1}{3}\left(\mu_1\,v_\rho^2\right)^\frac{2}{3},\quad
    m_{h_3}^2\simeq\frac{1}{2}\left(\lambda_\rho-8\lambda_1\right)v_\rho^2
\end{equation}
The Higgs resonance condition $2m_\chi\sim m_{h_3}$ provides an approximate relation of the self coupling of $\rho$,
\begin{equation}\label{eq:resonancecond}
    \lambda_\rho\sim 72\,\lambda_1+32\,\lambda_\eta^{-\frac{1}{3}}\left(\frac{\mu_1}{v_\rho}\right)^\frac{4}{3}\sim 72\,\lambda_1\quad\text{for }\mu_1/v_\rho\ll1
\end{equation}

\section{Boundedness condition on Scalar potential}\label{app:bounded}
The scalar potential is bounded from below if the following conditions \cite{Kannike:2012pe} are satisfied.
\begin{align}
    \lambda_H \geq 0, \,\,\,\, \lambda_\rho + 8\lambda_1 \geq 0, \,\,\,\, \lambda_\eta \geq 0, \,\,\,\, \lambda_\rho -8 \lambda_1 \geq 0 \nonumber \\
    2\lambda_{H\rho} + \sqrt{\lambda_H (\lambda_\rho+8\lambda_1)} \geq 0, \,\,\,\, 2\lambda_{\rho \eta} + \sqrt{\lambda_\eta (\lambda_\rho + 8\lambda_1)} \geq 0, \,\,\,\, 2\lambda_{H \eta} +\sqrt{\lambda_H \lambda_\eta} \geq 0 \nonumber \\ 
    \left (\frac{\lambda_\rho}{2}-4\lambda_1 \right) \sqrt{\frac{\lambda_\eta}{4}} + \lambda_{\rho \eta} \sqrt{\frac{\lambda_{\rho}}{4}+2\lambda_1} + \sqrt{2 \left(\frac{\lambda_\rho}{2}-4\lambda_1 \right) \left(\frac{\lambda_{\rho \eta}}{2}+\sqrt{\frac{\lambda_\eta}{4} \left ( \frac{\lambda_\rho}{4}+2\lambda_1 \right)}\right)^2 } \geq 0 \nonumber \\
    \left (\frac{\lambda_\rho}{2}-4\lambda_1 \right) \sqrt{\frac{\lambda_H}{4}} + \lambda_{H\rho} \sqrt{\frac{\lambda_{\rho}}{4}+2\lambda_1} + \sqrt{2 \left(\frac{\lambda_\rho}{2}-4\lambda_1 \right) \left(\frac{\lambda_{H\rho}}{2}+\sqrt{\frac{\lambda_H}{4} \left ( \frac{\lambda_\rho}{4}+2\lambda_1 \right)}\right)^2 } \geq 0 \nonumber
\end{align}

\section{Beta functions at one-loop
}
\label{app:beta}
The renormalisation group evolution (RGE) of a parameter $p_i$ can be written as
\begin{equation}
    \frac{dp_i}{dt}=\frac{1}{16\pi^2} \beta_{p_i}
\end{equation}
where $\beta_{p_i}$ is the corresponding beta function and $t=\ln{\left ( \frac{\mu}{\rm GeV} \right )}$ is the variable related to the renormalisation scale $\mu$.
\subsection{Gauge Couplings}
As there are no new gauge multiplets in the model, the beta functions of gauge couplings remain same as those in the SM given by
{\allowdisplaybreaks  \begin{align} 
\beta_{g_1}^{(1)} & =  
\frac{41}{10} g_{1}^{3}, \,\,\,\, 
\beta_{g_2}^{(1)} =  
-\frac{19}{6} g_{2}^{3}, \,\,\,\, 
\beta_{g_3}^{(1)} =  
-7 g_{3}^{3}.
\end{align}} 

\subsection{Quartic scalar couplings}
The beta function for quartic coupling of the SM Higgs is 
\begin{align}
\beta_{\lambda_H}^{(1)} & =  
+\frac{27}{50} g_{1}^{4} +\frac{9}{5} g_{1}^{2} g_{2}^{2} +\frac{9}{2} g_{2}^{4} -\frac{9}{5} g_{1}^{2} \lambda_H -9 g_{2}^{2} \lambda_H +6 \lambda_{H}^{2} +8 \lambda_{H\eta}^{2} +4 \lambda_{H\rho}^{2} +12 \lambda_H \mbox{Tr}\Big({Y_d  Y_{d}^{\dagger}}\Big) \nonumber \\ 
 &+4 \lambda_H \mbox{Tr}\Big({Y_e  Y_{e}^{\dagger}}\Big) +4 \lambda_H \mbox{Tr}\Big({Y_{L}  Y_{L}^{\dagger}}\Big) +12 \lambda_H \mbox{Tr}\Big({Y_u  Y_{u}^{\dagger}}\Big) -24 \mbox{Tr}\Big({Y_d  Y_{d}^{\dagger}  Y_d  Y_{d}^{\dagger}}\Big) \nonumber \\ 
 &-8 \mbox{Tr}\Big({Y_e  Y_{e}^{\dagger}  Y_e  Y_{e}^{\dagger}}\Big) -8 \mbox{Tr}\Big({Y_{L}  Y_{L}^{\dagger}  Y_{L}  Y_{L}^{\dagger}}\Big) -24 \mbox{Tr}\Big({Y_u  Y_{u}^{\dagger}  Y_u  Y_{u}^{\dagger}}\Big) \nonumber \\
 & = \beta^{\rm SM}_{\lambda_H} + \beta^{\rm new}_{\lambda_H},
\end{align}
where the new contribution in our model is 
\begin{align}
    \beta^{\rm new}_{\lambda_H}= 8 \lambda_{H\eta}^{2} +4 \lambda_{H\rho}^{2} + 4 \lambda_H \mbox{Tr}\Big({Y_{L}  Y_{L}^{\dagger}}\Big) -8 \mbox{Tr}\Big({Y_{L}  Y_{L}^{\dagger}  Y_{L}  Y_{L}^{\dagger}}\Big).
\end{align}
The beta functions for the new quartic couplings are
\begin{align} 
\beta_{\lambda_{1}}^{(1)} & =  
6 \lambda_{1} \lambda_{\rho}, \\ 
\beta_{\lambda_{\eta}}^{(1)} & =  
18 \lambda_{\eta}^{2}  + 4 \lambda_{\rho\eta}^{2}  + 8 \lambda_{\eta} \mbox{Tr}\Big({Y_{R}  Y_{R}^{\dagger}}\Big)  + 8 \lambda_{H\eta}^{2}  -8 \mbox{Tr}\Big({Y_{R}  Y_{R}^{\dagger}  Y_{R}  Y_{R}^{\dagger}}\Big), \\ 
\beta_{\lambda_{\rho\eta}}^{(1)} & =  
2 \lambda_{\rho\eta} \Big(3 \lambda_{\eta}  + 4 \lambda_{\rho\eta}  + \lambda_{\rho}\Big) + 4 \lambda_{H\eta} \lambda_{H\rho}  + 4 \lambda_{\rho\eta} \mbox{Tr}\Big({Y_{R}  Y_{R}^{\dagger}}\Big), \\ 
\beta_{\lambda_{H\eta}}^{(1)} & =  
-\frac{9}{10} g_{1}^{2} \lambda_{H\eta} -\frac{9}{2} g_{2}^{2} \lambda_{H\eta} +6 \lambda_{\eta} \lambda_{H\eta} +3 \lambda_H \lambda_{H\eta} +8 \lambda_{H\eta}^{2} +2 \lambda_{H\rho} \lambda_{\rho\eta} +6 \lambda_{H\eta} \mbox{Tr}\Big({Y_d  Y_{d}^{\dagger}}\Big)  \nonumber \\ 
 &+2 \lambda_{H\eta} \mbox{Tr}\Big({Y_e  Y_{e}^{\dagger}}\Big)+2 \lambda_{H\eta} \mbox{Tr}\Big({Y_{L}  Y_{L}^{\dagger}}\Big) +4 \lambda_{H\eta} \mbox{Tr}\Big({Y_{R}  Y_{R}^{\dagger}}\Big) +6 \lambda_{H\eta} \mbox{Tr}\Big({Y_u  Y_{u}^{\dagger}}\Big), \\ 
\beta_{\lambda_{H\rho}}^{(1)} & =  
-\frac{9}{10} g_{1}^{2} \lambda_{H\rho} -\frac{9}{2} g_{2}^{2} \lambda_{H\rho} +3 \lambda_H \lambda_{H\rho} +4 \lambda_{H\rho}^{2} +2 \lambda_{H\rho} \lambda_{\rho} +4 \lambda_{H\eta} \lambda_{\rho\eta} +6 \lambda_{H\rho} \mbox{Tr}\Big({Y_d  Y_{d}^{\dagger}}\Big)\nonumber \\ 
 & +2 \lambda_{H\rho} \mbox{Tr}\Big({Y_e  Y_{e}^{\dagger}}\Big) +2 \lambda_{H\rho} \mbox{Tr}\Big({Y_{L}  Y_{L}^{\dagger}}\Big) +6 \lambda_{H\rho} \mbox{Tr}\Big({Y_u  Y_{u}^{\dagger}}\Big), \\ 
\beta_{\lambda_{\rho}}^{(1)} & =  
576 \lambda_{1}^{2}  + 5 \lambda_{\rho}^{2}  + 8 \lambda_{H\rho}^{2}  + 8 \lambda_{\rho\eta}^{2} 
\end{align}

\subsection{Yukawa Couplings}
The beta functions of the new and SM Yukawa couplings are given by
{\allowdisplaybreaks  \begin{align} 
\beta_{Y_{L}}^{(1)} & =  
+\frac{3}{2} \Big(- {Y_{e}^{T}  Y_e^*  Y_{L}}  + {Y_{L}  Y_{L}^{\dagger}  Y_{L}}\Big)\nonumber \\ 
 &+Y_{L} \Big(3 \mbox{Tr}\Big({Y_d  Y_{d}^{\dagger}}\Big)  + 3 \mbox{Tr}\Big({Y_u  Y_{u}^{\dagger}}\Big)  -\frac{9}{20} g_{1}^{2}  -\frac{9}{4} g_{2}^{2}  + \mbox{Tr}\Big({Y_e  Y_{e}^{\dagger}}\Big) + \mbox{Tr}\Big({Y_{L}  Y_{L}^{\dagger}}\Big)\Big),\\
\beta_{Y_u}^{(1)} & =  
-\frac{3}{2} \Big(- {Y_u  Y_{u}^{\dagger}  Y_u}  + {Y_u  Y_{d}^{\dagger}  Y_d}\Big)\nonumber \\ 
 &+Y_u \Big(3 \mbox{Tr}\Big({Y_d  Y_{d}^{\dagger}}\Big)  + 3 \mbox{Tr}\Big({Y_u  Y_{u}^{\dagger}}\Big)  -8 g_{3}^{2}  -\frac{17}{20} g_{1}^{2}  -\frac{9}{4} g_{2}^{2}  + \mbox{Tr}\Big({Y_e  Y_{e}^{\dagger}}\Big) + \mbox{Tr}\Big({Y_{L}  Y_{L}^{\dagger}}\Big)\Big) \nonumber \\ 
& = \beta^{\rm SM}_{Y_u} + Y_u \mbox{Tr}\Big({Y_{L}  Y_{L}^{\dagger}}\Big), \\
\beta_{Y_{R}}^{(1)} & =  
2 Y_{R} \mbox{Tr}\Big({Y_{R}  Y_{R}^{\dagger}}\Big)  + 3 {Y_{R}  Y_{R}^{\dagger}  Y_{R}}, \\ 
\beta_{Y_d}^{(1)} & =  
+\frac{3}{2} \Big(- {Y_d  Y_{u}^{\dagger}  Y_u}  + {Y_d  Y_{d}^{\dagger}  Y_d}\Big)\nonumber \\ 
 &+Y_d \Big(3 \mbox{Tr}\Big({Y_d  Y_{d}^{\dagger}}\Big)  + 3 \mbox{Tr}\Big({Y_u  Y_{u}^{\dagger}}\Big)  -8 g_{3}^{2}  -\frac{1}{4} g_{1}^{2}  -\frac{9}{4} g_{2}^{2}  + \mbox{Tr}\Big({Y_e  Y_{e}^{\dagger}}\Big) + \mbox{Tr}\Big({Y_{L}  Y_{L}^{\dagger}}\Big)\Big) \nonumber \\ 
 & = \beta^{\rm SM}_{Y_d} + Y_d \mbox{Tr}\Big({Y_{L}  Y_{L}^{\dagger}}\Big), \\
\beta_{Y_e}^{(1)} & =  
\frac{1}{4} \Big(6 \Big(- {Y_e  Y_{L}^*  Y_{L}^{T}}  + {Y_e  Y_{e}^{\dagger}  Y_e}\Big)\nonumber \\ 
 &+Y_e \Big(12 \mbox{Tr}\Big({Y_d  Y_{d}^{\dagger}}\Big)  + 12 \mbox{Tr}\Big({Y_u  Y_{u}^{\dagger}}\Big)  + 4 \mbox{Tr}\Big({Y_e  Y_{e}^{\dagger}}\Big)  + 4 \mbox{Tr}\Big({Y_{L}  Y_{L}^{\dagger}}\Big)  -9 g_{1}^{2}  -9 g_{2}^{2} \Big)\Big) \nonumber \\
 & = \beta^{\rm SM}_{Y_e} - \frac{3}{2} Y_e Y^*_L Y^T_L + Y_e \mbox{Tr}\Big({Y_{L}  Y_{L}^{\dagger}}\Big). \\
\end{align}} 

\subsection{Running of the couplings}
For numerical analysis, we simplify the relevant beta functions by ignoring all SM couplings except the gauge couplings, top quark Yukawa and the Higgs quartic coupling. For simplicity, the Yukawa couplings of heavy fermions of quasi-degenerate mass are assumed to be diagonal with same numerical value $Y_{L,R}$. The simplified beta functions are given below.
{\allowdisplaybreaks  \begin{align} 
\beta_{g_1}^{(1)} & =  
\frac{41}{10} g_{1}^{3}, \,\,\,\, 
\beta_{g_2}^{(1)} =  
-\frac{19}{6} g_{2}^{3}, \,\,\,\, 
\beta_{g_3}^{(1)} =  
-7 g_{3}^{3},
\end{align}} 
\begin{align}
\beta_{\lambda_H}^{(1)} & =  
+\frac{27}{50} g_{1}^{4} +\frac{9}{5} g_{1}^{2} g_{2}^{2} +\frac{9}{2} g_{2}^{4} -\frac{9}{5} g_{1}^{2} \lambda_H -9 g_{2}^{2} \lambda_H +6 \lambda_{H}^{2} +8 \lambda_{H\eta}^{2} +4 \lambda_{H\rho}^{2} \nonumber \\ 
 &+(12 \lambda_H Y^2_L-24 Y^4_L) \Theta(\mu-M_N) +12 \lambda_H Y^2_t -24 Y^4_t,
\end{align}
\begin{align} 
\beta_{\lambda_{1}}^{(1)} & =  
6 \lambda_{1} \lambda_{\rho}, \\ 
\beta_{\lambda_{\eta}}^{(1)} & =  
18 \lambda_{\eta}^{2}  + 4 \lambda_{\rho\eta}^{2}  + 24 \lambda_{\eta} Y^2_R \Theta(\mu-M_N) + 8 \lambda_{H\eta}^{2}  -24 Y^4_R \Theta(\mu-M_N), \\ 
\beta_{\lambda_{\rho\eta}}^{(1)} & =  
2 \lambda_{\rho\eta} \Big(3 \lambda_{\eta}  + 4 \lambda_{\rho\eta}  + \lambda_{\rho}\Big) + 4 \lambda_{H\eta} \lambda_{H\rho}  + 12 \lambda_{\rho\eta} Y^2_R \Theta(\mu-M_N), \\ 
\beta_{\lambda_{H\eta}}^{(1)} & =  
-\frac{9}{10} g_{1}^{2} \lambda_{H\eta} -\frac{9}{2} g_{2}^{2} \lambda_{H\eta} +6 \lambda_{\eta} \lambda_{H\eta} +3 \lambda_H \lambda_{H\eta} +8 \lambda_{H\eta}^{2} +2 \lambda_{H\rho} \lambda_{\rho\eta}   \nonumber \\ 
 &+6 \lambda_{H\eta} Y^2_L \Theta(\mu-M_N) +12 \lambda_{H\eta} Y^2_R \Theta(\mu-M_N) +6 \lambda_{H\eta} Y^2_t, \\ 
\beta_{\lambda_{H\rho}}^{(1)} & =  
-\frac{9}{10} g_{1}^{2} \lambda_{H\rho} -\frac{9}{2} g_{2}^{2} \lambda_{H\rho} +3 \lambda_H \lambda_{H\rho} +4 \lambda_{H\rho}^{2} +2 \lambda_{H\rho} \lambda_{\rho} +4 \lambda_{H\eta} \lambda_{\rho\eta} \nonumber \\ 
 & +6 \lambda_{H\rho} Y^2_L \Theta(\mu-M_N) +6 \lambda_{H\rho} Y^2_t, \\ 
\beta_{\lambda_{\rho}}^{(1)} & =  
576 \lambda_{1}^{2}  + 5 \lambda_{\rho}^{2}  + 8 \lambda_{H\rho}^{2}  + 8 \lambda_{\rho\eta}^{2}, 
\end{align}
{\allowdisplaybreaks  \begin{align} 
\beta_{Y_{L}}^{(1)} & =  
+\frac{3}{2} Y^3_{L}  \Theta(\mu-M_N)
 +Y_{L} \Big(3 Y^2_t  -\frac{9}{20} g_{1}^{2}  -\frac{9}{4} g_{2}^{2} + 3Y^2_L\Big) \Theta(\mu-M_N),\\
\beta_{Y_t}^{(1)} & =  
\frac{3}{2} Y^3_t
 +Y_t \Big(3 Y^2_t  -8 g_{3}^{2}  -\frac{17}{20} g_{1}^{2}  -\frac{9}{4} g_{2}^{2} + 3Y^2_L \Theta(\mu-M_N)\Big), \\
\beta_{Y_{R}}^{(1)} & =  
9 Y^3_{R} \Theta(\mu-M_N). 
\end{align}}
\begin{figure}[!ht]
    \centering
    \includegraphics[width=0.48\linewidth]{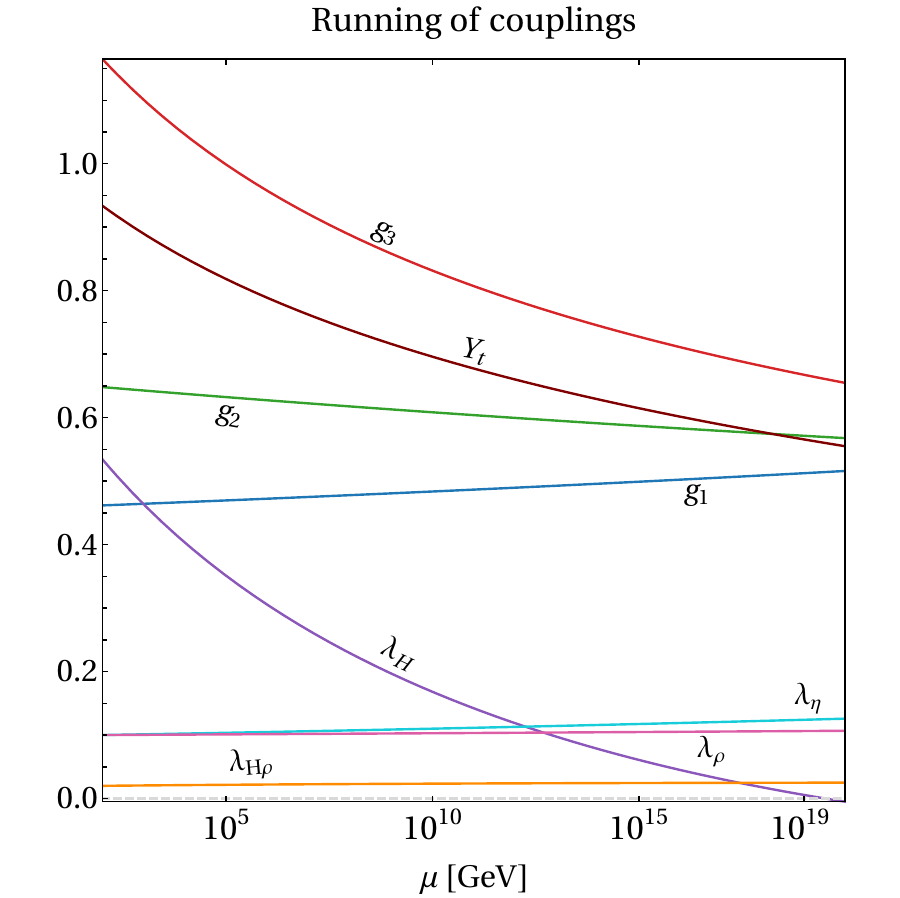}
    \includegraphics[width=0.48\linewidth]{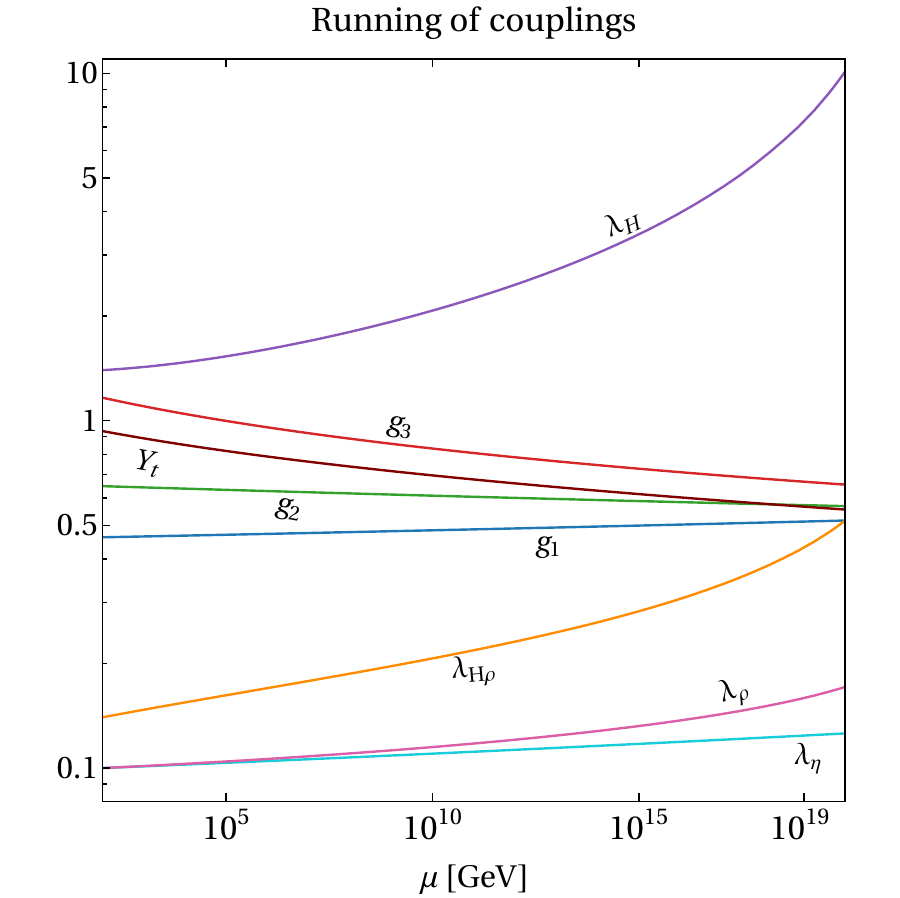}
    \caption{Running of the couplings for benchmark points with $\lambda_{H\rho}=0.02$ \textbf{(Left)} and $\lambda_{H\rho}=0.14$ \textbf{(Right)}.}
    \label{fig:RGE}
\end{figure}
In Fig.~\ref{fig:RGE} we show the running of the couplings by solving the RGE equations in a bottom-up manner considering the following input values at electroweak scale,
\begin{align}
   \nonumber M_i=10^9\,&\text{GeV},\;\lambda_\rho=0.1,\;\lambda_1=\frac{\lambda_\rho}{72},\;\lambda_{\rho\eta}=\lambda_{H\eta}\sim0,\\
    \lambda_\eta&=0.1,\;Y_L=0.0016,\;Y_R=0.008.
\end{align}
We consider $\lambda_{H \rho}=0.02$ in the \textbf{Left} plot and $\lambda_{H\rho}=0.14$ in the \textbf{Right} plot. The values of the SM gauge couplings at electroweak scale are~\cite{ParticleDataGroup:2024cfk},
\begin{equation}
    g_1=\sqrt{\frac{5}{3}}\times0.35767,\;g_2=0.64815,\;g_3= 1.16499,\;Y_t=0.93378.
\end{equation}
The value of $\lambda_H$ at electroweak scale is calculated using Eq.~\eqref{eq:lamH}. We find that for $\lambda_{H\rho}\in[0.02,0.14]$ with other values fixed as above, the Higgs quartic coupling remains positive upto the Planck scale implying stability of the scalar potential, and no coupling hits the Landau pole upto the Planck scale. We use this range of $\lambda_{H\rho}$ in the numerical analysis considered in the main text. We also find an upper bound on the elements of the Yukawa matrices from the requirement of stability of the scalar potential $|Y_L|<0.3,|Y_R|< 0.5$. It should be noted that we are considering degenerate seesaw scale in the RGE equations here $M_1=M_2=M_3$. For non-degenerate seesaw scales $M_2=10 M_1, M_3=20 M_1$, as considered in the numerical analysis in the main text, one needs to integrate out heavy fermions one at a time leading to different running at intermediate scales separating the heavy fermion masses \cite{Antusch:2002rr}. Since heavy fermion masses differ by only one order of magnitude in our setup and we stick to a regime where heavy fermion effects on RGE are always negligible, we do not expect to get any significant change in running of couplings for non-degenerate seesaw scales from what is shown in Fig. \ref{fig:RGE} above.

We also check the running of mass parameters of the potential. The corresponding beta functions are given by
{\allowdisplaybreaks  \begin{align} 
\beta_{\mu_1}^{(1)} &= \mu_1 \left( -24 \lambda_1 + 6 Y_R^2 \Theta(\mu - M_N) + 8 \lambda_{\rho \eta} + \lambda_\rho \right),\\
\beta_{\mu_\eta^2}^{(1)} &= -2 \lambda_{\rho \eta} \mu_\rho^2 - 4 \lambda_{H\eta} \mu_H^2 - 12  M_N^2 Y_R^2 \Theta(\mu - M_N)  + 4 \mu_1^2 \nonumber \\
&\quad + 12 \mu_\eta^2 Y_R^2  \Theta(\mu - M_N) + 6 \lambda_\eta \mu_\eta^2, \\
\beta_{\mu_H^2}^{(1)} &= -4 \lambda_{H\eta} \mu_\eta^2 - \frac{9}{10} g_1^2 \mu_H^2 - \frac{9}{2} g_2^2 \mu_H^2 + 3 \lambda_H \mu_H^2 + 2 \lambda_{H\rho} \mu_\rho^2 \nonumber \\
&\quad + 6 \mu_H^2 Y_L^2 \Theta(\mu - M_N) + 6 \mu_H^2 Y_t^2  + 12 M_N^2 Y_L^2 \Theta(\mu - M_N), \\
\beta_{\mu_\rho^2}^{(1)} &= 2 \lambda_\rho \mu_\rho^2 + 4 \lambda_{H\rho} \mu_H^2 - 4 \lambda_{\rho \eta} \mu_\eta^2 - 8 \mu_1^2.
\end{align}}
Clearly, $\mu^2_H, \mu^2_\eta$ receive large corrections from heavy fermions for $\mu > M_N$. Ignoring the contribution from other terms compared to the one proportional to $M^2_N$, we get \cite{Brdar:2019iem}
\begin{equation}
    \Delta \mu^2_H = \frac{1}{8\pi^2} 3M^2_N Y^2_L \left (\ln{\frac{M^2_N}{\mu^2}} +\frac{1}{2} \right).
\end{equation}
This gives a threshold correction at heavy fermion decoupling scale $\mu=M_N$ \cite{Brdar:2019iem}
\begin{equation}
\Delta \mu^2_H=\frac{1}{16\pi^2} 3M^2_N Y^2_L.
\end{equation}
Similar correction is also obtained for $\mu^2_\eta$,
\begin{equation}
    \Delta \mu_\eta^2
 = - \frac{1}{16 \pi^2} 3M_N^2Y_R^2.
 \end{equation}

We choose the initial bare mass parameters at $\mu=M_N$ appropriately to get the desired bare mass parameters at respective symmetry breaking scales consistent with them minimisation conditions given in Eq. \eqref{eq:veta}. As the threshold corrections to $\mu^2_\eta, \mu^2_H$ are large at $\mu=M_N$, appropriate fine-tuning\footnote{We show the bare mass parameter values only upto the fifth decimal places in table~\ref{tab:threshold}, though we have considered more fine-tuning for $\mu^2_\eta$ in order to get the desired values at lower scales.} is required in order to get their desired values at lower scales, consistent with symmetry breaking patterns. The values of the mass parameters at different scales for two different choices of $\lambda_{H\rho}$ are given in table~\ref{tab:threshold}. The values of other relevant parameters are taken according to their respective running shown in Fig. \ref{fig:RGE}.

\begin{table}[h!]
\centering
\tiny
\captionsetup{justification=centering}
\begin{tabular}{|c|c|c|c|c|c|c|}
\toprule
\textbf{Parameter} & \multicolumn{3}{c|}{$\lambda_{H\rho} = 0.02$} & \multicolumn{3}{c|}{$\lambda_{H\rho} = 0.14$} \\
 & $\mu =  v$ & $\mu =  v_\rho$ & $\mu = M_N$ & $\mu =  v$ &  $\mu = v_\rho$ &  $\mu = M_N$ \\
\midrule
$\mu_1$ (GeV) & 0.00998 & 0.01 & 0.01002 & 0.00999 & 0.01 & 0.01002 \\
$\mu_\eta^2$ (GeV$^2$) & 9.8845 & 10 &  $1.21585 \times 10^{12}$ & 9.8845 & $10$ & $1.21585 \times 10^{12}$ \\
$\mu_H^2$ (GeV$^2$) & $1.00008 \times 10^{8}$ & $1.07749\times 10^{8}$ & $-4.85342 \times 10^{10}$ & $7.00021 \times 10^{8}$ & $7.99334\times10^8$ & $-4.79341 \times 10^{10}$ \\
$\mu_\rho^2$ (GeV$^2$) & $2.2124\times 10^8$ & $2.2225 \times 10^8$&  $2.23634 \times 10^8$ & $2.10257\times 10^8$ & $2.1969 \times 10^8$ & $2.3673 \times 10^8$ \\
\bottomrule
\end{tabular}
\caption{Running of mass parameters with threshold corrections at $\mu=M_N$. The scales are chosen as $ M_N = 10^9$ GeV, $v_\rho = 10^5$ GeV while heavy fermion Yukawa couplings are $Y_L = 0.0016$ and $Y_R = 0.008$.} \label{tab:threshold}
\end{table}

\section{Quantum gravity origin of DW bias and DM instability}\label{app:D}
In section~\ref{sec:grav}, we mentioned that the bias term breaking $Z_4$ explicitly can arise naturally in theories of quantum gravity as higher dimensional operators suppressed by some powers of the scale of QG, $\Lambda_{\rm QG}$. Since a theory of QG is supposed to break all global symmetries, it will also break the $Z_2^{\rm CP}$ symmetry in our model that stabilizes the DM. Therefore, requirement of the correct bias term for domain walls and lifetime of dark matter, can constrain the scale of QG or the dimension of such QG scale suppressed operators. One possible leading order $\Lambda_{\rm QG}$ suppressed operator breaking $Z_2^{\rm CP}$ is the dimension 5 operator $c_1(H^\dagger H)^2\rho/\Lambda_{\rm QG}+{\rm h.c.}$ where $c_1$ is a $\mathcal{O}(1)$ complex number. Below electroweak symmetry breaking, this operator can lead to DM decay into two Higgs via the interaction
\begin{equation}
    \mathcal{L}_{\rm int}\supset 2\,\text{Im}[c_1]\frac{3iv^2}{2\sqrt{2}\Lambda_{\rm QG}}h^2\chi.
\end{equation}
We can consider the factor $2\,\text{Im}[c_1]=1$. The decay width of the DM candidate $\chi$ into two Higgs is given by,
\begin{equation}
    \Gamma(\chi\rightarrow hh)=\frac{|g_{\chi hh}|^2}{32\pi m_\chi}\left(1-\frac{4 m_{h_1}^2}{m_\chi^2}\right)^{1/2},
\end{equation}
where $|g_{\chi hh}|=\frac{3v^2}{2\sqrt{2}\Lambda_{\rm QG}}$ is the effective coupling. The lifetime of $\chi$ should be larger than the lifetime of the Universe,
\begin{equation}
    \tau_{\chi}=\frac{1}{\Gamma}\geq t_{\rm universe}\simeq4.36\times 10^{17}\,\text{s},
\end{equation}
which gives a lower bound on the scale of QG,
\begin{equation}
    \Lambda_{\rm QG}\gtrsim 10^{23}\,\text{GeV} \left(\frac{1\text{ TeV}}{m_\chi}\right)^{1/2}
\end{equation}
As mentioned earlier, the same QG scale is responsible for breaking the $Z_4$ symmetry of our model via terms like Eq.~\eqref{eq:bias} originating from . This implies $\epsilon\sim\frac{v_\rho}{\Lambda_{\rm QG}}$ or $\epsilon v_\rho^4\sim \frac{v_\rho^5}{\Lambda_{\rm QG}}=\frac{(m_\chi/\sqrt{8\lambda_1})^5}{\Lambda_{\rm QG}}\lesssim 10^{-3}$ GeV$^{4}$ for $m_\chi=1$ TeV. This is compatible with Fig.~\ref{fig:SNRmDM}, implying compatibility of the bias with DM lifetime assuming both have a QG origin. Similarly, we verified that other dimension-5 operators such as $c_2(H^\dagger H) \eta^2\rho/\Lambda_{\rm QG}$ + h.c. or $c_3(H^\dagger H) \rho^3/\Lambda_{\rm QG}$ + h.c. are also consistent with Fig.~\ref{fig:SNRmDM}.

\providecommand{\href}[2]{#2}\begingroup\raggedright\endgroup


\begin{thebibliography}{100}

\bibitem{ParticleDataGroup:2024cfk}
{\bf Particle Data Group} Collaboration, S.~Navas {\em et~al.}, {\it {Review of particle physics}},  {\em Phys. Rev. D} {\bf 110} (2024), no.~3 030001.

\bibitem{Dolinski:2019nrj}
M.~J. Dolinski, A.~W.~P. Poon, and W.~Rodejohann, {\it {Neutrinoless Double-Beta Decay: Status and Prospects}},  {\em Ann. Rev. Nucl. Part. Sci.} {\bf 69} (2019) 219--251, [\href{http://arxiv.org/abs/1902.04097}{{\tt arXiv:1902.04097}}].

\bibitem{Majorana:2022udl}
{\bf Majorana} Collaboration, I.~J. Arnquist {\em et~al.}, {\it {Final Result of the Majorana Demonstrator\textquoteright{}s Search for Neutrinoless Double-\ensuremath{\beta} Decay in Ge76}},  {\em Phys. Rev. Lett.} {\bf 130} (2023), no.~6 062501, [\href{http://arxiv.org/abs/2207.07638}{{\tt arXiv:2207.07638}}].

\bibitem{Minkowski:1977sc}
P.~Minkowski, {\it {$\mu \to e\gamma$ at a Rate of One Out of $10^{9}$ Muon Decays?}},  {\em Phys. Lett. B} {\bf 67} (1977) 421--428.

\bibitem{Gell-Mann:1979vob}
M.~Gell-Mann, P.~Ramond, and R.~Slansky, {\it {Complex Spinors and Unified Theories}},  {\em Conf. Proc. C} {\bf 790927} (1979) 315--321, [\href{http://arxiv.org/abs/1306.4669}{{\tt arXiv:1306.4669}}].

\bibitem{Mohapatra:1979ia}
R.~N. Mohapatra and G.~Senjanovic, {\it {Neutrino Mass and Spontaneous Parity Nonconservation}},  {\em Phys. Rev. Lett.} {\bf 44} (1980) 912.

\bibitem{Sawada:1979dis}
O.~Sawada and A.~Sugamoto, eds., {\em {Proceedings: Workshop on the Unified Theories and the Baryon Number in the Universe}: {Tsukuba, Japan, February 13-14, 1979}}, (Tsukuba, Japan), Natl.Lab.High Energy Phys., 1979.

\bibitem{Yanagida:1980xy}
T.~Yanagida, {\it {Horizontal Symmetry and Masses of Neutrinos}},  {\em Prog. Theor. Phys.} {\bf 64} (1980) 1103.

\bibitem{Schechter:1980gr}
J.~Schechter and J.~W.~F. Valle, {\it {Neutrino Masses in SU(2) x U(1) Theories}},  {\em Phys. Rev. D} {\bf 22} (1980) 2227.

\bibitem{Mohapatra:1980yp}
R.~N. Mohapatra and G.~Senjanovic, {\it {Neutrino Masses and Mixings in Gauge Models with Spontaneous Parity Violation}},  {\em Phys. Rev. D} {\bf 23} (1981) 165.

\bibitem{Schechter:1981cv}
J.~Schechter and J.~W.~F. Valle, {\it {Neutrino Decay and Spontaneous Violation of Lepton Number}},  {\em Phys. Rev. D} {\bf 25} (1982) 774.

\bibitem{Wetterich:1981bx}
C.~Wetterich, {\it {Neutrino Masses and the Scale of B-L Violation}},  {\em Nucl. Phys. B} {\bf 187} (1981) 343--375.

\bibitem{Lazarides:1980nt}
G.~Lazarides, Q.~Shafi, and C.~Wetterich, {\it {Proton Lifetime and Fermion Masses in an SO(10) Model}},  {\em Nucl. Phys. B} {\bf 181} (1981) 287--300.

\bibitem{Brahmachari:1997cq}
B.~Brahmachari and R.~N. Mohapatra, {\it {Unified explanation of the solar and atmospheric neutrino puzzles in a minimal supersymmetric SO(10) model}},  {\em Phys. Rev. D} {\bf 58} (1998) 015001, [\href{http://arxiv.org/abs/hep-ph/9710371}{{\tt hep-ph/9710371}}].

\bibitem{Foot:1988aq}
R.~Foot, H.~Lew, X.~G. He, and G.~C. Joshi, {\it {Seesaw Neutrino Masses Induced by a Triplet of Leptons}},  {\em Z. Phys. C} {\bf 44} (1989) 441.

\bibitem{Roncadelli:1983ty}
M.~Roncadelli and D.~Wyler, {\it {Naturally Light Dirac Neutrinos in Gauge Theories}},  {\em Phys. Lett. B} {\bf 133} (1983) 325--329.

\bibitem{Roy:1983be}
P.~Roy and O.~U. Shanker, {\it {Observable Neutrino Dirac Mass and Supergrand Unification}},  {\em Phys. Rev. Lett.} {\bf 52} (1984) 713--716. [Erratum: Phys.Rev.Lett. 52, 2190 (1984)].

\bibitem{Babu:1988yq}
K.~S. Babu and X.~G. He, {\it {DIRAC NEUTRINO MASSES AS TWO LOOP RADIATIVE CORRECTIONS}},  {\em Mod. Phys. Lett. A} {\bf 4} (1989) 61.

\bibitem{Peltoniemi:1992ss}
J.~T. Peltoniemi, D.~Tommasini, and J.~W.~F. Valle, {\it {Reconciling dark matter and solar neutrinos}},  {\em Phys. Lett. B} {\bf 298} (1993) 383--390.

\bibitem{Chulia:2016ngi}
S.~Centelles~Chuliá, E.~Ma, R.~Srivastava, and J.~W.~F. Valle, {\it {Dirac Neutrinos and Dark Matter Stability from Lepton Quarticity}},  {\em Phys. Lett.} {\bf B767} (2017) 209--213, [\href{http://arxiv.org/abs/1606.04543}{{\tt arXiv:1606.04543}}].

\bibitem{Aranda:2013gga}
A.~Aranda, C.~Bonilla, S.~Morisi, E.~Peinado, and J.~W.~F. Valle, {\it {Dirac neutrinos from flavor symmetry}},  {\em Phys. Rev. D} {\bf 89} (2014), no.~3 033001, [\href{http://arxiv.org/abs/1307.3553}{{\tt arXiv:1307.3553}}].

\bibitem{Chen:2015jta}
P.~Chen, G.-J. Ding, A.~D. Rojas, C.~A. Vaquera-Araujo, and J.~W.~F. Valle, {\it {Warped flavor symmetry predictions for neutrino physics}},  {\em JHEP} {\bf 01} (2016) 007, [\href{http://arxiv.org/abs/1509.06683}{{\tt arXiv:1509.06683}}].

\bibitem{Ma:2015mjd}
E.~Ma, N.~Pollard, R.~Srivastava, and M.~Zakeri, {\it {Gauge $B-L$ Model with Residual $Z_3$ Symmetry}},  {\em Phys. Lett. B} {\bf 750} (2015) 135--138, [\href{http://arxiv.org/abs/1507.03943}{{\tt arXiv:1507.03943}}].

\bibitem{Reig:2016ewy}
M.~Reig, J.~W.~F. Valle, and C.~A. Vaquera-Araujo, {\it {Realistic $\mathrm{SU(3)_c \otimes SU(3)_L \otimes U(1)_X}$ model with a type II Dirac neutrino seesaw mechanism}},  {\em Phys. Rev. D} {\bf 94} (2016), no.~3 033012, [\href{http://arxiv.org/abs/1606.08499}{{\tt arXiv:1606.08499}}].

\bibitem{Wang:2016lve}
W.~Wang and Z.-L. Han, {\it {Naturally Small Dirac Neutrino Mass with Intermediate $SU(2)_{L}$ Multiplet Fields}},  \href{http://arxiv.org/abs/1611.03240}{{\tt arXiv:1611.03240}}. [JHEP04,166(2017)].

\bibitem{Wang:2017mcy}
W.~Wang, R.~Wang, Z.-L. Han, and J.-Z. Han, {\it {The $B-L$ Scotogenic Models for Dirac Neutrino Masses}},  {\em Eur. Phys. J.} {\bf C77} (2017), no.~12 889, [\href{http://arxiv.org/abs/1705.00414}{{\tt arXiv:1705.00414}}].

\bibitem{Wang:2006jy}
F.~Wang, W.~Wang, and J.~M. Yang, {\it {Split two-Higgs-doublet model and neutrino condensation}},  {\em Europhys. Lett.} {\bf 76} (2006) 388--394, [\href{http://arxiv.org/abs/hep-ph/0601018}{{\tt hep-ph/0601018}}].

\bibitem{Gabriel:2006ns}
S.~Gabriel and S.~Nandi, {\it {A New two Higgs doublet model}},  {\em Phys. Lett.} {\bf B655} (2007) 141--147, [\href{http://arxiv.org/abs/hep-ph/0610253}{{\tt hep-ph/0610253}}].

\bibitem{Davidson:2009ha}
S.~M. Davidson and H.~E. Logan, {\it {Dirac neutrinos from a second Higgs doublet}},  {\em Phys. Rev. D} {\bf 80} (2009) 095008, [\href{http://arxiv.org/abs/0906.3335}{{\tt arXiv:0906.3335}}].

\bibitem{Davidson:2010sf}
S.~M. Davidson and H.~E. Logan, {\it {LHC phenomenology of a two-Higgs-doublet neutrino mass model}},  {\em Phys. Rev.} {\bf D82} (2010) 115031, [\href{http://arxiv.org/abs/1009.4413}{{\tt arXiv:1009.4413}}].

\bibitem{Bonilla:2016zef}
C.~Bonilla and J.~W.~F. Valle, {\it {Naturally light neutrinos in $Diracon$ model}},  {\em Phys. Lett.} {\bf B762} (2016) 162--165, [\href{http://arxiv.org/abs/1605.08362}{{\tt arXiv:1605.08362}}].

\bibitem{Farzan:2012sa}
Y.~Farzan and E.~Ma, {\it {Dirac neutrino mass generation from dark matter}},  {\em Phys. Rev.} {\bf D86} (2012) 033007, [\href{http://arxiv.org/abs/1204.4890}{{\tt arXiv:1204.4890}}].

\bibitem{Bonilla:2016diq}
C.~Bonilla, E.~Ma, E.~Peinado, and J.~W.~F. Valle, {\it {Two-loop Dirac neutrino mass and WIMP dark matter}},  {\em Phys. Lett.} {\bf B762} (2016) 214--218, [\href{http://arxiv.org/abs/1607.03931}{{\tt arXiv:1607.03931}}].

\bibitem{Ma:2016mwh}
E.~Ma and O.~Popov, {\it {Pathways to Naturally Small Dirac Neutrino Masses}},  {\em Phys. Lett.} {\bf B764} (2017) 142--144, [\href{http://arxiv.org/abs/1609.02538}{{\tt arXiv:1609.02538}}].

\bibitem{Ma:2017kgb}
E.~Ma and U.~Sarkar, {\it {Radiative Left-Right Dirac Neutrino Mass}},  {\em Phys. Lett.} {\bf B776} (2018) 54--57, [\href{http://arxiv.org/abs/1707.07698}{{\tt arXiv:1707.07698}}].

\bibitem{Borah:2016lrl}
D.~Borah, {\it {Light sterile neutrino and dark matter in left-right symmetric models without a Higgs bidoublet}},  {\em Phys. Rev.} {\bf D94} (2016), no.~7 075024, [\href{http://arxiv.org/abs/1607.00244}{{\tt arXiv:1607.00244}}].

\bibitem{Borah:2016zbd}
D.~Borah and A.~Dasgupta, {\it {Common Origin of Neutrino Mass, Dark Matter and Dirac Leptogenesis}},  {\em JCAP} {\bf 12} (2016), no.~12 034, [\href{http://arxiv.org/abs/1608.03872}{{\tt arXiv:1608.03872}}].

\bibitem{Borah:2016hqn}
D.~Borah and A.~Dasgupta, {\it {Observable Lepton Number Violation with Predominantly Dirac Nature of Active Neutrinos}},  {\em JHEP} {\bf 01} (2017) 072, [\href{http://arxiv.org/abs/1609.04236}{{\tt arXiv:1609.04236}}].

\bibitem{Borah:2017leo}
D.~Borah and A.~Dasgupta, {\it {Naturally Light Dirac Neutrino in Left-Right Symmetric Model}},  {\em JCAP} {\bf 06} (2017), no.~06 003, [\href{http://arxiv.org/abs/1702.02877}{{\tt arXiv:1702.02877}}].

\bibitem{CentellesChulia:2017koy}
S.~Centelles~Chuli\'a, R.~Srivastava, and J.~W.~F. Valle, {\it {Generalized Bottom-Tau unification, neutrino oscillations and dark matter: predictions from a lepton quarticity flavor approach}},  {\em Phys. Lett. B} {\bf 773} (2017) 26--33, [\href{http://arxiv.org/abs/1706.00210}{{\tt arXiv:1706.00210}}].

\bibitem{Bonilla:2017ekt}
C.~Bonilla, J.~M. Lamprea, E.~Peinado, and J.~W.~F. Valle, {\it {Flavour-symmetric type-II Dirac neutrino seesaw mechanism}},  {\em Phys. Lett. B} {\bf 779} (2018) 257--261, [\href{http://arxiv.org/abs/1710.06498}{{\tt arXiv:1710.06498}}].

\bibitem{Memenga:2013vc}
N.~Memenga, W.~Rodejohann, and H.~Zhang, {\it {$A_4$ flavor symmetry model for Dirac neutrinos and sizable $U_{e3}$}},  {\em Phys. Rev. D} {\bf 87} (2013), no.~5 053021, [\href{http://arxiv.org/abs/1301.2963}{{\tt arXiv:1301.2963}}].

\bibitem{Borah:2017dmk}
D.~Borah and B.~Karmakar, {\it {$A_4$ flavour model for Dirac neutrinos: Type I and inverse seesaw}},  {\em Phys. Lett. B} {\bf 780} (2018) 461--470, [\href{http://arxiv.org/abs/1712.06407}{{\tt arXiv:1712.06407}}].

\bibitem{CentellesChulia:2018gwr}
S.~Centelles~Chuli\'a, R.~Srivastava, and J.~W.~F. Valle, {\it {Seesaw roadmap to neutrino mass and dark matter}},  {\em Phys. Lett. B} {\bf 781} (2018) 122--128, [\href{http://arxiv.org/abs/1802.05722}{{\tt arXiv:1802.05722}}].

\bibitem{CentellesChulia:2018bkz}
S.~Centelles~Chuli\'a, R.~Srivastava, and J.~W.~F. Valle, {\it {Seesaw Dirac neutrino mass through dimension-six operators}},  {\em Phys. Rev. D} {\bf 98} (2018), no.~3 035009, [\href{http://arxiv.org/abs/1804.03181}{{\tt arXiv:1804.03181}}].

\bibitem{Han:2018zcn}
Z.-L. Han and W.~Wang, {\it {$Z'$ Portal Dark Matter in $B-L$ Scotogenic Dirac Model}},  \href{http://arxiv.org/abs/1805.02025}{{\tt arXiv:1805.02025}}.

\bibitem{Borah:2018gjk}
D.~Borah, B.~Karmakar, and D.~Nanda, {\it {Common Origin of Dirac Neutrino Mass and Freeze-in Massive Particle Dark Matter}},  {\em JCAP} {\bf 07} (2018) 039, [\href{http://arxiv.org/abs/1805.11115}{{\tt arXiv:1805.11115}}].

\bibitem{Borah:2018nvu}
D.~Borah and B.~Karmakar, {\it {Linear seesaw for Dirac neutrinos with $A_4$ flavour symmetry}},  {\em Phys. Lett. B} {\bf 789} (2019) 59--70, [\href{http://arxiv.org/abs/1806.10685}{{\tt arXiv:1806.10685}}].

\bibitem{Borah:2019bdi}
D.~Borah, D.~Nanda, and A.~K. Saha, {\it {Common origin of modified chaotic inflation, nonthermal dark matter, and Dirac neutrino mass}},  {\em Phys. Rev. D} {\bf 101} (2020), no.~7 075006, [\href{http://arxiv.org/abs/1904.04840}{{\tt arXiv:1904.04840}}].

\bibitem{CentellesChulia:2019xky}
S.~Centelles~Chuli\'a, R.~Cepedello, E.~Peinado, and R.~Srivastava, {\it {Systematic classification of two loop $d$ = 4 Dirac neutrino mass models and the Diracness-dark matter stability connection}},  {\em JHEP} {\bf 10} (2019) 093, [\href{http://arxiv.org/abs/1907.08630}{{\tt arXiv:1907.08630}}].

\bibitem{Jana:2019mgj}
S.~Jana, P.~K. Vishnu, and S.~Saad, {\it {Minimal realizations of Dirac neutrino mass from generic one-loop and two-loop topologies at $d = 5$}},  {\em JCAP} {\bf 04} (2020) 018, [\href{http://arxiv.org/abs/1910.09537}{{\tt arXiv:1910.09537}}].

\bibitem{Nanda:2019nqy}
D.~Nanda and D.~Borah, {\it {Connecting Light Dirac Neutrinos to a Multi-component Dark Matter Scenario in Gauged $B-L$ Model}},  {\em Eur. Phys. J. C} {\bf 80} (2020), no.~6 557, [\href{http://arxiv.org/abs/1911.04703}{{\tt arXiv:1911.04703}}].

\bibitem{Guo:2020qin}
S.-Y. Guo and Z.-L. Han, {\it {Observable Signatures of Scotogenic Dirac Model}},  {\em JHEP} {\bf 12} (2020) 062, [\href{http://arxiv.org/abs/2005.08287}{{\tt arXiv:2005.08287}}].

\bibitem{Bernal:2021ezl}
N.~Bernal, J.~Calle, and D.~Restrepo, {\it {Anomaly-free Abelian gauge symmetries with Dirac scotogenic models}},  {\em Phys. Rev. D} {\bf 103} (2021), no.~9 095032, [\href{http://arxiv.org/abs/2102.06211}{{\tt arXiv:2102.06211}}].

\bibitem{Borah:2022obi}
D.~Borah, S.~Mahapatra, D.~Nanda, and N.~Sahu, {\it {Type II Dirac seesaw with observable \ensuremath{\Delta}Neff in the light of W-mass anomaly}},  {\em Phys. Lett. B} {\bf 833} (2022) 137297, [\href{http://arxiv.org/abs/2204.08266}{{\tt arXiv:2204.08266}}].

\bibitem{Li:2022chc}
S.-P. Li, X.-Q. Li, X.-S. Yan, and Y.-D. Yang, {\it {Scotogenic Dirac neutrino mass models embedded with leptoquarks: one pathway to address the flavor anomalies and the neutrino masses together}},  {\em Eur. Phys. J. C} {\bf 82} (2022), no.~11 1078, [\href{http://arxiv.org/abs/2204.09201}{{\tt arXiv:2204.09201}}].

\bibitem{Dey:2024ctx}
M.~Dey and S.~Roy, {\it {Revisiting the Dirac Nature of Neutrinos}},  \href{http://arxiv.org/abs/2403.12461}{{\tt arXiv:2403.12461}}.

\bibitem{Singh:2024imk}
L.~Singh, M.~Kashav, and S.~Verma, {\it {Minimal type-I Dirac seesaw and leptogenesis under A4 modular invariance}},  {\em Nucl. Phys. B} {\bf 1007} (5, 2024) 116666, [\href{http://arxiv.org/abs/2405.07165}{{\tt arXiv:2405.07165}}].

\bibitem{Borah:2024gql}
D.~Borah, P.~Das, B.~Karmakar, and S.~Mahapatra, {\it {Discrete dark matter with light Dirac neutrinos}},  \href{http://arxiv.org/abs/2406.17861}{{\tt arXiv:2406.17861}}.

\bibitem{Abazajian:2019oqj}
K.~N. Abazajian and J.~Heeck, {\it {Observing Dirac neutrinos in the cosmic microwave background}},  {\em Phys. Rev. D} {\bf 100} (2019) 075027, [\href{http://arxiv.org/abs/1908.03286}{{\tt arXiv:1908.03286}}].

\bibitem{FileviezPerez:2019cyn}
P.~Fileviez~P\'erez, C.~Murgui, and A.~D. Plascencia, {\it {Neutrino-Dark Matter Connections in Gauge Theories}},  {\em Phys. Rev. D} {\bf 100} (2019), no.~3 035041, [\href{http://arxiv.org/abs/1905.06344}{{\tt arXiv:1905.06344}}].

\bibitem{Han:2020oet}
C.~Han, M.~L. L\'opez-Ib\'a\~nez, B.~Peng, and J.~M. Yang, {\it {Dirac dark matter in $U(1)_{B-L}$ with the Stueckelberg mechanism}},  {\em Nucl. Phys. B} {\bf 959} (2020) 115154, [\href{http://arxiv.org/abs/2001.04078}{{\tt arXiv:2001.04078}}].

\bibitem{Luo:2020sho}
X.~Luo, W.~Rodejohann, and X.-J. Xu, {\it {Dirac neutrinos and $N_{{\rm eff}}$}},  {\em JCAP} {\bf 06} (2020) 058, [\href{http://arxiv.org/abs/2005.01629}{{\tt arXiv:2005.01629}}].

\bibitem{Borah:2020boy}
D.~Borah, A.~Dasgupta, C.~Majumdar, and D.~Nanda, {\it {Observing left-right symmetry in the cosmic microwave background}},  {\em Phys. Rev. D} {\bf 102} (2020), no.~3 035025, [\href{http://arxiv.org/abs/2005.02343}{{\tt arXiv:2005.02343}}].

\bibitem{Adshead:2020ekg}
P.~Adshead, Y.~Cui, A.~J. Long, and M.~Shamma, {\it {Unraveling the Dirac neutrino with cosmological and terrestrial detectors}},  {\em Phys. Lett. B} {\bf 823} (2021) 136736, [\href{http://arxiv.org/abs/2009.07852}{{\tt arXiv:2009.07852}}].

\bibitem{Luo:2020fdt}
X.~Luo, W.~Rodejohann, and X.-J. Xu, {\it {Dirac neutrinos and N$_{eff}$. Part II. The freeze-in case}},  {\em JCAP} {\bf 03} (2021) 082, [\href{http://arxiv.org/abs/2011.13059}{{\tt arXiv:2011.13059}}].

\bibitem{Mahanta:2021plx}
D.~Mahanta and D.~Borah, {\it {Low scale Dirac leptogenesis and dark matter with observable $\Delta N_{\mathrm{eff}}$}},  {\em Eur. Phys. J. C} {\bf 82} (1, 2022) 495, [\href{http://arxiv.org/abs/2101.02092}{{\tt arXiv:2101.02092}}].

\bibitem{Du:2021idh}
Y.~Du and J.-H. Yu, {\it {Neutrino non-standard interactions meet precision measurements of N$_{eff}$}},  {\em JHEP} {\bf 05} (2021) 058, [\href{http://arxiv.org/abs/2101.10475}{{\tt arXiv:2101.10475}}].

\bibitem{Biswas:2021kio}
A.~Biswas, D.~Borah, and D.~Nanda, {\it {Light Dirac neutrino portal dark matter with observable \ensuremath{\Delta}Neff}},  {\em JCAP} {\bf 10} (3, 2021) 002, [\href{http://arxiv.org/abs/2103.05648}{{\tt arXiv:2103.05648}}].

\bibitem{Borah:2022qln}
D.~Borah, S.~Jyoti~Das, and N.~Okada, {\it {Affleck-Dine cogenesis of baryon and dark matter}},  {\em JHEP} {\bf 05} (2023) 004, [\href{http://arxiv.org/abs/2212.04516}{{\tt arXiv:2212.04516}}].

\bibitem{Li:2022yna}
S.-P. Li, X.-Q. Li, X.-S. Yan, and Y.-D. Yang, {\it {Cosmological imprints of Dirac neutrinos in a keV-vacuum 2HDM*}},  {\em Chin. Phys. C} {\bf 47} (2023), no.~4 043109, [\href{http://arxiv.org/abs/2202.10250}{{\tt arXiv:2202.10250}}].

\bibitem{Biswas:2022fga}
A.~Biswas, D.~K. Ghosh, and D.~Nanda, {\it {Concealing Dirac neutrinos from cosmic microwave background}},  {\em JCAP} {\bf 10} (2022) 006, [\href{http://arxiv.org/abs/2206.13710}{{\tt arXiv:2206.13710}}].

\bibitem{Adshead:2022ovo}
P.~Adshead, P.~Ralegankar, and J.~Shelton, {\it {Dark radiation constraints on portal interactions with hidden sectors}},  {\em JCAP} {\bf 09} (2022) 056, [\href{http://arxiv.org/abs/2206.13530}{{\tt arXiv:2206.13530}}].

\bibitem{Borah:2023dhk}
D.~Borah, S.~Mahapatra, D.~Nanda, S.~K. Sahoo, and N.~Sahu, {\it {Singlet-doublet fermion Dark Matter with Dirac neutrino mass, (g \ensuremath{-} 2)$_{\mu}$ and \ensuremath{\Delta}N$_{eff}$}},  {\em JHEP} {\bf 05} (2024) 096, [\href{http://arxiv.org/abs/2310.03721}{{\tt arXiv:2310.03721}}].

\bibitem{Borah:2022enh}
D.~Borah, P.~Das, and D.~Nanda, {\it {Observable $\Delta {\textrm{N}}_{\textrm{eff}}$ in Dirac scotogenic model}},  {\em Eur. Phys. J. C} {\bf 84} (2024), no.~2 140, [\href{http://arxiv.org/abs/2211.13168}{{\tt arXiv:2211.13168}}].

\bibitem{Das:2023oph}
N.~Das, S.~Jyoti~Das, and D.~Borah, {\it {Thermalized dark radiation in the presence of a PBH: \ensuremath{\Delta}Neff and gravitational waves complementarity}},  {\em Phys. Rev. D} {\bf 108} (2023), no.~9 095052, [\href{http://arxiv.org/abs/2306.00067}{{\tt arXiv:2306.00067}}].

\bibitem{Esseili:2023ldf}
H.~Esseili and G.~D. Kribs, {\it {Cosmological implications of gauged U(1)$_{B-L}$ on \ensuremath{\Delta}N $_{eff}$ in the CMB and BBN}},  {\em JCAP} {\bf 05} (2024) 110, [\href{http://arxiv.org/abs/2308.07955}{{\tt arXiv:2308.07955}}].

\bibitem{Das:2023yhv}
N.~Das and D.~Borah, {\it {Light Dirac neutrino portal dark matter with gauged U(1)B-L symmetry}},  {\em Phys. Rev. D} {\bf 109} (2024), no.~7 075045, [\href{http://arxiv.org/abs/2312.06777}{{\tt arXiv:2312.06777}}].

\bibitem{Borah:2024twm}
D.~Borah, N.~Das, S.~Jahedi, and B.~Thacker, {\it {Collider and CMB complementarity of leptophilic dark matter with light Dirac neutrinos}},  \href{http://arxiv.org/abs/2408.14548}{{\tt arXiv:2408.14548}}.

\bibitem{Barman:2022yos}
B.~Barman, D.~Borah, A.~Dasgupta, and A.~Ghoshal, {\it {Probing high scale Dirac leptogenesis via gravitational waves from domain walls}},  {\em Phys. Rev. D} {\bf 106} (2022), no.~1 015007, [\href{http://arxiv.org/abs/2205.03422}{{\tt arXiv:2205.03422}}].

\bibitem{Barman:2023fad}
B.~Barman, D.~Borah, S.~Jyoti~Das, and I.~Saha, {\it {Scale of Dirac leptogenesis and left-right symmetry in the light of recent PTA results}},  {\em JCAP} {\bf 10} (7, 2023) 053, [\href{http://arxiv.org/abs/2307.00656}{{\tt arXiv:2307.00656}}].

\bibitem{Fukugita:1986hr}
M.~Fukugita and T.~Yanagida, {\it {Baryogenesis Without Grand Unification}},  {\em Phys. Lett. B} {\bf 174} (1986) 45--47.

\bibitem{Dick:1999je}
K.~Dick, M.~Lindner, M.~Ratz, and D.~Wright, {\it {Leptogenesis with Dirac neutrinos}},  {\em Phys. Rev. Lett.} {\bf 84} (2000) 4039--4042, [\href{http://arxiv.org/abs/hep-ph/9907562}{{\tt hep-ph/9907562}}].

\bibitem{Murayama:2002je}
H.~Murayama and A.~Pierce, {\it {Realistic Dirac leptogenesis}},  {\em Phys. Rev. Lett.} {\bf 89} (2002) 271601, [\href{http://arxiv.org/abs/hep-ph/0206177}{{\tt hep-ph/0206177}}].

\bibitem{Boz:2004ga}
M.~Boz and N.~K. Pak, {\it {Dirac Leptogenesis and anomalous U(1)}},  {\em Eur. Phys. J. C} {\bf 37} (2004) 507--510.

\bibitem{Thomas:2005rs}
B.~Thomas and M.~Toharia, {\it {Phenomenology of Dirac neutrinogenesis in split supersymmetry}},  {\em Phys. Rev. D} {\bf 73} (2006) 063512, [\href{http://arxiv.org/abs/hep-ph/0511206}{{\tt hep-ph/0511206}}].

\bibitem{Gu:2006dc}
P.-H. Gu and H.-J. He, {\it {Neutrino Mass and Baryon Asymmetry from Dirac Seesaw}},  {\em JCAP} {\bf 12} (2006) 010, [\href{http://arxiv.org/abs/hep-ph/0610275}{{\tt hep-ph/0610275}}].

\bibitem{Bechinger:2009qk}
A.~Bechinger and G.~Seidl, {\it {Resonant Dirac leptogenesis on throats}},  {\em Phys. Rev. D} {\bf 81} (2010) 065015, [\href{http://arxiv.org/abs/0907.4341}{{\tt arXiv:0907.4341}}].

\bibitem{Chen:2011sb}
M.-C. Chen, J.~Huang, and W.~Shepherd, {\it {Dirac Leptogenesis with a Non-anomalous $U(1)^{\prime}$ Family Symmetry}},  {\em JHEP} {\bf 11} (2012) 059, [\href{http://arxiv.org/abs/1111.5018}{{\tt arXiv:1111.5018}}].

\bibitem{Choi:2012ba}
K.-Y. Choi, E.~J. Chun, and C.~S. Shin, {\it {Dark matter asymmetry in supersymmetric Dirac leptogenesis}},  {\em Phys. Lett. B} {\bf 723} (2013) 90--94, [\href{http://arxiv.org/abs/1211.5409}{{\tt arXiv:1211.5409}}].

\bibitem{Gu:2016hxh}
P.-H. Gu, {\it {Peccei-Quinn symmetry for Dirac seesaw and leptogenesis}},  {\em JCAP} {\bf 07} (2016) 004, [\href{http://arxiv.org/abs/1603.05070}{{\tt arXiv:1603.05070}}].

\bibitem{Narendra:2017uxl}
N.~Narendra, N.~Sahoo, and N.~Sahu, {\it {Dark matter assisted Dirac leptogenesis and neutrino mass}},  {\em Nucl. Phys. B} {\bf 936} (2018) 76--90, [\href{http://arxiv.org/abs/1712.02960}{{\tt arXiv:1712.02960}}].

\bibitem{Zwicky:1933gu}
F.~Zwicky, {\it {Die Rotverschiebung von extragalaktischen Nebeln}},  {\em Helv. Phys. Acta} {\bf 6} (1933) 110--127. [Gen. Rel. Grav.41,207(2009)].

\bibitem{Rubin:1970zza}
V.~C. Rubin and W.~K. Ford, Jr., {\it {Rotation of the Andromeda Nebula from a Spectroscopic Survey of Emission Regions}},  {\em Astrophys. J.} {\bf 159} (1970) 379--403.

\bibitem{Clowe:2006eq}
D.~Clowe, M.~Bradac, A.~H. Gonzalez, M.~Markevitch, S.~W. Randall, C.~Jones, and D.~Zaritsky, {\it {A direct empirical proof of the existence of dark matter}},  {\em Astrophys. J. Lett.} {\bf 648} (2006) L109--L113, [\href{http://arxiv.org/abs/astro-ph/0608407}{{\tt astro-ph/0608407}}].

\bibitem{Planck:2018vyg}
{\bf Planck} Collaboration, N.~Aghanim {\em et~al.}, {\it {Planck 2018 results. VI. Cosmological parameters}},  {\em Astron. Astrophys.} {\bf 641} (2020) A6, [\href{http://arxiv.org/abs/1807.06209}{{\tt arXiv:1807.06209}}]. [Erratum: Astron.Astrophys. 652, C4 (2021)].

\bibitem{Arcadi:2017kky}
G.~Arcadi, M.~Dutra, P.~Ghosh, M.~Lindner, Y.~Mambrini, M.~Pierre, S.~Profumo, and F.~S. Queiroz, {\it {The waning of the WIMP? A review of models, searches, and constraints}},  {\em Eur. Phys. J. C} {\bf 78} (2018), no.~3 203, [\href{http://arxiv.org/abs/1703.07364}{{\tt arXiv:1703.07364}}].

\bibitem{Arcadi:2024ukq}
G.~Arcadi, D.~Cabo-Almeida, M.~Dutra, P.~Ghosh, M.~Lindner, Y.~Mambrini, J.~P. Neto, M.~Pierre, S.~Profumo, and F.~S. Queiroz, {\it {The Waning of the WIMP: Endgame?}},  \href{http://arxiv.org/abs/2403.15860}{{\tt arXiv:2403.15860}}.

\bibitem{LZ:2022lsv}
{\bf LZ} Collaboration, J.~Aalbers {\em et~al.}, {\it {First Dark Matter Search Results from the LUX-ZEPLIN (LZ) Experiment}},  {\em Phys. Rev. Lett.} {\bf 131} (2023), no.~4 041002, [\href{http://arxiv.org/abs/2207.03764}{{\tt arXiv:2207.03764}}].

\bibitem{Craig:2022eqo}
N.~Craig, {\it {Naturalness: past, present, and future}},  {\em Eur. Phys. J. C} {\bf 83} (2023), no.~9 825, [\href{http://arxiv.org/abs/2205.05708}{{\tt arXiv:2205.05708}}].

\bibitem{Wu:2022tpe}
Y.~Wu, K.-P. Xie, and Y.-L. Zhou, {\it {Classification of Abelian domain walls}},  {\em Phys. Rev. D} {\bf 106} (2022), no.~7 075019, [\href{http://arxiv.org/abs/2205.11529}{{\tt arXiv:2205.11529}}].

\bibitem{Coito:2021fgo}
L.~Coito, C.~Faubel, J.~Herrero-Garcia, and A.~Santamaria, {\it {Dark matter from a complex scalar singlet: the role of dark CP and other discrete symmetries}},  {\em JHEP} {\bf 11} (2021) 202, [\href{http://arxiv.org/abs/2106.05289}{{\tt arXiv:2106.05289}}].

\bibitem{Pham:2024vso}
H.~T. Pham and E.~Senaha, {\it {Gravitational waves from domain wall collapses and dark matter in the SM with a complex scalar}},  \href{http://arxiv.org/abs/2403.16568}{{\tt arXiv:2403.16568}}.

\bibitem{Sakurai:2021ipp}
K.~Sakurai and W.~Yin, {\it {Phenomenology of CP-even ALP}},  {\em JHEP} {\bf 04} (2022) 113, [\href{http://arxiv.org/abs/2111.03653}{{\tt arXiv:2111.03653}}].

\bibitem{Peskin:1990zt}
M.~E. Peskin and T.~Takeuchi, {\it {A New constraint on a strongly interacting Higgs sector}},  {\em Phys. Rev. Lett.} {\bf 65} (1990) 964--967.

\bibitem{Peskin:1991sw}
M.~E. Peskin and T.~Takeuchi, {\it {Estimation of oblique electroweak corrections}},  {\em Phys. Rev. D} {\bf 46} (1992) 381--409.

\bibitem{Kennedy:1990ib}
D.~C. Kennedy and P.~Langacker, {\it {Precision electroweak experiments and heavy physics: A Global analysis}},  {\em Phys. Rev. Lett.} {\bf 65} (1990) 2967--2970. [Erratum: Phys.Rev.Lett. 66, 395 (1991)].

\bibitem{Han:2000gp}
T.~Han, D.~Marfatia, and R.-J. Zhang, {\it {Oblique parameter constraints on large extra dimensions}},  {\em Phys. Rev. D} {\bf 62} (2000) 125018, [\href{http://arxiv.org/abs/hep-ph/0001320}{{\tt hep-ph/0001320}}].

\bibitem{Porod:2003um}
W.~Porod, {\it {SPheno, a program for calculating supersymmetric spectra, SUSY particle decays and SUSY particle production at e+ e- colliders}},  {\em Comput. Phys. Commun.} {\bf 153} (2003) 275--315, [\href{http://arxiv.org/abs/hep-ph/0301101}{{\tt hep-ph/0301101}}].

\bibitem{Thomas:2006gr}
B.~Thomas and M.~Toharia, {\it {Lepton flavor violation and supersymmetric Dirac leptogenesis}},  {\em Phys. Rev. D} {\bf 75} (2007) 013013, [\href{http://arxiv.org/abs/hep-ph/0607285}{{\tt hep-ph/0607285}}].

\bibitem{Cerdeno:2006ha}
D.~G. Cerdeno, A.~Dedes, and T.~E.~J. Underwood, {\it {The Minimal Phantom Sector of the Standard Model: Higgs Phenomenology and Dirac Leptogenesis}},  {\em JHEP} {\bf 09} (2006) 067, [\href{http://arxiv.org/abs/hep-ph/0607157}{{\tt hep-ph/0607157}}].

\bibitem{Gu:2007mc}
P.-H. Gu, H.-J. He, and U.~Sarkar, {\it {Realistic neutrinogenesis with radiative vertex correction}},  {\em Phys. Lett. B} {\bf 659} (2008) 634--639, [\href{http://arxiv.org/abs/0709.1019}{{\tt arXiv:0709.1019}}].

\bibitem{Chun:2008pg}
E.~J. Chun and P.~Roy, {\it {Dirac Leptogenesis in extended nMSSM}},  {\em JHEP} {\bf 06} (2008) 089, [\href{http://arxiv.org/abs/0803.1720}{{\tt arXiv:0803.1720}}].

\bibitem{Babu:2024glr}
K.~S. Babu and A.~Kaladharan, {\it {Dirac Leptogenesis in Left-Right Symmetric Models}},  \href{http://arxiv.org/abs/2410.24125}{{\tt arXiv:2410.24125}}.

\bibitem{Heeck:2013vha}
J.~Heeck, {\it {Leptogenesis with Lepton-Number-Violating Dirac Neutrinos}},  {\em Phys. Rev. D} {\bf 88} (2013) 076004, [\href{http://arxiv.org/abs/1307.2241}{{\tt arXiv:1307.2241}}].

\bibitem{Gu:2019yvw}
P.-H. Gu, {\it {Leptogenesis with testable Dirac neutrino mass generation}},  {\em Phys. Lett. B} {\bf 805} (2020) 135411, [\href{http://arxiv.org/abs/1907.09443}{{\tt arXiv:1907.09443}}].

\bibitem{Casas:2001sr}
J.~A. Casas and A.~Ibarra, {\it {Oscillating neutrinos and $\mu \to e, \gamma$}},  {\em Nucl. Phys. B} {\bf 618} (2001) 171--204, [\href{http://arxiv.org/abs/hep-ph/0103065}{{\tt hep-ph/0103065}}].

\bibitem{Esteban:2024eli}
I.~Esteban, M.~C. Gonzalez-Garcia, M.~Maltoni, I.~Martinez-Soler, J.~a.~P. Pinheiro, and T.~Schwetz, {\it {NuFit-6.0: updated global analysis of three-flavor neutrino oscillations}},  {\em JHEP} {\bf 12} (2024) 216, [\href{http://arxiv.org/abs/2410.05380}{{\tt arXiv:2410.05380}}].

\bibitem{Buchmuller:2004nz}
W.~Buchmuller, P.~Di~Bari, and M.~Plumacher, {\it {Leptogenesis for pedestrians}},  {\em Annals Phys.} {\bf 315} (2005) 305--351, [\href{http://arxiv.org/abs/hep-ph/0401240}{{\tt hep-ph/0401240}}].

\bibitem{Davidson:2002qv}
S.~Davidson and A.~Ibarra, {\it {A Lower bound on the right-handed neutrino mass from leptogenesis}},  {\em Phys. Lett. B} {\bf 535} (2002) 25--32, [\href{http://arxiv.org/abs/hep-ph/0202239}{{\tt hep-ph/0202239}}].

\bibitem{Degrassi:2012ry}
G.~Degrassi, S.~Di~Vita, J.~Elias-Miro, J.~R. Espinosa, G.~F. Giudice, G.~Isidori, and A.~Strumia, {\it {Higgs mass and vacuum stability in the Standard Model at NNLO}},  {\em JHEP} {\bf 08} (2012) 098, [\href{http://arxiv.org/abs/1205.6497}{{\tt arXiv:1205.6497}}].

\bibitem{Gonderinger:2009jp}
M.~Gonderinger, Y.~Li, H.~Patel, and M.~J. Ramsey-Musolf, {\it {Vacuum Stability, Perturbativity, and Scalar Singlet Dark Matter}},  {\em JHEP} {\bf 01} (2010) 053, [\href{http://arxiv.org/abs/0910.3167}{{\tt arXiv:0910.3167}}].

\bibitem{Gondolo:1990dk}
P.~Gondolo and G.~Gelmini, {\it {Cosmic abundances of stable particles: Improved analysis}},  {\em Nucl. Phys.} {\bf B360} (1991) 145--179.

\bibitem{Duerr:2015aka}
M.~Duerr, P.~Fileviez~P\'erez, and J.~Smirnov, {\it {Scalar Dark Matter: Direct vs. Indirect Detection}},  {\em JHEP} {\bf 06} (2016) 152, [\href{http://arxiv.org/abs/1509.04282}{{\tt arXiv:1509.04282}}].

\bibitem{Binder:2021bmg}
T.~Binder, T.~Bringmann, M.~Gustafsson, and A.~Hryczuk, {\it {Dark matter relic abundance beyond kinetic equilibrium}},  {\em Eur. Phys. J. C} {\bf 81} (2021) 577, [\href{http://arxiv.org/abs/2103.01944}{{\tt arXiv:2103.01944}}].

\bibitem{Staub:2015kfa}
F.~Staub, {\it {Exploring new models in all detail with SARAH}},  {\em Adv. High Energy Phys.} {\bf 2015} (2015) 840780, [\href{http://arxiv.org/abs/1503.04200}{{\tt arXiv:1503.04200}}].

\bibitem{Belyaev:2012qa}
A.~Belyaev, N.~D. Christensen, and A.~Pukhov, {\it {CalcHEP 3.4 for collider physics within and beyond the Standard Model}},  {\em Comput. Phys. Commun.} {\bf 184} (2013) 1729--1769, [\href{http://arxiv.org/abs/1207.6082}{{\tt arXiv:1207.6082}}].

\bibitem{Belanger:2013ywg}
G.~Belanger, F.~Boudjema, and A.~Pukhov, {\it {micrOMEGAs : a code for the calculation of Dark Matter properties in generic models of particle interaction}},  in {\em {Theoretical Advanced Study Institute in Elementary Particle Physics}: {The Dark Secrets of the Terascale}}, pp.~739--790, 2013.
\newblock \href{http://arxiv.org/abs/1402.0787}{{\tt arXiv:1402.0787}}.

\bibitem{XENON:2018voc}
{\bf XENON} Collaboration, E.~Aprile {\em et~al.}, {\it {Dark Matter Search Results from a One Ton-Year Exposure of XENON1T}},  {\em Phys. Rev. Lett.} {\bf 121} (2018), no.~11 111302, [\href{http://arxiv.org/abs/1805.12562}{{\tt arXiv:1805.12562}}].

\bibitem{XENON:2020kmp}
{\bf XENON} Collaboration, E.~Aprile {\em et~al.}, {\it {Projected WIMP sensitivity of the XENONnT dark matter experiment}},  {\em JCAP} {\bf 11} (2020) 031, [\href{http://arxiv.org/abs/2007.08796}{{\tt arXiv:2007.08796}}].

\bibitem{Goodman:1984dc}
M.~W. Goodman and E.~Witten, {\it {Detectability of Certain Dark Matter Candidates}},  {\em Phys. Rev. D} {\bf 31} (1985) 3059.

\bibitem{Conrad:2014tla}
J.~Conrad, {\it {Indirect Detection of WIMP Dark Matter: a compact review}},  in {\em {Interplay between Particle and Astroparticle physics}}, 11, 2014.
\newblock \href{http://arxiv.org/abs/1411.1925}{{\tt arXiv:1411.1925}}.

\bibitem{Fermi-LAT:2015bhf}
{\bf Fermi-LAT} Collaboration, F.~Acero {\em et~al.}, {\it {Fermi Large Area Telescope Third Source Catalog}},  {\em Astrophys. J. Suppl.} {\bf 218} (2015), no.~2 23, [\href{http://arxiv.org/abs/1501.02003}{{\tt arXiv:1501.02003}}].

\bibitem{Fermi-LAT:2015kyq}
{\bf Fermi-LAT} Collaboration, M.~Ackermann {\em et~al.}, {\it {Updated search for spectral lines from Galactic dark matter interactions with pass 8 data from the Fermi Large Area Telescope}},  {\em Phys. Rev. D} {\bf 91} (2015), no.~12 122002, [\href{http://arxiv.org/abs/1506.00013}{{\tt arXiv:1506.00013}}].

\bibitem{Foster:2022nva}
J.~W. Foster, Y.~Park, B.~R. Safdi, Y.~Soreq, and W.~L. Xu, {\it {Search for dark matter lines at the Galactic Center with 14~years of Fermi data}},  {\em Phys. Rev. D} {\bf 107} (2023), no.~10 103047, [\href{http://arxiv.org/abs/2212.07435}{{\tt arXiv:2212.07435}}].

\bibitem{HESS:2016mib}
{\bf H.E.S.S.} Collaboration, H.~Abdallah {\em et~al.}, {\it {Search for dark matter annihilations towards the inner Galactic halo from 10 years of observations with H.E.S.S}},  {\em Phys. Rev. Lett.} {\bf 117} (2016), no.~11 111301, [\href{http://arxiv.org/abs/1607.08142}{{\tt arXiv:1607.08142}}].

\bibitem{HESS:2018cbt}
{\bf HESS} Collaboration, H.~Abdallah {\em et~al.}, {\it {Search for $\gamma$-Ray Line Signals from Dark Matter Annihilations in the Inner Galactic Halo from 10 Years of Observations with H.E.S.S.}},  {\em Phys. Rev. Lett.} {\bf 120} (2018), no.~20 201101, [\href{http://arxiv.org/abs/1805.05741}{{\tt arXiv:1805.05741}}].

\bibitem{CTAConsortium:2017dvg}
{\bf CTA Consortium} Collaboration, B.~S. Acharya {\em et~al.}, {\em {Science with the Cherenkov Telescope Array}}.
\newblock WSP, 11, 2018.

\bibitem{CTAO:2024wvb}
{\bf CTAO} Collaboration, S.~Abe {\em et~al.}, {\it {Dark matter line searches with the Cherenkov Telescope Array}},  {\em JCAP} {\bf 07} (2024) 047, [\href{http://arxiv.org/abs/2403.04857}{{\tt arXiv:2403.04857}}].

\bibitem{Fermi-LAT:2011vow}
{\bf Fermi-LAT} Collaboration, M.~Ackermann {\em et~al.}, {\it {Constraining Dark Matter Models from a Combined Analysis of Milky Way Satellites with the Fermi Large Area Telescope}},  {\em Phys. Rev. Lett.} {\bf 107} (2011) 241302, [\href{http://arxiv.org/abs/1108.3546}{{\tt arXiv:1108.3546}}].

\bibitem{Zeldovich:1974uw}
Y.~B. Zeldovich, I.~Y. Kobzarev, and L.~B. Okun, {\it {Cosmological Consequences of the Spontaneous Breakdown of Discrete Symmetry}},  {\em Zh. Eksp. Teor. Fiz.} {\bf 67} (1974) 3--11.

\bibitem{Kibble:1976sj}
T.~W.~B. Kibble, {\it {Topology of Cosmic Domains and Strings}},  {\em J. Phys. A} {\bf 9} (1976) 1387--1398.

\bibitem{Vilenkin:1981zs}
A.~Vilenkin, {\it {Gravitational Field of Vacuum Domain Walls and Strings}},  {\em Phys. Rev. D} {\bf 23} (1981) 852--857.

\bibitem{Coulson:1995nv}
D.~Coulson, Z.~Lalak, and B.~A. Ovrut, {\it {Biased domain walls}},  {\em Phys. Rev. D} {\bf 53} (1996) 4237--4246.

\bibitem{Krajewski:2021jje}
T.~Krajewski, J.~H. Kwapisz, Z.~Lalak, and M.~Lewicki, {\it {Stability of domain walls in models with asymmetric potentials}},  {\em Phys. Rev. D} {\bf 104} (2021), no.~12 123522, [\href{http://arxiv.org/abs/2103.03225}{{\tt arXiv:2103.03225}}].

\bibitem{Borboruah:2022eex}
Z.~A. Borboruah and U.~A. Yajnik, {\it {Left-right symmetry breaking and gravitational waves: A tale of two phase transitions}},  {\em Phys. Rev. D} {\bf 110} (2024), no.~4 043016, [\href{http://arxiv.org/abs/2212.05829}{{\tt arXiv:2212.05829}}].

\bibitem{Yajnik:1998sw}
U.~A. Yajnik, H.~Widyan, D.~Choudhari, S.~Mahajan, and A.~Mukherjee, {\it {Topological defects in the left-right symmetric model and their relevance to cosmology}},  {\em Phys. Rev. D} {\bf 59} (1999) 103508, [\href{http://arxiv.org/abs/hep-ph/9812406}{{\tt hep-ph/9812406}}].

\bibitem{Hiramatsu:2012sc}
T.~Hiramatsu, M.~Kawasaki, K.~Saikawa, and T.~Sekiguchi, {\it {Axion cosmology with long-lived domain walls}},  {\em JCAP} {\bf 01} (2013) 001, [\href{http://arxiv.org/abs/1207.3166}{{\tt arXiv:1207.3166}}].

\bibitem{Borah:2022wdy}
D.~Borah and A.~Dasgupta, {\it {Probing left-right symmetry via gravitational waves from domain walls}},  {\em Phys. Rev. D} {\bf 106} (2022), no.~3 035016, [\href{http://arxiv.org/abs/2205.12220}{{\tt arXiv:2205.12220}}].

\bibitem{Sikivie:1982qv}
P.~Sikivie, {\it {Of Axions, Domain Walls and the Early Universe}},  {\em Phys. Rev. Lett.} {\bf 48} (1982) 1156--1159.

\bibitem{Gelmini:1988sf}
G.~B. Gelmini, M.~Gleiser, and E.~W. Kolb, {\it {Cosmology of Biased Discrete Symmetry Breaking}},  {\em Phys. Rev. D} {\bf 39} (1989) 1558.

\bibitem{Larsson:1996sp}
S.~E. Larsson, S.~Sarkar, and P.~L. White, {\it {Evading the cosmological domain wall problem}},  {\em Phys. Rev. D} {\bf 55} (1997) 5129--5135, [\href{http://arxiv.org/abs/hep-ph/9608319}{{\tt hep-ph/9608319}}].

\bibitem{Bai:2023cqj}
Y.~Bai, T.-K. Chen, and M.~Korwar, {\it {QCD-collapsed domain walls: QCD phase transition and gravitational wave spectroscopy}},  {\em JHEP} {\bf 12} (6, 2023) 194, [\href{http://arxiv.org/abs/2306.17160}{{\tt arXiv:2306.17160}}].

\bibitem{Kallosh:1995hi}
R.~Kallosh, A.~D. Linde, D.~A. Linde, and L.~Susskind, {\it {Gravity and global symmetries}},  {\em Phys. Rev. D} {\bf 52} (1995) 912--935, [\href{http://arxiv.org/abs/hep-th/9502069}{{\tt hep-th/9502069}}].

\bibitem{Witten:2017hdv}
E.~Witten, {\it {Symmetry and Emergence}},  {\em Nature Phys.} {\bf 14} (2018), no.~2 116--119, [\href{http://arxiv.org/abs/1710.01791}{{\tt arXiv:1710.01791}}].

\bibitem{Rai:1992xw}
B.~Rai and G.~Senjanovic, {\it {Gravity and domain wall problem}},  {\em Phys. Rev. D} {\bf 49} (1994) 2729--2733, [\href{http://arxiv.org/abs/hep-ph/9301240}{{\tt hep-ph/9301240}}].

\bibitem{King:2023ayw}
S.~F. King, R.~Roshan, X.~Wang, G.~White, and M.~Yamazaki, {\it {Quantum gravity effects on dark matter and gravitational waves}},  {\em Phys. Rev. D} {\bf 109} (8, 2024) 024057, [\href{http://arxiv.org/abs/2308.03724}{{\tt arXiv:2308.03724}}].

\bibitem{King:2023ztb}
S.~F. King, R.~Roshan, X.~Wang, G.~White, and M.~Yamazaki, {\it {Quantum Gravity Effects on Fermionic Dark Matter and Gravitational Waves}},  \href{http://arxiv.org/abs/2311.12487}{{\tt arXiv:2311.12487}}.

\bibitem{Borah:2024kfn}
D.~Borah, N.~Das, and R.~Roshan, {\it {Observable gravitational waves and \ensuremath{\Delta}Neff with global lepton number symmetry and dark matter}},  {\em Phys. Rev. D} {\bf 110} (6, 2024) 075042, [\href{http://arxiv.org/abs/2406.04404}{{\tt arXiv:2406.04404}}].

\bibitem{Gouttenoire:2025ofv}
Y.~Gouttenoire, S.~F. King, R.~Roshan, X.~Wang, G.~White, and M.~Yamazaki, {\it {Cosmological Consequences of Domain Walls Biased by Quantum Gravity}},  \href{http://arxiv.org/abs/2501.16414}{{\tt arXiv:2501.16414}}.

\bibitem{Barr:1992qq}
S.~M. Barr and D.~Seckel, {\it {Planck scale corrections to axion models}},  {\em Phys. Rev. D} {\bf 46} (1992) 539--549.

\bibitem{Dias:2002gg}
A.~G. Dias, V.~Pleitez, and M.~D. Tonasse, {\it {Naturally light invisible axion in models with large local discrete symmetries}},  {\em Phys. Rev. D} {\bf 67} (2003) 095008, [\href{http://arxiv.org/abs/hep-ph/0211107}{{\tt hep-ph/0211107}}].

\bibitem{Carpenter:2009zs}
L.~M. Carpenter, M.~Dine, and G.~Festuccia, {\it {Dynamics of the Peccei Quinn Scale}},  {\em Phys. Rev. D} {\bf 80} (2009) 125017, [\href{http://arxiv.org/abs/0906.1273}{{\tt arXiv:0906.1273}}].

\bibitem{Hiramatsu:2010yz}
T.~Hiramatsu, M.~Kawasaki, and K.~Saikawa, {\it {Gravitational Waves from Collapsing Domain Walls}},  {\em JCAP} {\bf 05} (2010) 032, [\href{http://arxiv.org/abs/1002.1555}{{\tt arXiv:1002.1555}}].

\bibitem{Hiramatsu:2013qaa}
T.~Hiramatsu, M.~Kawasaki, and K.~Saikawa, {\it {On the estimation of gravitational wave spectrum from cosmic domain walls}},  {\em JCAP} {\bf 02} (2014) 031, [\href{http://arxiv.org/abs/1309.5001}{{\tt arXiv:1309.5001}}].

\bibitem{Kadota:2015dza}
K.~Kadota, M.~Kawasaki, and K.~Saikawa, {\it {Gravitational waves from domain walls in the next-to-minimal supersymmetric standard model}},  {\em JCAP} {\bf 10} (2015) 041, [\href{http://arxiv.org/abs/1503.06998}{{\tt arXiv:1503.06998}}].

\bibitem{Paul:2020wbz}
A.~Paul, U.~Mukhopadhyay, and D.~Majumdar, {\it {Gravitational Wave Signatures from Domain Wall and Strong First-Order Phase Transitions in a Two Complex Scalar extension of the Standard Model}},  {\em JHEP} {\bf 05} (10, 2021) 223, [\href{http://arxiv.org/abs/2010.03439}{{\tt arXiv:2010.03439}}].

\bibitem{Maggiore:1999vm}
M.~Maggiore, {\it {Gravitational wave experiments and early universe cosmology}},  {\em Phys. Rept.} {\bf 331} (2000) 283--367, [\href{http://arxiv.org/abs/gr-qc/9909001}{{\tt gr-qc/9909001}}].

\bibitem{Cyburt:2015mya}
R.~H. Cyburt, B.~D. Fields, K.~A. Olive, and T.-H. Yeh, {\it {Big Bang Nucleosynthesis: 2015}},  {\em Rev. Mod. Phys.} {\bf 88} (2016) 015004, [\href{http://arxiv.org/abs/1505.01076}{{\tt arXiv:1505.01076}}].

\bibitem{CMB-HD:2022bsz}
{\bf CMB-HD} Collaboration, S.~Aiola {\em et~al.}, {\it {Snowmass2021 CMB-HD White Paper}},  \href{http://arxiv.org/abs/2203.05728}{{\tt arXiv:2203.05728}}.

\bibitem{CMB-bharat}
{\bf CMB-Bharat Collaboration} Collaboration, C.-B. A. I.~C. Consortium, {\it {Exploring Cosmic History and Origin: A proposal for a next generation space mission for near-ultimate measurements of the Cosmic Microwave Background (CMB) polarization and discovery of global CMB spectral distortions}}, .

\bibitem{doi:10.1146/annurev-nucl-102014-021908}
K.~N. Abazajian and M.~Kaplinghat, {\it Neutrino physics from the cosmic microwave background and large-scale structure},  {\em Annual Review of Nuclear and Particle Science} {\bf 66} (2016), no.~1 401--420, [\href{http://arxiv.org/abs/https://doi.org/10.1146/annurev-nucl-102014-021908}{{\tt https://doi.org/10.1146/annurev-nucl-102014-021908}}].

\bibitem{Alvarez:2019rhd}
M.~Alvarez {\em et~al.}, {\it {PICO: Probe of Inflation and Cosmic Origins}},  {\em Bull. Am. Astron. Soc.} {\bf 51} (2019), no.~7 194, [\href{http://arxiv.org/abs/1908.07495}{{\tt arXiv:1908.07495}}].

\bibitem{CORE:2017oje}
{\bf CORE} Collaboration, J.~Delabrouille {\em et~al.}, {\it {Exploring cosmic origins with CORE: Survey requirements and mission design}},  {\em JCAP} {\bf 04} (2018) 014, [\href{http://arxiv.org/abs/1706.04516}{{\tt arXiv:1706.04516}}].

\bibitem{SPT-3G:2014dbx}
{\bf SPT-3G} Collaboration, B.~A. Benson {\em et~al.}, {\it {SPT-3G: A Next-Generation Cosmic Microwave Background Polarization Experiment on the South Pole Telescope}},  {\em Proc. SPIE Int. Soc. Opt. Eng.} {\bf 9153} (2014) 91531P, [\href{http://arxiv.org/abs/1407.2973}{{\tt arXiv:1407.2973}}].

\bibitem{SimonsObservatory:2018koc}
{\bf Simons Observatory} Collaboration, P.~Ade {\em et~al.}, {\it {The Simons Observatory: Science goals and forecasts}},  {\em JCAP} {\bf 02} (2019) 056, [\href{http://arxiv.org/abs/1808.07445}{{\tt arXiv:1808.07445}}].

\bibitem{Yagi:2011wg}
K.~Yagi and N.~Seto, {\it {Detector configuration of DECIGO/BBO and identification of cosmological neutron-star binaries}},  {\em Phys. Rev. D} {\bf 83} (2011) 044011, [\href{http://arxiv.org/abs/1101.3940}{{\tt arXiv:1101.3940}}]. [Erratum: Phys.Rev.D 95, 109901 (2017)].

\bibitem{2017arXiv170200786A}
{\bf LISA} Collaboration, P.~Amaro-Seoane~et al, {\it {Laser Interferometer Space Antenna}},  {\em arXiv e-prints} (feb, 2017) arXiv:1702.00786, [\href{http://arxiv.org/abs/1702.00786}{{\tt arXiv:1702.00786}}].

\bibitem{Punturo_2010}
{\bf ET Collaboration} Collaboration, M.~Punturo~et al, {\it The einstein telescope: a third-generation gravitational wave observatory},  {\em Classical and Quantum Gravity} {\bf 27} (sep, 2010) 194002.

\bibitem{Garcia-Bellido:2021zgu}
J.~Garcia-Bellido, H.~Murayama, and G.~White, {\it {Exploring the early Universe with Gaia and Theia}},  {\em JCAP} {\bf 12} (4, 2021) 023, [\href{http://arxiv.org/abs/2104.04778}{{\tt arXiv:2104.04778}}].

\bibitem{Sesana:2019vho}
A.~Sesana {\em et~al.}, {\it {Unveiling the gravitational universe at $\mu$-Hz frequencies}},  {\em Exper. Astron.} {\bf 51} (2021), no.~3 1333--1383, [\href{http://arxiv.org/abs/1908.11391}{{\tt arXiv:1908.11391}}].

\bibitem{Antoniadis:2023ott}
J.~Antoniadis {\em et~al.}, {\it {The second data release from the European Pulsar Timing Array III. Search for gravitational wave signals}},  \href{http://arxiv.org/abs/2306.16214}{{\tt arXiv:2306.16214}}.

\bibitem{Weltman:2018zrl}
A.~Weltman {\em et~al.}, {\it {Fundamental physics with the Square Kilometre Array}},  {\em Publ. Astron. Soc. Austral.} {\bf 37} (2020) e002, [\href{http://arxiv.org/abs/1810.02680}{{\tt arXiv:1810.02680}}].

\bibitem{NANOGrav:2023gor}
{\bf NANOGrav} Collaboration, G.~Agazie {\em et~al.}, {\it {The NANOGrav 15 yr Data Set: Evidence for a Gravitational-wave Background}},  {\em Astrophys. J. Lett.} {\bf 951} (2023), no.~1 L8, [\href{http://arxiv.org/abs/2306.16213}{{\tt arXiv:2306.16213}}].

\bibitem{Kogut:2024vbi}
A.~Kogut {\em et~al.}, {\it {The Primordial Inflation Explorer (PIXIE): Mission Design and Science Goals}},  \href{http://arxiv.org/abs/2405.20403}{{\tt arXiv:2405.20403}}.

\bibitem{Basu:2019rzm}
K.~Basu {\em et~al.}, {\it {A space mission to map the entire observable universe using the CMB as a backlight: Voyage 2050 science white paper}},  {\em Exper. Astron.} {\bf 51} (2021), no.~3 1555--1591, [\href{http://arxiv.org/abs/1909.01592}{{\tt arXiv:1909.01592}}].

\bibitem{Kitajima:2023cek}
N.~Kitajima, J.~Lee, K.~Murai, F.~Takahashi, and W.~Yin, {\it {Gravitational waves from domain wall collapse, and application to nanohertz signals with QCD-coupled axions}},  {\em Phys. Lett. B} {\bf 851} (6, 2024) 138586, [\href{http://arxiv.org/abs/2306.17146}{{\tt arXiv:2306.17146}}].

\bibitem{Ferreira:2023jbu}
R.~Z. Ferreira, S.~Gasparotto, T.~Hiramatsu, I.~Obata, and O.~Pujolas, {\it {Axionic defects in the CMB: birefringence and gravitational waves}},  {\em JCAP} {\bf 05} (2024) 066, [\href{http://arxiv.org/abs/2312.14104}{{\tt arXiv:2312.14104}}].

\bibitem{Dankovsky:2024zvs}
I.~Dankovsky, E.~Babichev, D.~Gorbunov, S.~Ramazanov, and A.~Vikman, {\it {Revisiting evolution of domain walls and their gravitational radiation with CosmoLattice}},  {\em JCAP} {\bf 09} (2024) 047, [\href{http://arxiv.org/abs/2406.17053}{{\tt arXiv:2406.17053}}].

\bibitem{Dunsky:2021tih}
D.~I. Dunsky, A.~Ghoshal, H.~Murayama, Y.~Sakakihara, and G.~White, {\it {Gravitational Wave Gastronomy}},  {\em Phys. Rev. D} {\bf 106} (11, 2021) 075030, [\href{http://arxiv.org/abs/2111.08750}{{\tt arXiv:2111.08750}}].

\bibitem{Schmitz:2020syl}
K.~Schmitz, {\it {New Sensitivity Curves for Gravitational-Wave Signals from Cosmological Phase Transitions}},  {\em JHEP} {\bf 01} (2, 2021) 097, [\href{http://arxiv.org/abs/2002.04615}{{\tt arXiv:2002.04615}}].

\bibitem{Heeck:2023soj}
J.~Heeck, J.~Heisig, and A.~Thapa, {\it {Testing Dirac leptogenesis with the cosmic microwave background and proton decay}},  {\em Phys. Rev. D} {\bf 108} (2023), no.~3 035014, [\href{http://arxiv.org/abs/2304.09893}{{\tt arXiv:2304.09893}}].

\bibitem{Griest:1989wd}
K.~Griest and M.~Kamionkowski, {\it {Unitarity Limits on the Mass and Radius of Dark Matter Particles}},  {\em Phys. Rev. Lett.} {\bf 64} (1990) 615.

\bibitem{Hattori:2015xla}
H.~Hattori, T.~Kobayashi, N.~Omoto, and O.~Seto, {\it {Entropy production by domain wall decay in the NMSSM}},  {\em Phys. Rev. D} {\bf 92} (2015), no.~10 103518, [\href{http://arxiv.org/abs/1510.03595}{{\tt arXiv:1510.03595}}].

\bibitem{Kawasaki:2004rx}
M.~Kawasaki and F.~Takahashi, {\it {Late-time entropy production due to the decay of domain walls}},  {\em Phys. Lett. B} {\bf 618} (2005) 1--6, [\href{http://arxiv.org/abs/hep-ph/0410158}{{\tt hep-ph/0410158}}].

\bibitem{Sato:2017dkf}
S.~Sato {\em et~al.}, {\it {The status of DECIGO}},  {\em J. Phys. Conf. Ser.} {\bf 840} (2017), no.~1 012010.

\bibitem{Ishikawa:2020hlo}
T.~Ishikawa {\em et~al.}, {\it {Improvement of the target sensitivity in DECIGO by optimizing its parameters for quantum noise including the effect of diffraction loss}},  {\em Galaxies} {\bf 9} (2021), no.~1 14, [\href{http://arxiv.org/abs/2012.11859}{{\tt arXiv:2012.11859}}].

\bibitem{Kannike:2012pe}
K.~Kannike, {\it {Vacuum Stability Conditions From Copositivity Criteria}},  {\em Eur. Phys. J. C} {\bf 72} (2012) 2093, [\href{http://arxiv.org/abs/1205.3781}{{\tt arXiv:1205.3781}}].

\bibitem{Antusch:2002rr}
S.~Antusch, J.~Kersten, M.~Lindner, and M.~Ratz, {\it {Neutrino mass matrix running for nondegenerate seesaw scales}},  {\em Phys. Lett. B} {\bf 538} (2002) 87--95, [\href{http://arxiv.org/abs/hep-ph/0203233}{{\tt hep-ph/0203233}}].

\bibitem{Brdar:2019iem}
V.~Brdar, A.~J. Helmboldt, S.~Iwamoto, and K.~Schmitz, {\it {Type-I Seesaw as the Common Origin of Neutrino Mass, Baryon Asymmetry, and the Electroweak Scale}},  {\em Phys. Rev. D} {\bf 100} (2019) 075029, [\href{http://arxiv.org/abs/1905.12634}{{\tt arXiv:1905.12634}}].

\end{thebibliography}
\end{document}